# Radial Velocity Prospects Current and Future

## A White Paper Report prepared by the Study Analysis Group 8 for the Exoplanet Program Analysis Group (ExoPAG)

**Contributing Authors:**

| | |
|---|---|
| Peter Plavchan | Missouri State University, |
| Dave Latham | Harvard-Smithsonian Center for Astrophysics, |

Radial Velocity Study Analysis Group co-chairs and
ExoPAG Executive Committee members,

| | |
|---|---|
| Scott Gaudi | Ohio State University, |

ExoPAG Executive Committee Chair,

| | |
|---|---|
| Justin Crepp | Notre Dame, |
| Dumusque Xavier | Harvard-Smithsonian Center for Astrophysics, |
| Gabor Furesz | MIT, |
| Andrew Vanderburg | Harvard University, |
| Cullen Blake | University of Pennsylvania, |
| Debra Fischer | Yale University, |
| Lisa Prato | Lowell Observatory, |
| Russel White | Georgia State University, |
| Valeri Makarov | U.S. Naval Observatory, |
| Geoff Marcy | University of California, Berkeley, |
| Karl Stapelfeldt | NASA Goddard, |
| Raphaëlle Haywood | St Andrews University, |
| Andrew Collier-Cameron | St Andrews University, |
| Andreas Quirrenbach | University of Heidelberg, |
| Suvrath Mahadevan | Penn State University, |
| Guillem Anglada | Queen Mary, University of London, |
| Philip Muirhead | Boston University |

# 1. Table of Contents

## 2. Executive Summary

In this white paper report, we present an assessment of the current capabilities and the future potential of the precise radial velocity (PRV) method to advance the NASA goal to “search for planetary bodies and Earth-like planets in orbit around other stars.” (U.S. National Space Policy, June 28, 2010). PRVs complement other exoplanet detection methods, for example offering a direct path to obtaining the bulk density and thus the structure and composition of transiting exoplanets.

PRVs will provide essential NASA mission support for:

| **Mission** | **Target identification for mission science yield optimization** | **Follow-up validation & characterization of low mass transiting exoplanets** | **Exoplanet mass & orbit determination** |
|---|---|---|---|
| **Kepler** | | ✔ | ✔ |
| **K2** | ✔ | ✔ | ✔ |
| **TESS** | ✔ | ✔ | ✔ |
| **JWST** | ✔ | ✔ | ✔ |
| **AFTA/probe Coronagraph or Starshade direct imaging** | ✔ | | ✔ |
| **Future Flagship direct imaging** | ✔ | | ✔ |

**Table 1.** Summary of PRV support for NASA mission science objectives.

Our analysis builds upon previous community input, including the ExoPlanet Community Report chapter on radial velocities in 2008, the 2010 Decadal Survey of Astronomy, the Penn State Precise Radial Velocities Workshop response to the Decadal Survey in 2010, and the NSF Portfolio Review in 2012. The radial-velocity detection of exoplanets is strongly endorsed by both the Astro 2010 Decadal Survey "New Worlds, New Horizons" and the NSF Portfolio Review, and the community has recommended robust investment in PRVs:

*"The first task on the ground is to improve the precision radial velocity method by which the majority of the close to 500 known exoplanets have been discovered. … Using existing large ground-based or new dedicated mid-size ground-based telescopes equipped with a new generation of high-resolution spectrometers in the optical and near-infrared, a velocity goal of 10 to 20 centimeters per second is realistic."* – page 7-8, Astro 2010 Decadal Survey "New Worlds, New Horizons"

The demands on telescope time for the above mission support, especially for systems of small planets, will exceed the number of nights available using instruments now in operation by a factor of at least several for TESS alone. Pushing down towards true Earth twins will require more photons (i.e. larger telescopes), more stable spectrographs than are currently available, better calibration, and better correction for stellar jitter.

We outline four hypothetical situations for PRV work necessary to meet NASA mission exoplanet science objectives:

1) Access to full-time long-term use of a 4-m class telescope equipped with a HARPS-like instrument dedicated to PRV observations, to supplement the 100 nights per year projected to be available to the TESS mission, and to support a long-term survey needed to identify optimum targets for an AFTA-type direct imaging mission.

2) Access to long-term use of a 10-m class telescope with a next generation spectrograph stable at the level of 5 cm/s, at the level of nominally 20% of the time, to push towards the confirmation and characterization of true Earth twins. This will also require significant advances in the ability to correct for stellar jitter.

3) Access to an IR spectrograph or second arm of a visible spectrograph on a 4-m class telescope equipped with a CARMENES-like instrument stable at the level of <1 m/s dedicated to PRV observations, to be available to the TESS mission, to assess the removal of wavelength-dependent stellar-jitter, and to survey nearby M dwarfs for habitable planets.

4) A NIR or visible diffraction-limited spectrograph for a large-aperture telescope with extreme AO to achieve unprecedented precision and compete with ESPRESSO.

These scenarios are all mid-scale projects that exceed the cost of a NASA individual investigator grant. Approximately, these efforts will each cost ~$2-5M/yr over a decade split between instrument development and operations, will require the procurement of available telescope resources, and will require significant investment in technology development and

multi-faceted data analysis.  With such efforts, the potential to achieve the Decadal Survey PRV prediction remains.

In Section 3 we present the current and future programmatic relevance of the radial-velocity (RV)  method for current and planned NASA missions, including TESS and JWST.  In Section 4 we present the analysis of potential facilities to meet NASA mission science objectives.   In Section 5, we present constraining astrophysical sources of RV noise, or “jitter,” and our current understanding thereof.  In Section 6, we present areas of technology development for advances in instrumental precision and data analysis.  In Section 7, we present a compilation of United States based community input on the PRV method from the past decade, and conclude with definitions, acknowledgments, and references cited.

# 3. NASA Programmatic Considerations

Ground-based PRV capabilities play a critical role in support of NASA exoplanet missions, past, present, and future. We live in a golden age for the study of exoplanets, largely due to progress enabled by the discovery of transiting exoplanets. PRV surveys identified hot Jupiters that were subsequently shown to transit and led to pioneering studies of exoplanet atmospheres with NASA's Hubble Space Telescope and Spitzer Space Telescope (e.g. Charbonneau et al. 2002, Deming et al. 2005, Knutson et al. 2007).

## 3.1 Kepler and K2

More recently, NASA's Kepler mission has shown that most stars host planets smaller than Neptune (4 Earth radii), often in compact systems with coplanar orbits (Latham et al. 2011, Howard et al. 2012, Fressin et al. 2013). PRVs are playing an essential role in confirming and characterizing the bulk properties of the most accessible Kepler planet candidates by providing orbital solutions and mass determinations to complement the sizes measured by Kepler, thus leading to the characterization of a handful of rocky planets with compact atmospheres.

Determining the masses for small Kepler planet candidates with sufficient accuracy to show that they are rocky has been a severe challenge to the best PRV instruments in the world. For example, more than 100 nights with HIRES on Keck 1 were dedicated to following up 22 stars hosting 49 Kepler planet candidates, based on the prediction that orbital motion could be detected for a planet with an Earth-like density. This effort (Marcy et al. 2013) yielded 28 mass determinations with accuracies between 1 and 3 sigma - good enough to confirm the planetary nature of the candidate, but not good enough to characterize the planet as rocky with a compact atmosphere.

Even before the launch of Kepler it was realized that the limited access to state-of-the-art PRV follow up would be a bottleneck to the confirmation and characterization of small Kepler planets. An international collaboration was established to build a northern copy of the highly successful HARPS on the 3.6-m telescope at the European Southern Observatory, with the primary goal of following up rocky planet candidates identified by Kepler. HARPS-N began science operations on the 3.6-m Telescopio Nazionale Galileo on La Palma in August 2012. There are now five confirmed rocky planets with mass determinations better than 5 sigma and with internal structure and composition remarkably similar to Earth and Venus (Dressing et al. 2015): CoRoT-7b from the original HARPS, Kepler-36b from a photo-dynamical model, Kepler-10b and 93b from HARPS-N, and Kepler-78b (Figure 1).

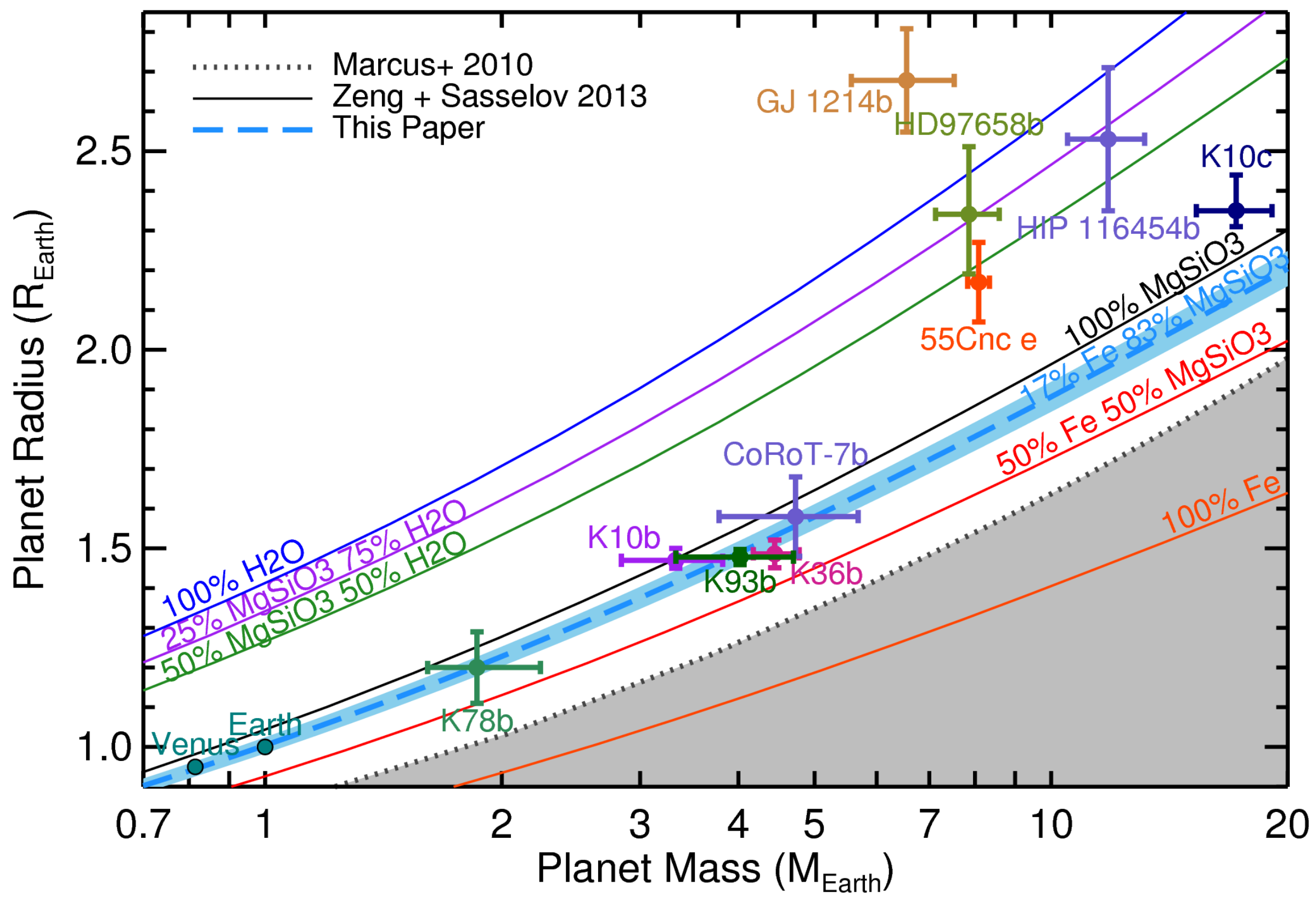

**Figure 1.** *Reproduced from Dressing et al. (2015, Figure 4): The mass-radius diagram for planets smaller than 2.7$R_{Earth}$ with masses measured to better than 20% precision (Dressing et al. 2015). The shaded gray region in the lower right indicates planets with iron content exceeding the maximum value predicted from models of collisional stripping (Marcus et al. 2010). The solid lines are theoretical mass-radius curves (Zeng & Sasselov 2013) for planets with compositions of 100% $H_2O$ (blue), 25% $MgSiO_3$ – 75% H2O (purple), 50% $MgSiO_3$ – 50% H2O (green), 100% $MgSiO_3$ (black), 50% Fe – 50% $MgSiO_3$ (red), and 100% Fe (orange). The best-fit relation based on the Zeng & Sasselov (2013) models is the dashed light blue line representing an Earth-like composition (modeled as 17% iron and 83% magnesium silicate using a fully-differentiated, two-component model). The shaded region surrounding the line indicates the 2% dispersion in radius expected from variation in Mg/Si and Fe/Si ratios (Grasset et al. 2009).*

With 77 observations from HIRES and 109 from HARPS-N, Kepler-78-b has the best mass determination for an Earth-sized rocky exoplanet to date, with a 17% mass uncertainty (Pepe et al. 2013, Howard et al. 2013). At V=12 mag, Kepler-78 represents a bright Kepler target. The semi-amplitude of its orbital motion, 2 m/s, is large enough to be determined with HIRES and HARPS-N because of the very short period, 8.5 hours. Determining the mass of a true Earth twin from the reflex motion of 10 cm/s due to a one-year orbit around a star like the Sun for a typical Kepler candidate at V=14 mag is hopelessly beyond the reach of present PRV instruments.

NASA's K2 mission, a two-reaction-wheel reincarnation of Kepler, has embarked on a two-year survey of 10 fields along the ecliptic. K2 is expected to provide a significant number of additional short-period rocky planet candidates suitable for follow up with existing PRV facilities in both the north and south. It remains to be seen whether the photometric precision achieved by K2 together with the limit of 3 months per campaign field will allow detection of interesting small planets at the same rate as the baseline Kepler mission, but several initial discoveries have been reported (Crossfield et al. 2015).

## 3.2 TESS and JWST

NASA's Transiting Exoplanet Survey Satellite (TESS), scheduled for launch in 2017, has a primary ground-based follow-up observation goal of measuring masses for at least 50 transiting planets smaller than 4 Earth radii. Nearby transiting systems discovered by TESS will provide prime targets for spectroscopic studies of exoplanet atmospheres with NASA's James Webb Space Telescope (JWST) and for follow-up observations by other missions under development, such as CHEOPS and PLATO.

Selection of the very best transiting-system targets for JWST and interpretation of JWST spectroscopy of a planetary atmosphere will require knowledge of the structure and composition of the planet's interior. This is one of the drivers for measuring masses for small TESS planets that are the most promising targets for JWST. The transiting systems identified by TESS will typically be 30 to 100 times brighter than the Kepler candidates. This will bring many more small planets within the reach of the PRV facilities that will be available for following up TESS candidates. The TESS Science Office is responsible for organizing the TESS Follow-up Observing Program (TFOP). Many of the top PRV teams are already on board, representing more than 4000 hours of telescope time for PRV follow up during the baseline mission (Ricker et al. 2014). While this may be sufficient to secure the 50 promised mass determinations, it will leave many interesting targets untouched for future follow up by the community

## 3.3 Future space-based direct imaging missions

The exoplanet science for future NASA space-based direct-imaging missions - e.g. AFTA-WFIRST with a coronagraph or starshade, a Probe class mission, or a >8 meter flagship mission - falls into two possible categories:

1) Spectroscopy and sini determinations for known RV planets; and
2) Searches for planets beyond RV limits and their spectral characterization.

Mission design requirements for each category necessitate an informed assessment of what we know already about exoplanet populations. In particular, what is the planet population around accessible direct imaging targets that RV currently has and has not detected? Additionally, what would the detected and undetected planet populations look like in the 2020s, when such a direct imaging mission may launch?

The goal of answering these questions is to ascertain how much "blind searching" the direct imaging space missions should be doing. If direct imaging target stars are being observed well with the RV method, then not much direct imaging mission time is needed doing blind searches for new exoplanets. However, if the RV survey completeness leave lots of room for new direct imaging discoveries, then substantial direct imaging mission time should be dedicated to such searches. Consequently, a detailed understanding of the RV survey completeness and sensitivity is crucial to future direct imaging mission science. The Exo-C interim report (http://exep.jpl.nasa.gov/stdt/Exo-C_InterimReport.pdf) recommended a detailed review of existing RV data for our best imaging search targets. AFTA-Starshade arrived at the preliminary conclusion that there is a need for a long-term decade-long RV survey for AFTA direct imaging targets, not only for identifying targets, but also for RV limit curves for targets which have been ruled out to have Jovian companions from RV surveys. There are currently less than a half-dozen known exoplanets, all gas giants, suitable for direct imaging (SAG 9 report).

Little research has been done to date exploring the overlap between direct imaging targets and RV surveys identifying long-term trends. The PRV method has demonstrated the ability to identify stellar and brown dwarf direct-imaging targets based on long-term RV trends around nearby bright stars. For example, the TRENDS survey (Crepp et al. 2014,2012) has taken the long-term RV variations from the California Planet Search and followed them up with high contrast imaging to look for low mass companions at large separations. Can this approach be extended to lower mass exoplanets at larger orbital separations?

Recognizing this knowledge gap, the NASA Exoplanet Exploration Program contracted with Andrew Howard to compare the direct imaging target lists for Exo-C, Exo-S and AFTA studies with RV survey results from the California Planet Search and eta Earth samples observed with the Hamilton Echelle and HIRES high-resolution spectrographs at the Lick and Keck Observatories, respectively. We briefly summarize the results of the report from Howard and Fulton (2014). The Lick and Keck RV surveys targeted mostly main sequence dwarfs in a sample of FGK spectral types and main sequence stars from the Bright Star Catalog and Gliese-Jahreiss catalog, in addition to other sources. These surveys did not include earlier spectral-type stars, evolved stars, or young stars, in some cases due to lack of absorption lines for RV information content, and in other cases because of RV jitter due to stellar activity and the rotational modulation of starspots, and to pulsations. Southern stars with declinations below -30$^{o}$, representing a quarter of the sky, were not targeted from the Lick and Keck Observatories. Many suitable southern targets have been observed with CORALIE (e.g. Pepe et al. 2002) and the original HARPS (e.g. Pepe et al. 2011), but were not addressed in the Howard and Fulton (2014) report.

Howard & Fulton (2014) identified 76 targets from the Lick and Keck RV surveys that overlap with the 312 stars in the Exo-C, Exo-S and AFTA studies target lists (Figure 2). In other words, approximately 75% of entries in direct-imaging target lists for proposed future NASA missions have not been observed with the Lick and Keck RV surveys. One-third of the direct-imaging targets that have not been observed with Lick and Keck can be attributed to being too far South on the sky (below -30°). The remaining two-thirds are either earlier-type stars or have active photospheres. Since direct-imaging exoplanet searches prefer brighter and younger stars, the direct-imaging target lists are biased towards earlier-type stars than targeted by RV surveys.

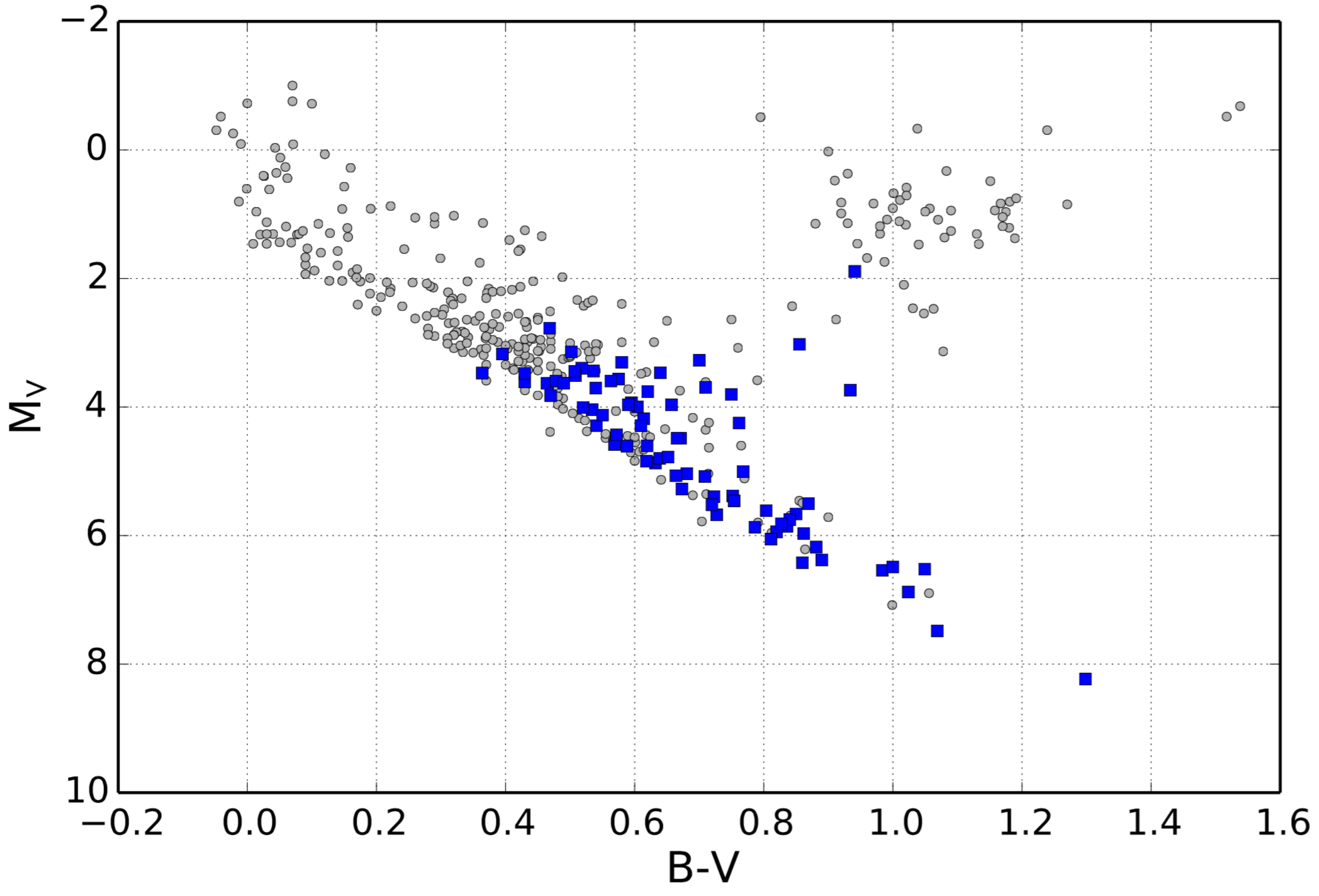

**Figure 2.** *Reproduced from Howard & Fulton (2014, Fig. 1). The color-magnitude diagram for stars in the Exo-S, Exo-C, and AFTA target lists with (blue squares) and without (gray circles) Doppler observations from the Lick and Keck Observatories. Doppler searches favor late-F through mid-M type dwarfs for planet detectability. Imaging searches typically prefer bright stars, which are dominated by early-type and evolved stars. The region of overlap encompasses primarily quiescent F8–K0 dwarfs.*

For the 76 RV targets in Howard & Fulton (2014) that overlap with direct-imaging targets, the median RV completeness limit is ~100 $M_{Earth}$ (~1/3 $M_{Jupiter}$) at 1 AU, decreasing at larger semi-major axes to ~1000 $M_{Earth}$ (~3 $M_{Jupiter}$) at ~8 AU, with a large variation of approximately one order of magnitude for individual stars (Figure 3). To summarize, current RV surveys are limited to the detection of Jovian-mass planets and sub-stellar companions at orbital separations amenable to direct imaging, for the targets that do overlap (~25%) between RV and direct-imaging samples.

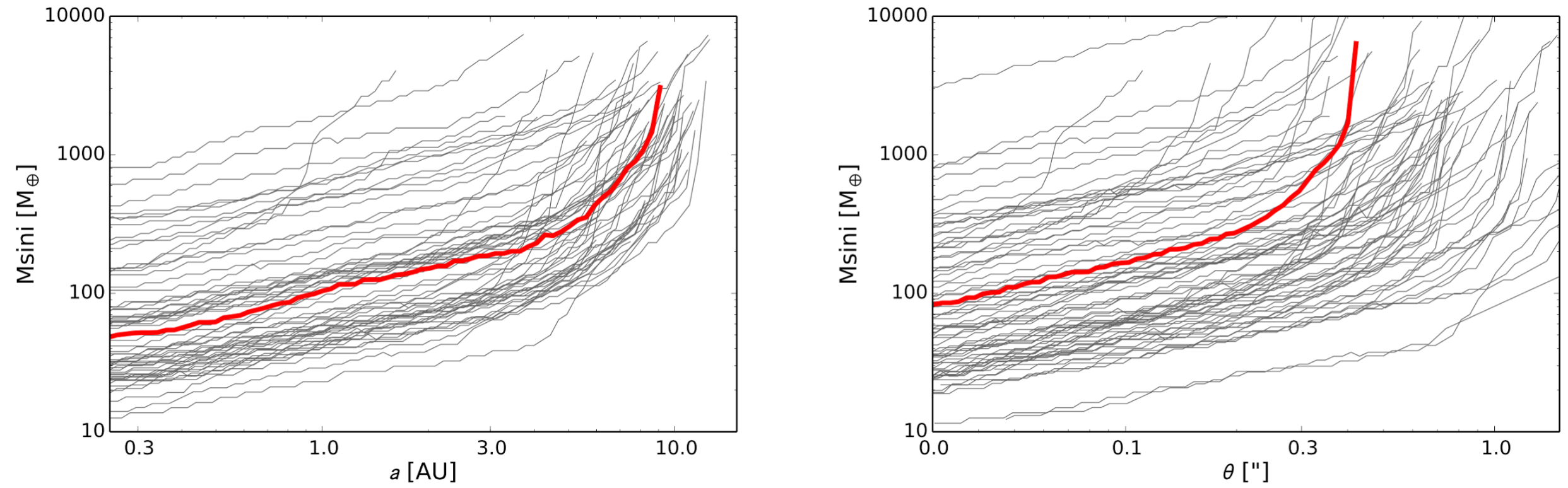

**Figure 3.** *Reproduced from Howard and Fulton (2014, Fig. 7): Contours showing 50% completeness limits for every star with Lick and Keck Doppler measurements (gray lines) and the median of those contours (red line). The panels show the same completeness curves as a function of semi-major axis (left) and projected separation (right).*

We compare these sensitivities to the expected sensitivities for an AFTA Coronagraph and starshade direct imaging mission. The period range of interest for a direct imaging mission is ~200 to 7000 days; planets with shorter periods will not be resolved from the star with current high-contrast imaging technology, whereas planets with longer periods will be too faint to detect with a 1-1.5 meter diameter telescope. Early estimates indicate that such a direct imaging mission should be sensitive to 2 $R_{Earth}$ objects out to 1.5 AU, 4 $R_{Earth}$ objects out to 3 AU, Saturn-sized objects to 7 AU, and Jupiter-sized objects out to 9 AU (Exo-C Interim Report).

We conclude that there is a clear phase space of exoplanets that could be discovered or characterized by direct imaging that have not been adequately probed by RV surveys to date. These include sub-Jovian planet masses around FGK stars, particularly for sub-Saturn planets between 1 and 7 AU, and exoplanets around A and early F stars in general. A task of this report is thus to investigate what might be feasible with the PRV method over the coming decade to detect sub-Saturn mass planets in this semi-major axis regime.

To extend these RV searches towards lower masses and longer orbital periods more suitable for direct-imaging mission goals (e.g. a Saturn at 5 AU), a NASA coordinated and sustained long-term (>10 yr) RV monitoring campaign of nearby stars is required. Indeed, it is not too early to get started on a PRV survey of potential targets for an eventual flagship direct-imaging mission, because the best targets will inevitably need to be at wider separations in orbits with longer periods around their host stars.

# 4 Analysis: Ground-Based Telescope Resources Necessary to meet NASA objectives

In this section, we analyze four hypothetical situations for PRV work necessary to meet NASA mission exoplanet science objectives, in particular follow-up of transit candidates from TESS and Kepler/K2, and precursor searches of direct imaging targets for a future direct imaging mission:

1) Access to full-time long-term use of a 4-m class telescope equipped with a HARPS-like instrument dedicated to PRV observations, to supplement the 100 nights per year projected to be available to the TESS mission, and to support a long-term survey needed to identify optimum targets for an AFTA-type direct imaging mission (e.g. Exo-C).

2) Access to long-term use of a 10-m class telescope with a next generation spectrograph stable at the level of 5 cm/s, at the level of nominally 20% of the time, to push towards the confirmation and characterization of true Earth twins. This will also require significant advances in the ability to correct for stellar jitter.

3) Access to an IR spectrograph or second arm of a visible spectrograph on a 4-m class telescope equipped with a CARMENES-like instrument stable at the level of <1 m/s dedicated to PRV observations, to be available to the TESS mission, to assess the removal of wavelength-dependent stellar-jitter, and to survey nearby M dwarfs for habitable planets.

4) Access to a NIR or visible diffraction-limited spectrograph for a large-aperture telescope with extreme AO to achieve unprecedented precision and compete with ESPRESSO.

All four scenarios we investigate are mid-scale projects that exceed the cost of a NASA individual investigator grant. Approximately, these efforts will cost ~$2-5M/yr each over a decade split between instrument development and operations, will require the procurement of available telescope resources, and will require significant investment in technology development and multi-faceted data analysis. With such efforts, the potential to achieve the Decadal Survey PRV prediction remains.

Dedicated facilities with stable instrumentation focused on a small number of targets with high cadence over a long time baseline is necessary to achieve NASA objectives, particularly on >4-meter class telescopes in the near-term (~20-50 cm/s), and >8 m class telescopes in the medium term (~10-20 cm/s) and >25 m telescopes in the long term (< 10 cm/s). These resources are required for primarily two reasons: the cadence necessary to model the effects of stellar jitter, and collecting a sufficient number of photons on nearby stars. The present state of the art for velocity stability is illustrated by the two HARPS spectrographs, which currently achieve a long-term RV zero point precision of ~70-80 cm/s.

Many of the advances in PRVs are primarily being led by European efforts (e.g. the original HARPS, and the performance expected for ESPRESSO, CARMENES, and CODEX). At over $10M apiece, next-generation PRV spectrographs have outgrown the PI-based grant model of

building an instrument at a single or small number of institutions for less than $5M. This can be addressed through programmatic means, both in terms of coordination, funding for technology development and the construction of major new PRV facilities, and the acquisition of suitable telescope resources.

In Section 4.1, we present a brief proposed standardization of the definition of Doppler Precision, which is a term that can often be taken to mean different things by the community. In Section 4.2, we quantify the expected survey capabilities of the first two spectrometer scenarios listed above in support of NASA mission objectives. In Sections 4.3 and 4.4, we compare the benefit of the latter two spectrometer scenarios listed above by quantifying the expected wavelength-dependent performance of visible and near-infrared PRV spectrometers, modeling both the photon noise and stellar jitter as a function of wavelength and stellar effective temperature, and the primary benefits of diffraction-limited spectrometers.

## 4.1 Definition of Doppler Precision

The community would benefit from a standardized definition of what is meant by precision in order to make accurate comparisons between instruments and Doppler analysis techniques. Ideally, future literature should include estimates of all three of these metrics.

- The single measurement precision is a function of SNR, analysis technique and spectral resolution. This metric reflects the floor of the Doppler precision for a given instrument and characterizes the instrumental precision. This is the only metric that can be accurately simulated.
- The short term RMS over a week or two of consecutive observations is a useful metric for characterizing instrumental precision because this timescale is long enough to capture instrument and analysis errors, but short compared to changes in fractional spot coverage or stellar magnetic fields for chromospherically inactive stars. Unfortunately, this metric cannot be simulated - it can only reflect the performance of an instrument after it is commissioned.
- The long term RMS (over a season or years) includes velocities from the stellar atmosphere as well as center of mass velocities from the star. This is the useable Doppler precision that limits what exoplanet semi-amplitudes a particular instrument can detect.

## 4.2 Survey Simulations

### 4.2.1 TESS and Exo-C Target Samples

We make use of the Exo-C target sample, and a simulated TESS candidate sample, to quantify the utility of NASA access to a PRV spectrometer, for either a precursor survey for Exo-C, or for dedicated follow-up observations of candidates for TESS. We show the apparent magnitude distribution of both samples in Figure 4. We restrict the Exo-C target sample to 172 stars with spectral types later than ~A4 (B-V>0.1) and a declination >-30 degrees. For TESS, we simulate a volume-scaled sample, with the brightest candidate at V=7.0, and 62 targets with V<10.0. This is consistent with the more rigorous simulations in Winn (2014). There is no overlap in apparent magnitude between the two target lists.

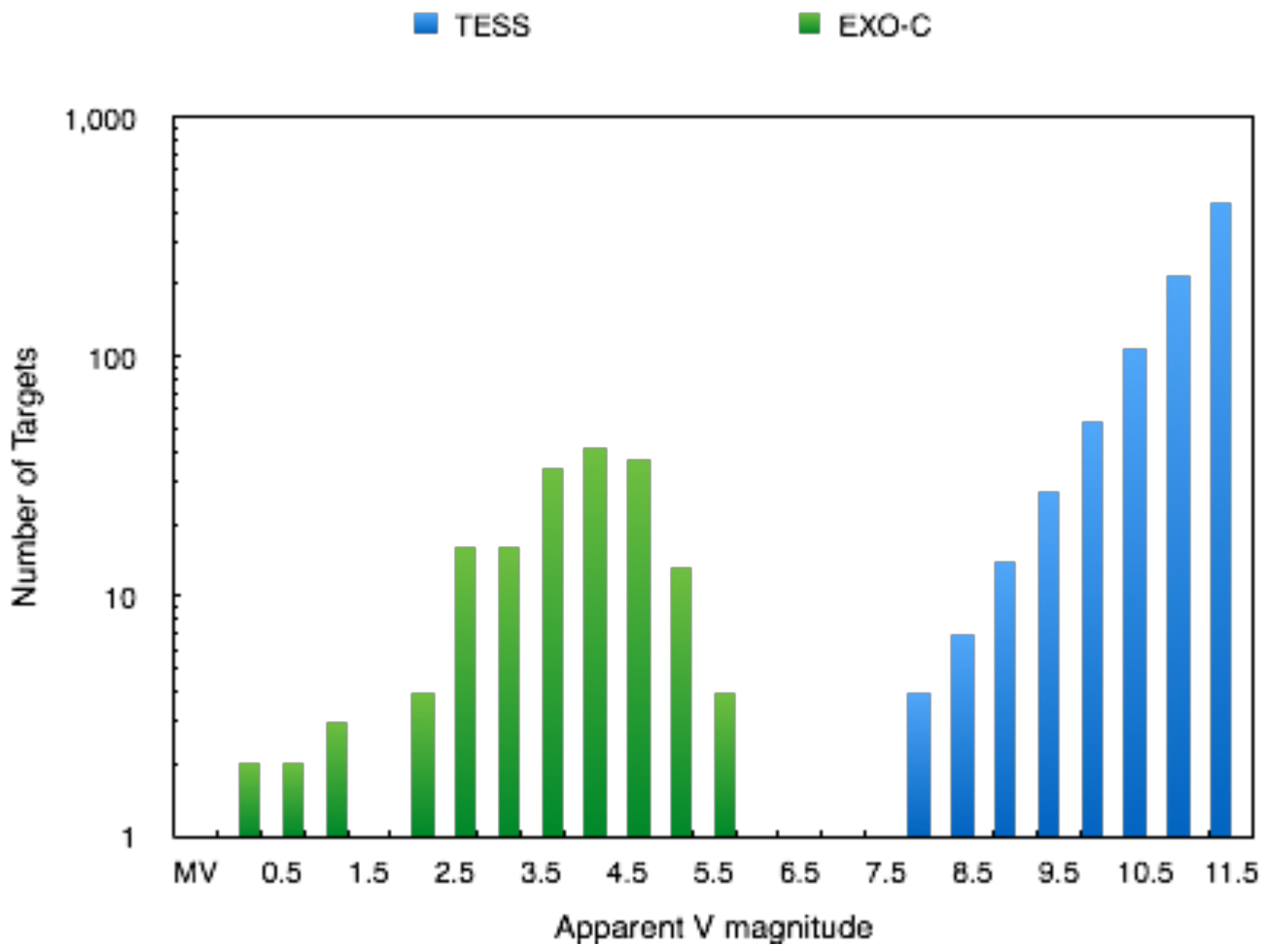

**Figure 4.** *Distribution of apparent V magnitudes for the Exo-C (green) targets with B-V>0.1 and a declination >-30 degrees, and simulated TESS (blue) targets.*

### 4.2.2 K Semi-amplitude Sensitivity

From Section 3.3, the Exo-C interim report (http://exep.jpl.nasa.gov/stdt/Exo-C_InterimReport.pdf) estimates a direct imaging detection sensitivity to approximately Saturn-sized objects out to ~7 AU. The RV semi-amplitude of a Saturn at the location of Jupiter in the Solar System is $K$=3.8 m/s. From the TESS Science Memo (Winn 2014), the expected RV semi-amplitudes of TESS candidate exoplanets range from ~1-4 m/s, with a typical value of 3 m/s. Consequently, we are interested in a survey that is able to reliably detect $K$=3 m/s and smaller signals. These are much larger signals than the $K$=9 cm/s RV semi-amplitude of an Earth-mass planet at 1 AU, but looking towards the future we extend our survey simulations down towards that detection threshold.

### 4.2.3 Choice of Telescope Aperture

For our analysis, we assume either dedicated (e.g. 100% time allocation) access to a 4-meter class telescope, or partial (e.g. 20% time allocation) access to a 10-meter class telescope. In the 4-m class, NASA already operates the IRTF 3-m, and NSF operates the KPNO 4-m and WIYN 3.6-m telescopes. In the 10-m class, NASA is already a 1/6th partner in the Keck I and II telescopes, and is a partner in the LBT. The NSF operates the Gemini N&S telescopes. Consequently, there are existing NASA and NSF telescope facilities that match our assumed telescope apertures.

### 4.2.4 Survey Simulation Parameters

For each target in each survey, we assume that we need a minimum of 10 observations to reliably detect an exoplanet orbiting companion. In Section 4.2.6 we consider the viability of this assumption. For each observation, we assume a telescope slew and target acquisition "overhead" time of 3 minutes (time to move the telescope to align the target with the spectrometer slit and start telescope guiding). We assume a minimum integration time per target per epoch of 5 minutes to average over stellar RV variability from p-mode oscillations (Section 5.2.1.1). We consider two survey options for both the Exo-C and TESS target lists:

1) A survey with a fixed measurement precision, which will result in longer integration times for fainter targets. In this survey, we assume the spectrometer instrument stability precision is smaller than the achievable precision from the photon and detector noise, and thus can be ignored. The photon noise is dependent on the brightness of the target star.

2) A survey with a fixed integration time of 3600 seconds per target per epoch, which will result in an apparent magnitude dependent single measurement precision. In this survey simulation, we include a systematic noise floor from instrument stability.

In both survey simulations, we do not include the impact of stellar jitter, which we will address in Sections 4.3 and 5.2. We also don't address how long the surveys have to last in order to cover the range of periods we are interested in. For Exo-C, the orbital periods will span on the order of a decade. Consequently, these simulations represent minimum survey durations in the idealized limit. Actual survey times will be longer. In the first survey option, there are cases where it takes an ideal integration time of $t > 1$ hour to reach a specified single measurement precision for $N$ desired observations at that precision. In those cases, we assume that the practical integration time upper limit will be 1 hour ($t_p$ = 1 hr), and that the actual number of observations $N_p$ will increase such that $N_p\, t_p = N\, t$. In other words, observations at lower precision are co-added to achieve the desired number of observations at the specified precision.

For the spectrometer, we assume an end-to-end throughput, including telescope mirrors, spectrograph optics, and detector quantum efficiency, of 10%. We assume a spectrograph resolution of R = 100,000, with a Nyquist pixel sampling in the spectral dimension and 4 pixels in the spatial direction, with no dark current, a detector read noise of 5 electrons per pixel, and a detector gain of 4. Using the standard CCD equation, we compute the SNR of an observation from the detector noise and photon noise as a function of the apparent V-band magnitude, telescope aperture, spectrograph parameters and exposure time *t* (Figures 5-8). We assume that an SNR per resolution element of 600 is required to reach a 10 cm/s photon noise precision, as is estimated for ESPRESSO's spectral grasp of 350-730 nm (Pepe et al. 2010).

Finally, for the execution of survey, we assume eight hours of observation per night on average. We assume a loss of 50% of the survey time due to poor weather conditions and other hardware and software failure modes. We do not include any survey time for dedicated follow-up of specific targets (see Section 4.2.5).

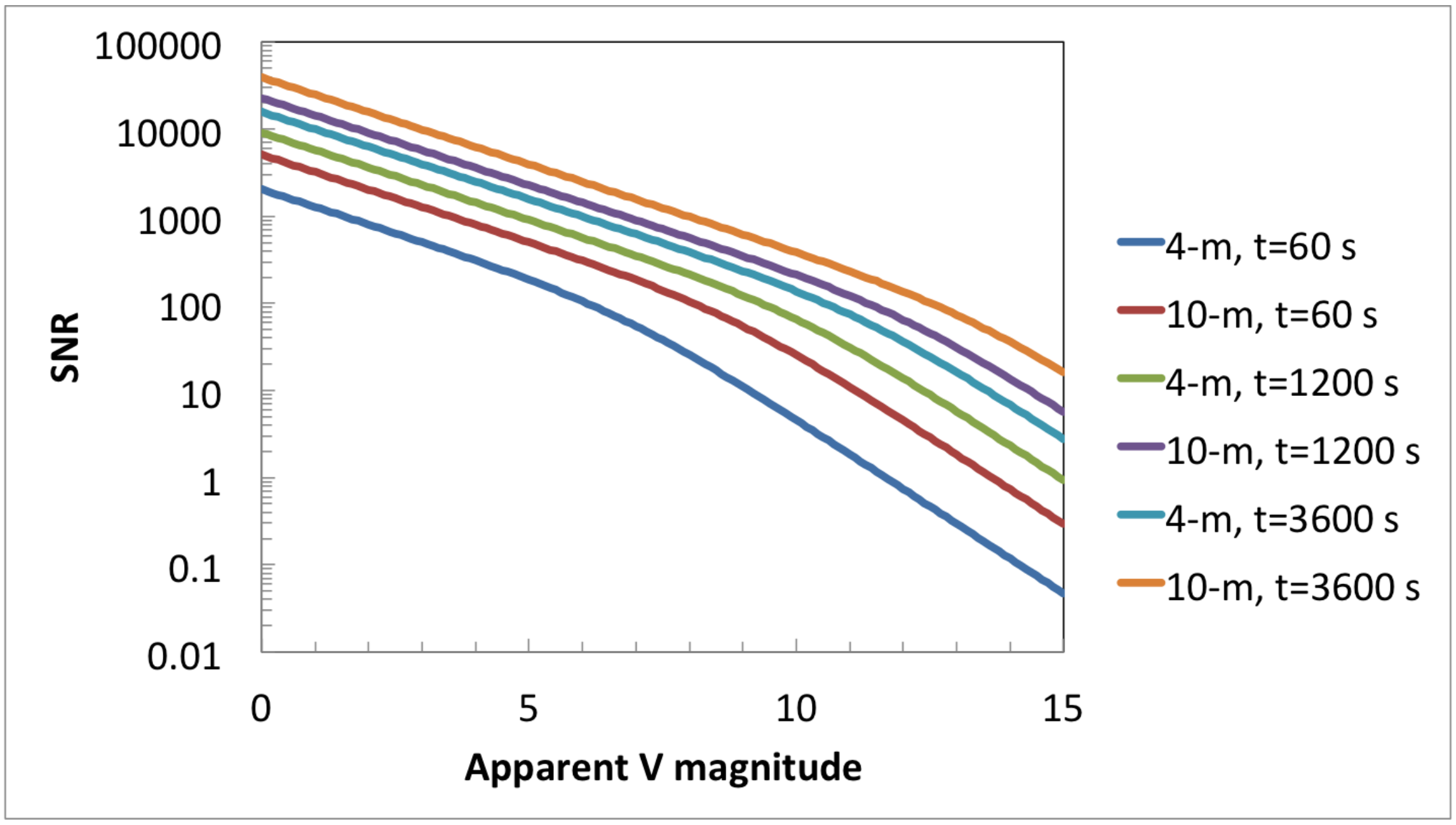

**Figure 5.** *The achievable SNR including photon and detector noise as a function of apparent magnitude for various integration times for 4-m and 10-m diameter telescopes with the spectrograph parameters specified in Section 4.1.4. A SNR=600 is the approximate SNR needed to reach a single measurement precision of 10 cm/s. The 'knee' in the curves is due to the transition between photon and read noise dominated regimes.*

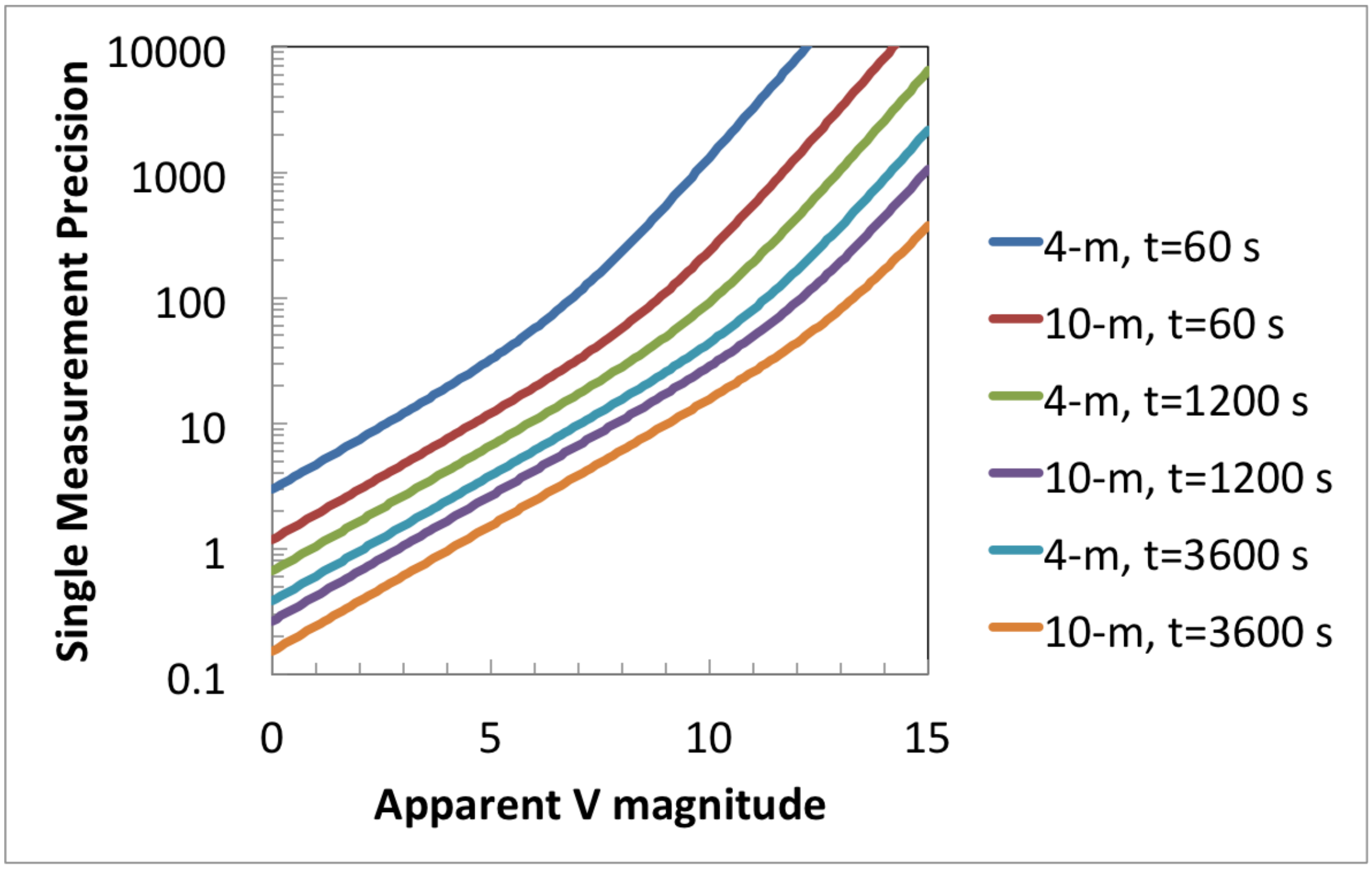

**Figure 6.** *The single measurement precision (cm/s vertical axis) obtainable as a function of apparent magnitude for various integration times for 4-m and 10-m diameter telescopes with the spectrograph parameters specified in Section 4.2.4. In this figure, no systematic instrumental noise floor is introduced.*

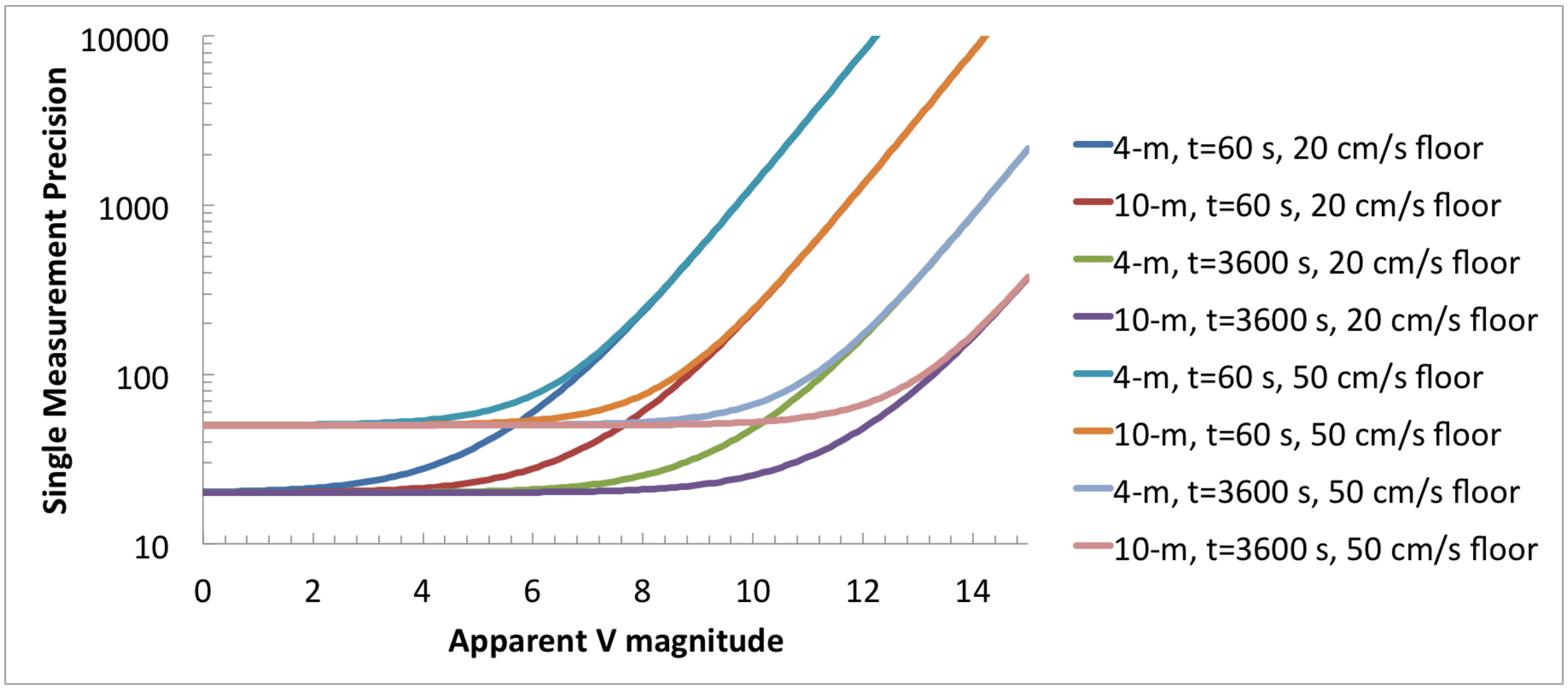

**Figure 7.** *The single measurement precision (cm/s vertical axis) obtainable as a function of apparent magnitude for various integration times for 4-m and 10-m diameter telescopes with the spectrograph parameters specified in Section 4.2.4. In this figure, two systematic instrumental noise floors of 20 and 50 cm/s are shown.*

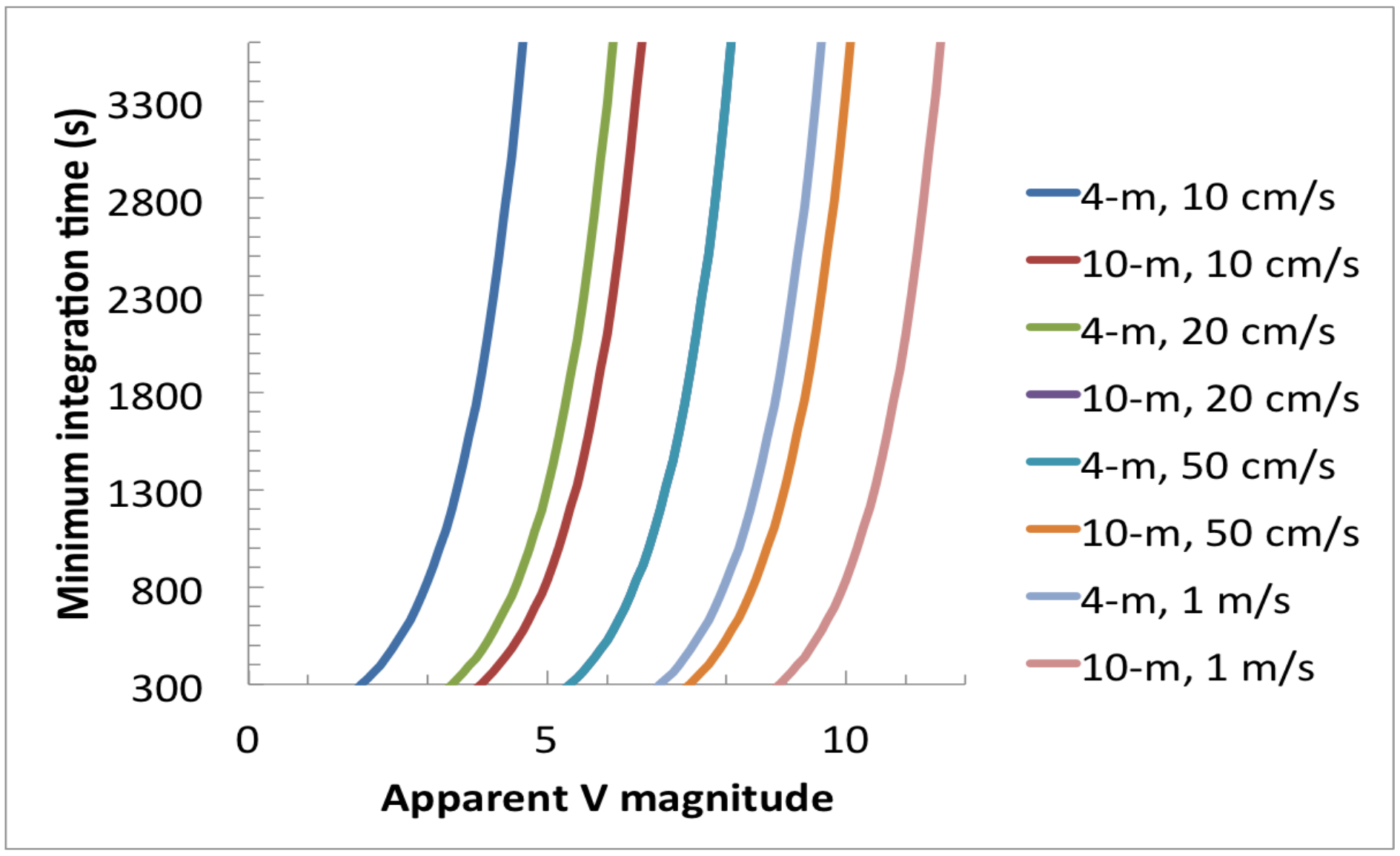

**Figure 8.** *The minimum integration time needed as a function of apparent magnitude for various single measurement precision goals for 4-m and 10-m diameter telescopes with the spectrograph parameters specified in Section 4.2.4. The vertical axis is restricted to 300-3600 seconds to represent practical integration times. In this figure, a systematic instrumental noise floor is not included, and the computation is based upon the photon and detector noise only.*

### 4.2.5 Is a minimum of 10 observations per target realistic?

With the radial velocity method, the minimum number of observations needed to detect an exoplanet with a stellar reflex motion semi-amplitude *K* and a single measurement precision of $\sigma$ is given by:

$$N_{obs} = 2\left(\frac{SNR\,\sigma}{K}\right)^2$$

where the SNR is the detection confidence (e.g. SNR=10 for a 10-sigma detection; Gaudi & Winn 2007). With a known orbit, such as for the confirmation of a TESS transiting planet candidate, the factor of 2 drops out (Winn 2014).

We investigate whether this theoretical limit is justified. We review the discovery paper literature listed at the NASA Exoplanet Archive for 120 radial-velocity discovered exoplanets orbiting a variety of host star spectral types and evolutionary states (Akeson et al. 2013). This sample represents ~20% of the exoplanets discovered with the radial velocity method to date, and excludes such systems that are radial velocity confirmation of known transiting planets.

In Figure 9, we plot $K/\sigma$ as a function of the number of observations reported in the discovery papers. We note that $\sigma$ is not the reported uncertainty on the semi-amplitude of the stellar reflex motion, but rather the typical single measurement uncertainty. In the absence of an explicitly documented typical single measurement uncertainty in the discovery paper, we also use the reported rms of the (O-C) residuals when available. We also keep track of whether a discovery paper announces the discovery of a single exoplanet, two exoplanets, or more than two exoplanets, or additional exoplanets in a system with a previously identified exoplanet or exoplanets. In the case of exoplanets in a system discovered non-simultaneously (e.g. in a series of papers with sequentially more observations), we categorize the first exoplanet discovery paper as a single exoplanet system (or two if the first paper announce two exoplanets, etc.), and the additional exoplanets reported in subsequent discovery papers are categorized as multiple exoplanet systems.

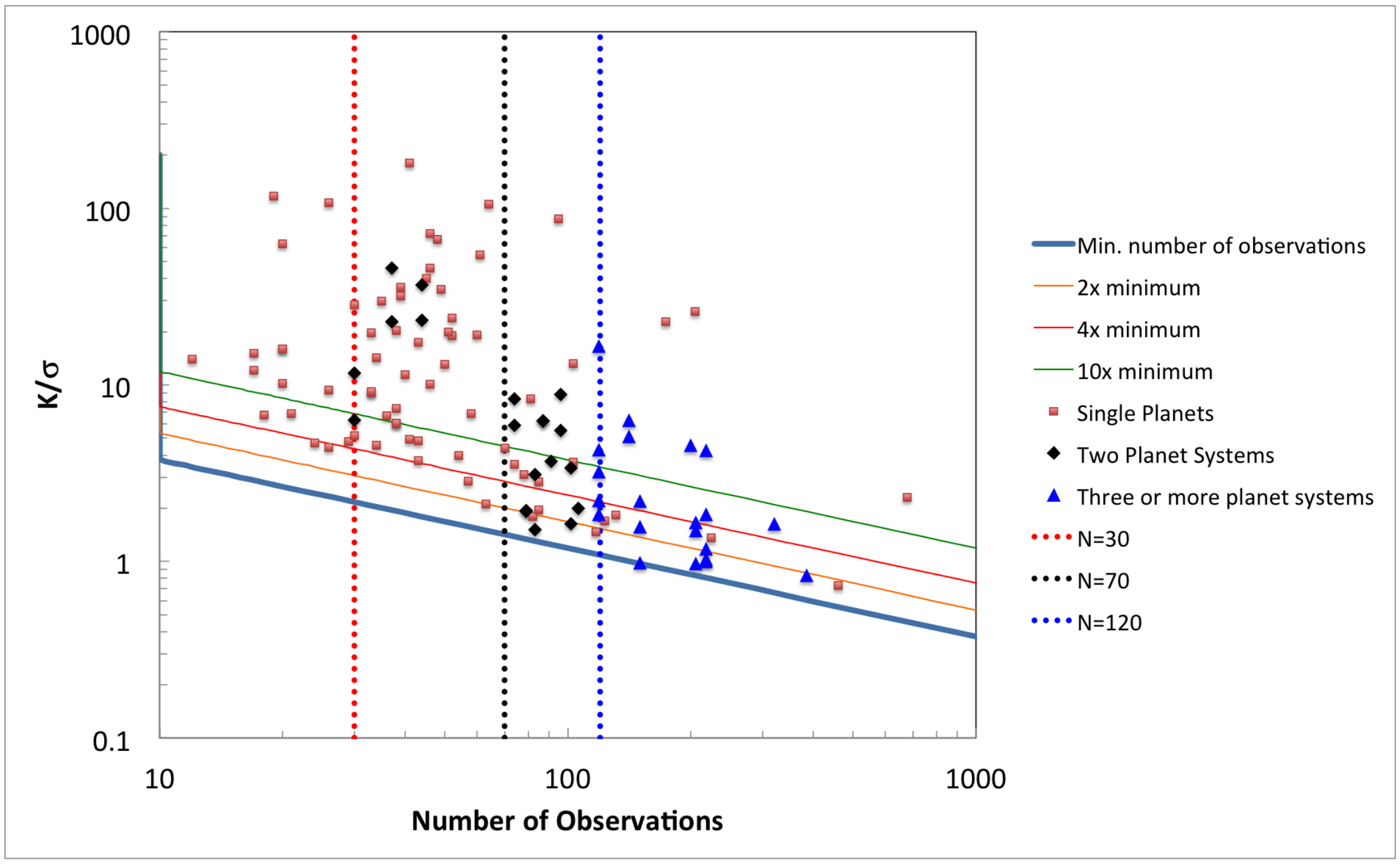

**Figure 9.** *We plot the number of observations reported in radial velocity discovery papers, compared to the expected theoretical limits for a given K/σ , where σ is not the reported uncertainty on the semi-amplitude of the stellar reflex motion, but rather the typical single measurement uncertainty. Brown squares: literature-reported K/σ values as a function of the number of observations reported in radial velocity discovery papers for single-exoplanet systems; black diamonds: two planet systems; blue triangles: three or more planet systems. The blue solid diagonal line represents the theoretical detection limit in K/σ as a function of the number of observations (Gaudi & Winn 2007), and the orange, red and green solid diagonal lines are two, four and ten times the theoretical minimum number of observations. The red, black and blue vertical dashed lines correspond to 30,70, and 120 observations.*

From Figure 9, there are several notable trends from discovery paper literature:

1) 3 or more exoplanet systems require a minimum of ~100 epochs;
2) 2 exoplanet systems require a minimum of ~30 epochs in some high SNR cases, and more than ~70 epochs otherwise.

3) For less than ~70 observations, discovery papers for any number of planets typically report >2-4 times the theoretical minimum number of observations necessary.
4) For greater than ~70 observations, radial velocity exoplanet discoveries are pushing towards the theoretical limit for the minimum number of observations, independent of the number of planets.

Consequently, our survey simulations represent a minimum survey duration. Detailed follow-up of exoplanet system candidates will require up to an additional 70-100 epochs or more per candidate for confirmation, more than ten times the minimum assumed for our survey. Given the known frequency of exoplanets from Kepler, we may expect more than 10% of survey targets will require additional follow-up. Thus, to include survey time for candidate follow-up, particularly multi-planet systems and systems at low $K/\sigma < 2$, actual survey durations will approximately double what is simulated here to be consistent with the discovery standards as reported in the current exoplanet literature. Furthermore, as noted above, we have not yet integrated into this assessment the contribution to the radial velocity precision from stellar jitter, which we will address in Sections 4.3 and 5.2.

### 4.2.6 Survey Simulation Results and Discussion

#### 4.2.6.1. Fixed Integration Time Surveys

For the surveys with a fixed integration time of 3600 sec, we calculate that it will take 168 nights (0.46 years) to observe 62 TESS targets, and 466 nights (1.28 years) to observe 172 Exo-C targets with the number of observations per target $N$=10, and overhead for bad weather. For a dedicated telescope access, this is a relatively short survey. For a telescope with ~20% of the year dedicated to the survey, these surveys would take 2.3 and 6.4 years for TESS and Exo-C respectively. As mentioned in 4.2.5, these represent minimum survey durations, since we do not account for stellar jitter and follow-up observations. Also, we have not taken into account the exoplanet orbital periods of interest, which can exceed one decade for Exo-C targets.

In Figures 10 and 11, we present the achievable measurement precision as a function of apparent V magnitude and instrument noise floor for 4 and 10-m telescopes and spectrograph parameters as described in the preceding sections. In Figures 12 and 13, we translate this into the minimum semi-amplitude $K$ detection limit.

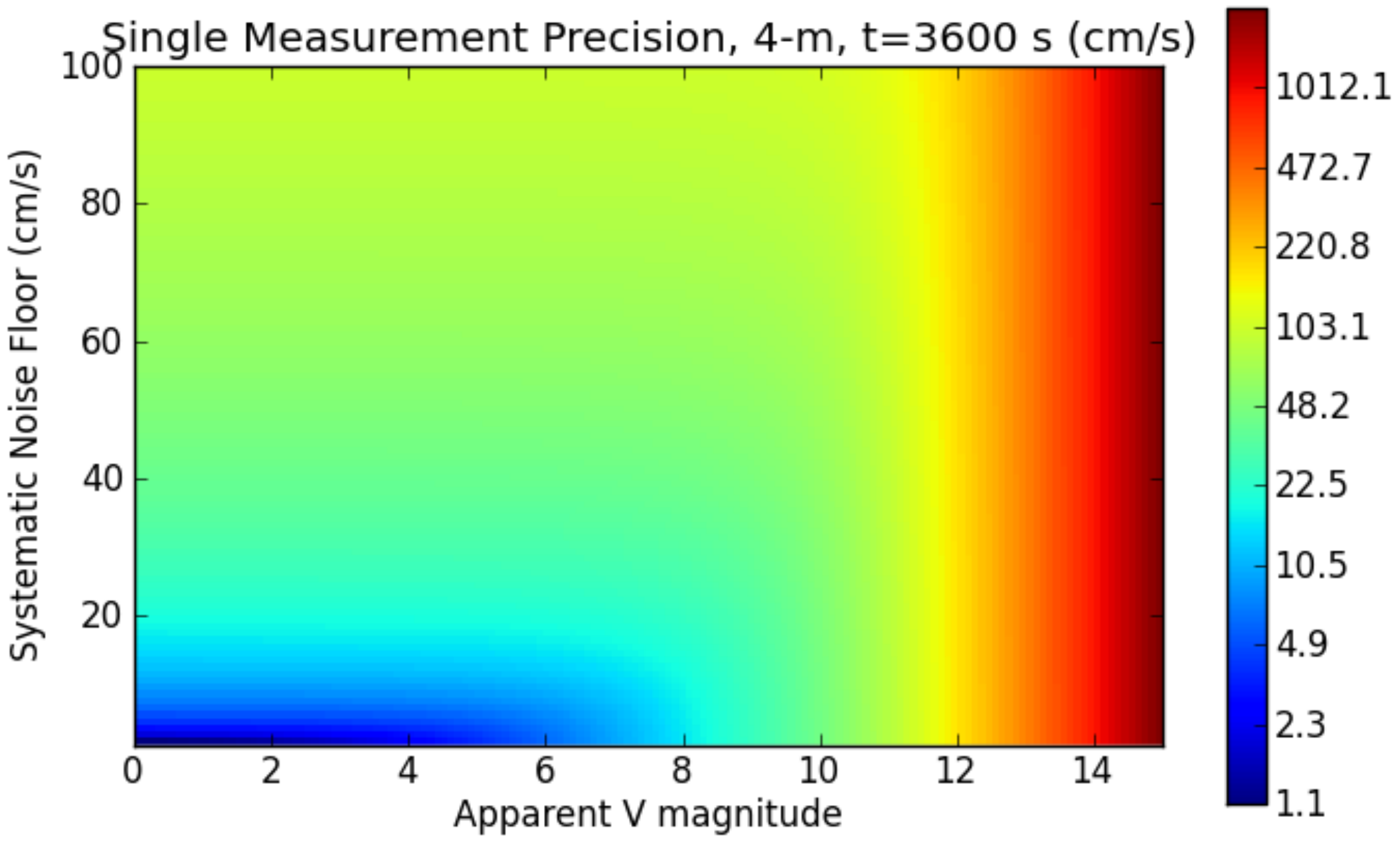

**Figure 10.** *A surface plot of the limiting single measurement precision in cm/s obtainable in 3600 seconds of integration with a 4-m telescope, as a function of the apparent V magnitude, and as a function of the systematic noise floor. The color scale is logarithmic.*

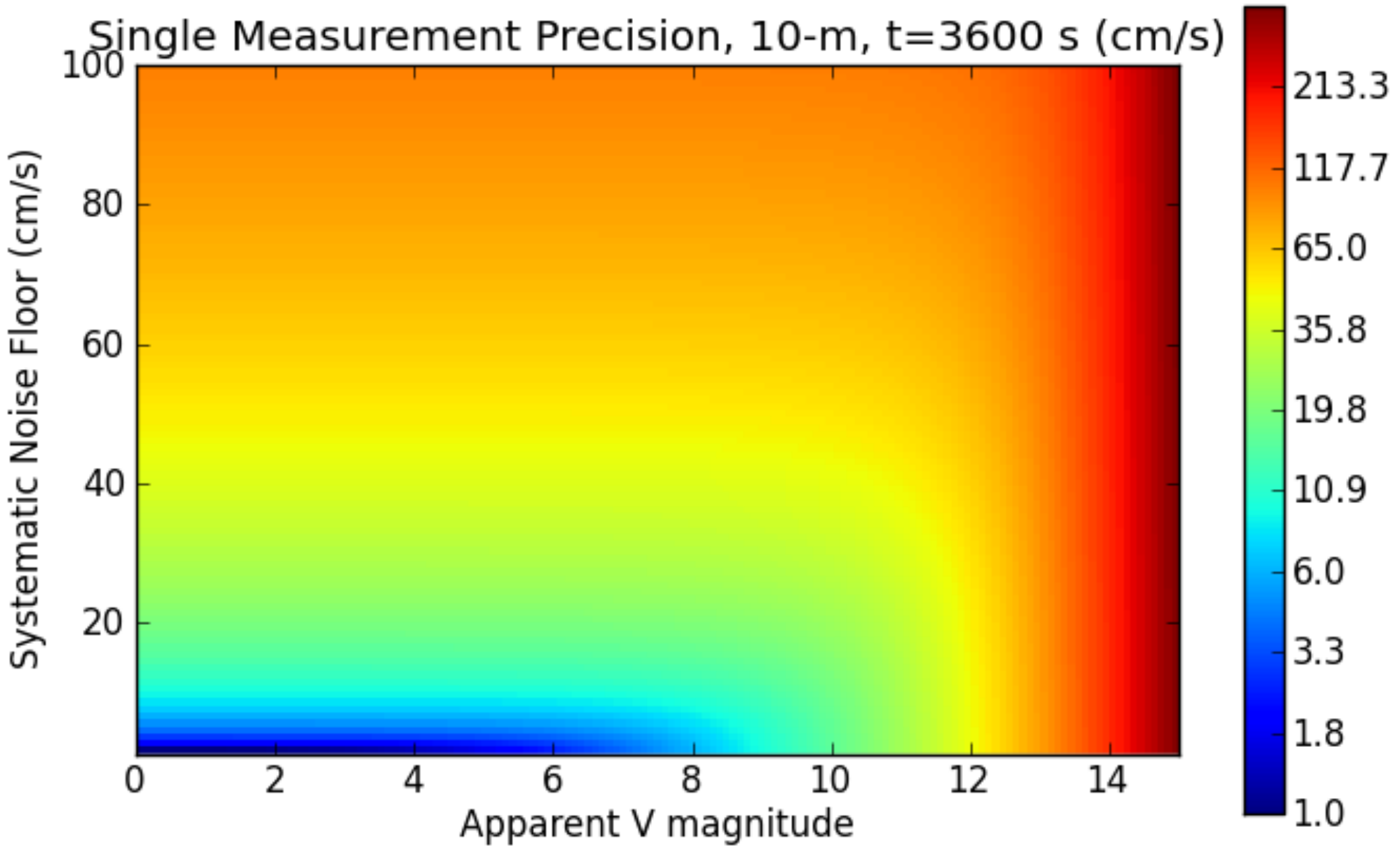

**Figure 11.** *The same as Figure 10, but for a 10-m telescope.*

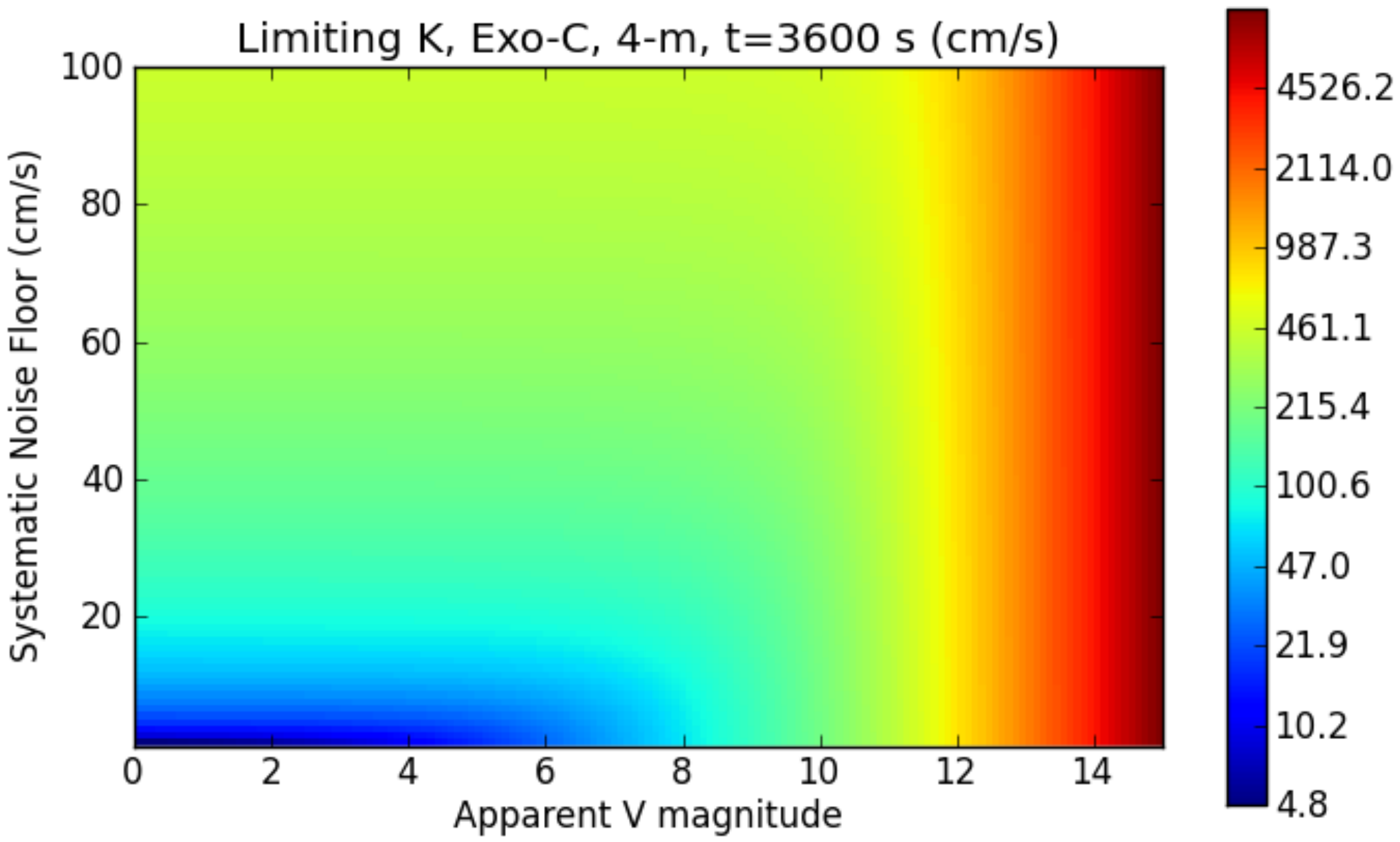

**Figure 12.** *The same as Figure 10, but showing the limiting* K *semi-amplitude for Exo-C targets. For TESS-targets, the limiting* K *semi-amplitude is smaller than the value shown here by a factor of* $\sqrt{2}$.

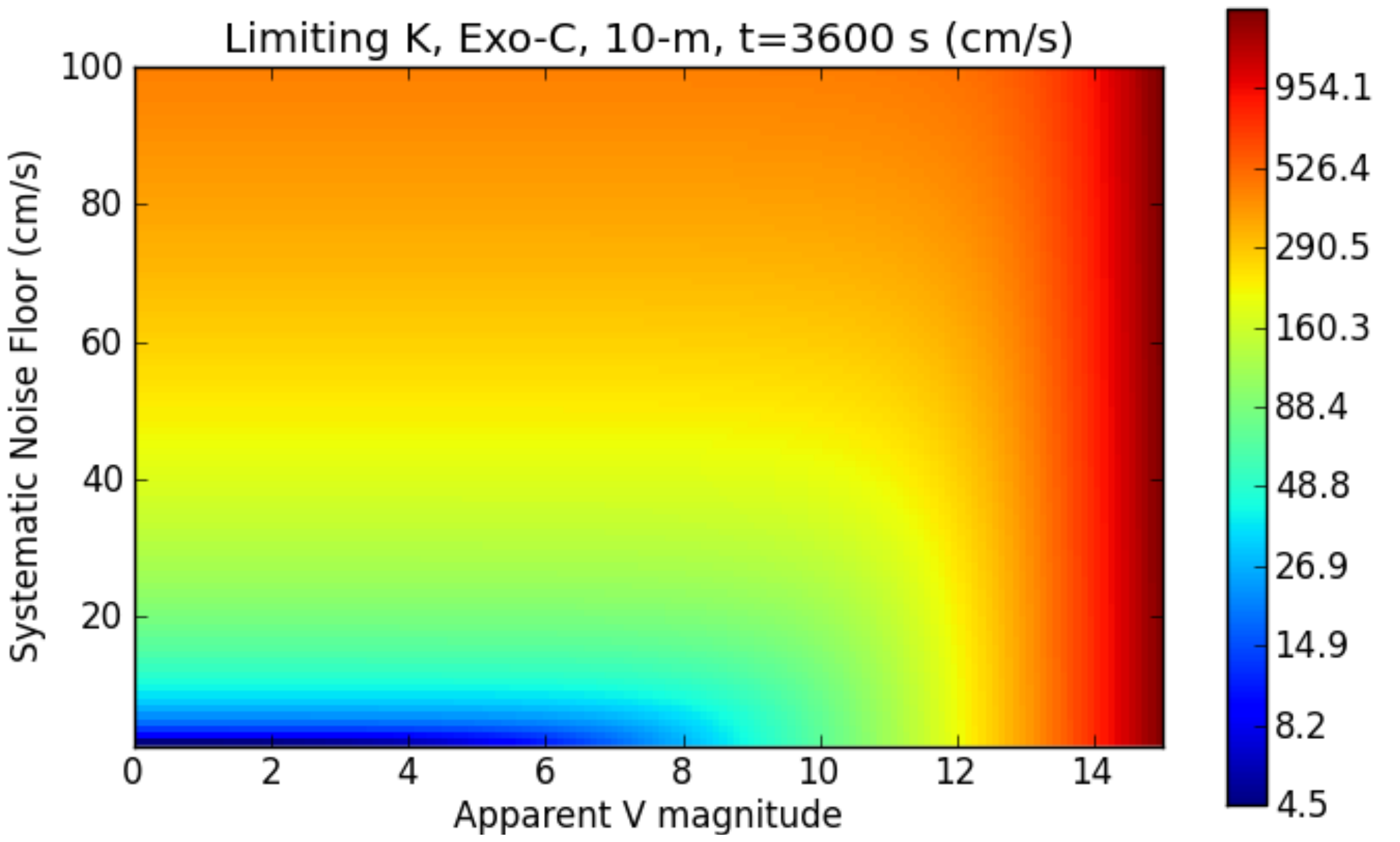

**Figure 13.** *The same as Figure 12, but for a 10-m telescope.*

From Figures 12 and 13, we draw the following conclusions:

1) For targets brighter than V~8, which encompasses nearly all of the Exo-C sample, PRVs are firmly in the systematic instrumental noise regime for both 4-m and 10-m telescopes with integration times of one hour, down to instrumental stability noise of less than ~10 cm/s.
2) TESS targets are primarily photon-noise limited on a 4-m telescope for V>11, and photon-noise limited on a 10-m telescope for V>13, with instrumental noise an increasingly important factor for brighter TESS candidates, particularly when the instrumental noise is greater than ~ 50 cm/s.
3) With a spectrometer that delivers better than 10 cm/s instrumental stability, Exo-C targets could readily be surveyed for Earth-mass companions (e.g. K~10-20 cm/s), with better survey performance on a 10-m aperture but still feasible on a 4-m telescope.
4) Saturn-mass exoplanets at 5 AU (e.g. K=3 m/s) could be surveyed down to V~10 on a 4-m telescope, and V~12 on a 10-m telescope, with a spectrometer instrumental stability of ~60 cm/s or better.

#### 4.2.6.2 Fixed Measurement Precision Survey

For the surveys with a fixed target measurement precision, we calculate the exposure times needed to reach the specified precision for a given apparent magnitude bin (Section 4.2.4). We use bins of 0.1 magnitudes. If the exposure time is less than 5 minutes, we set it equal to five minutes. Then we add an overhead time of 3 minutes for target acquisition. For each apparent magnitude bin, we calculate the number of observations needed to reach a specified semi-amplitude sensitivity, with a minimum of 10 epochs, as in Section 4.2.4. We then multiply the on-target time by the number of observations. Then we multiply this total integration time by the number of TESS or Exo-C targets in that apparent magnitude bin. We sum up this quantity, and multiply by the time lost due to weather and daytime to arrive at the minimum survey duration times. We present in Figures 14-25 the simulated minimum survey duration times for the TESS and Exo-C target samples, and for 4-m and 10-m telescopes.

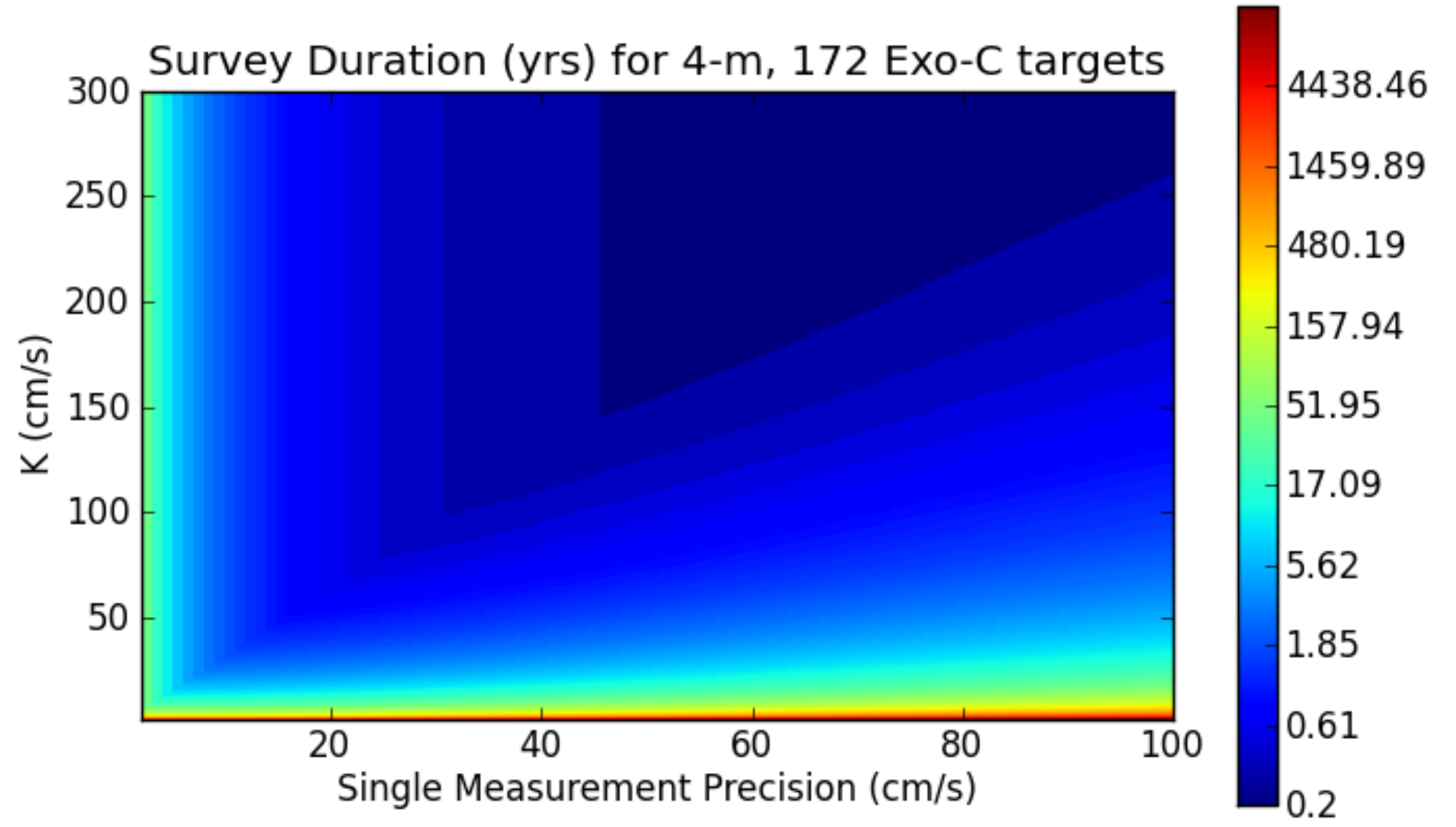

**Figure 14.** *A surface plot of the simulated survey time on a 4-m telescope for 172 Exo-C targets as a function of the single-measurement precision and semi-amplitude* K. *The color scale is logarithmic.*

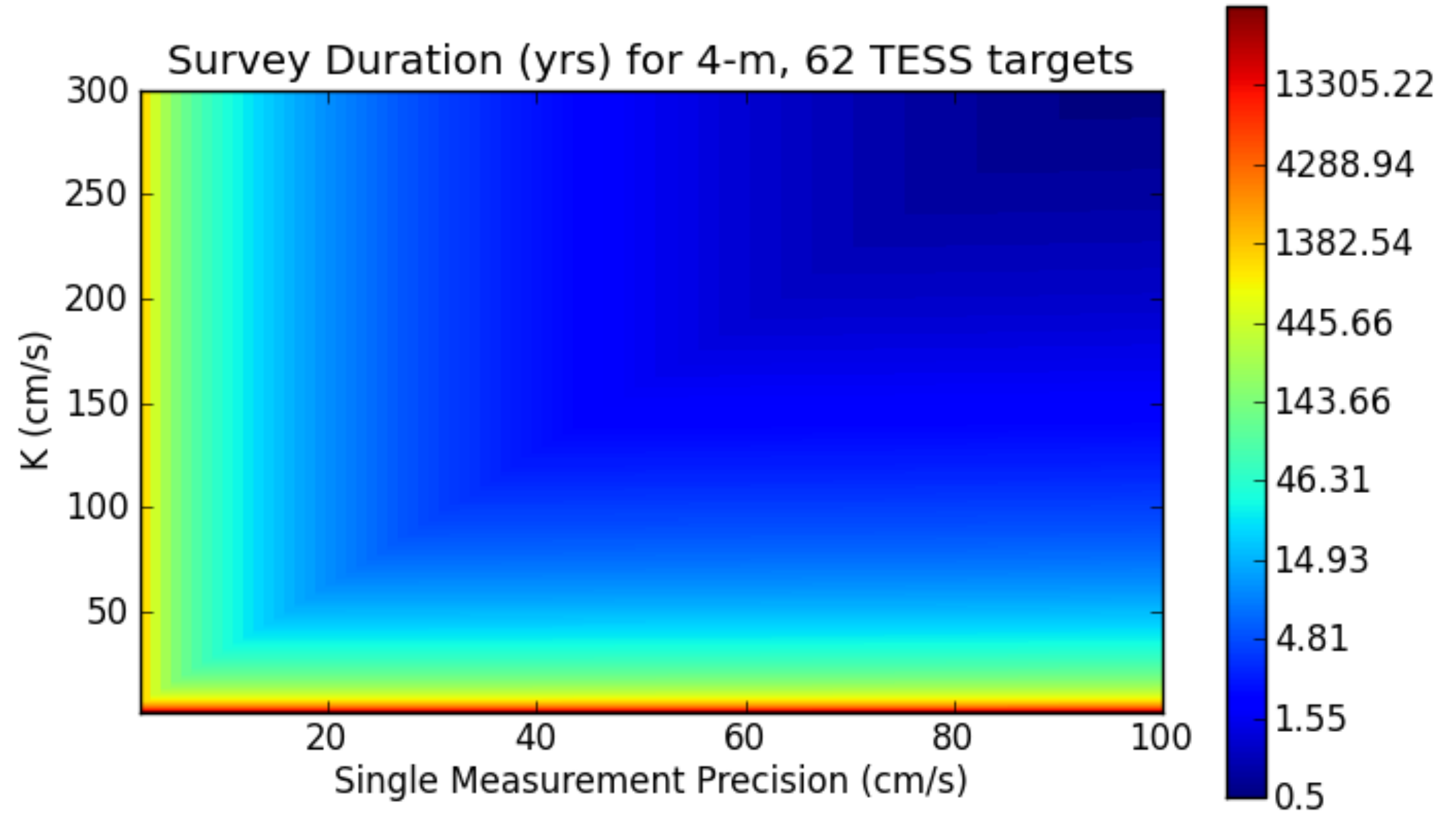

**Figure 15.** *The same as Figure 14, but for TESS targets.*

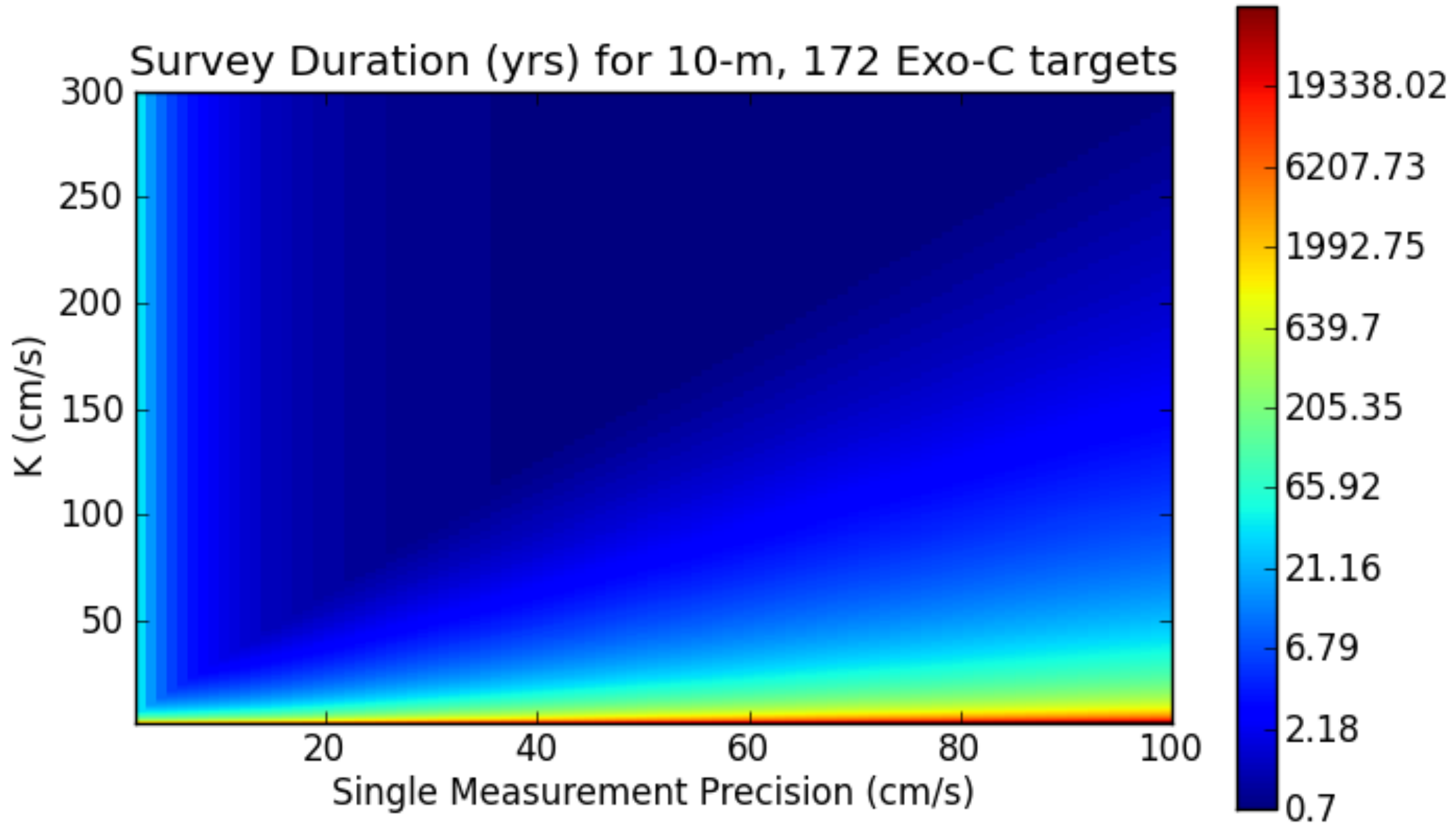

**Figure 16.** *The same as Figure 14, but for a 10-m telescope.*

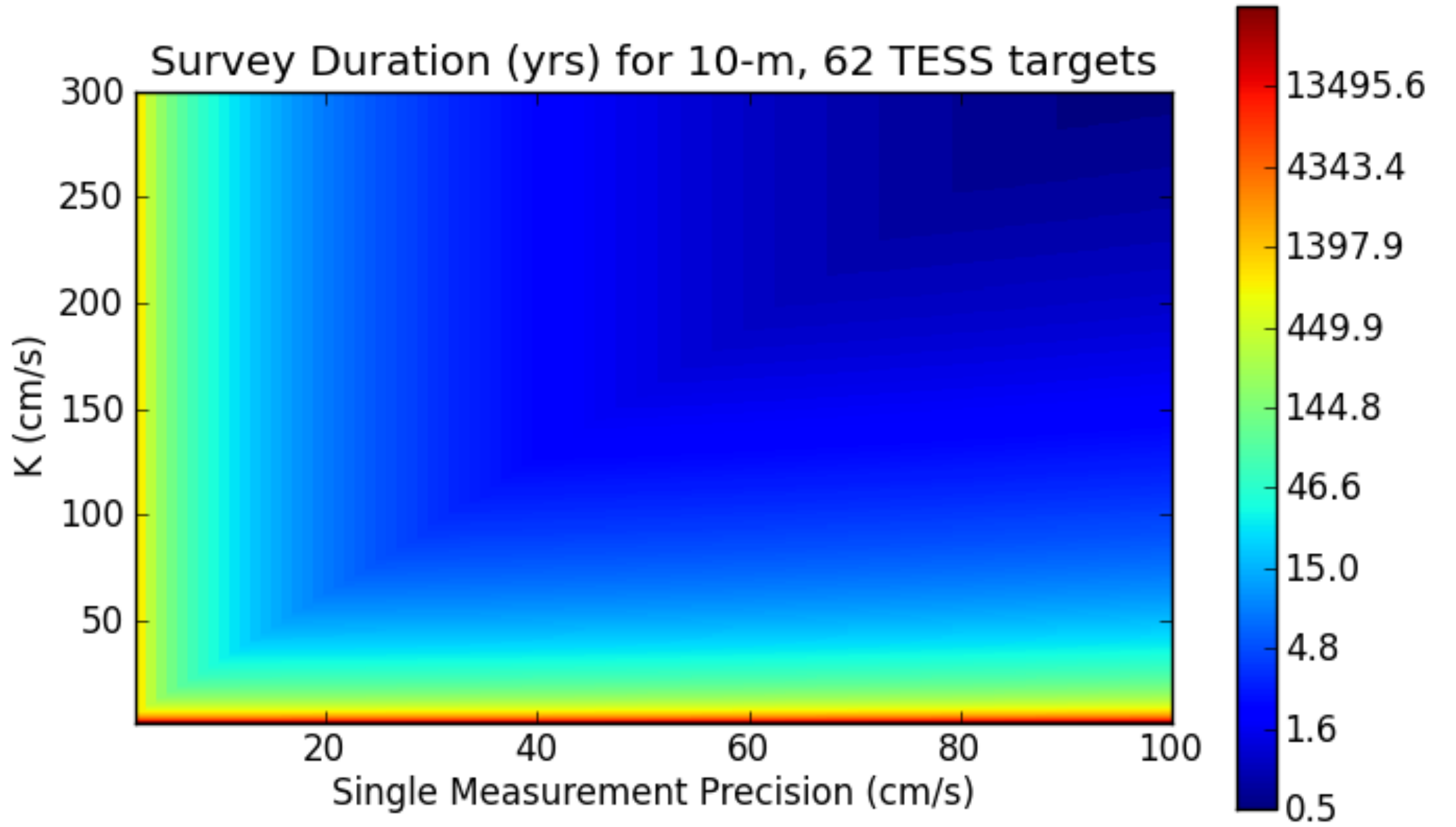

**Figure 17.** *The same as Figure 14, but for TESS targets and a 10-m telescope.*

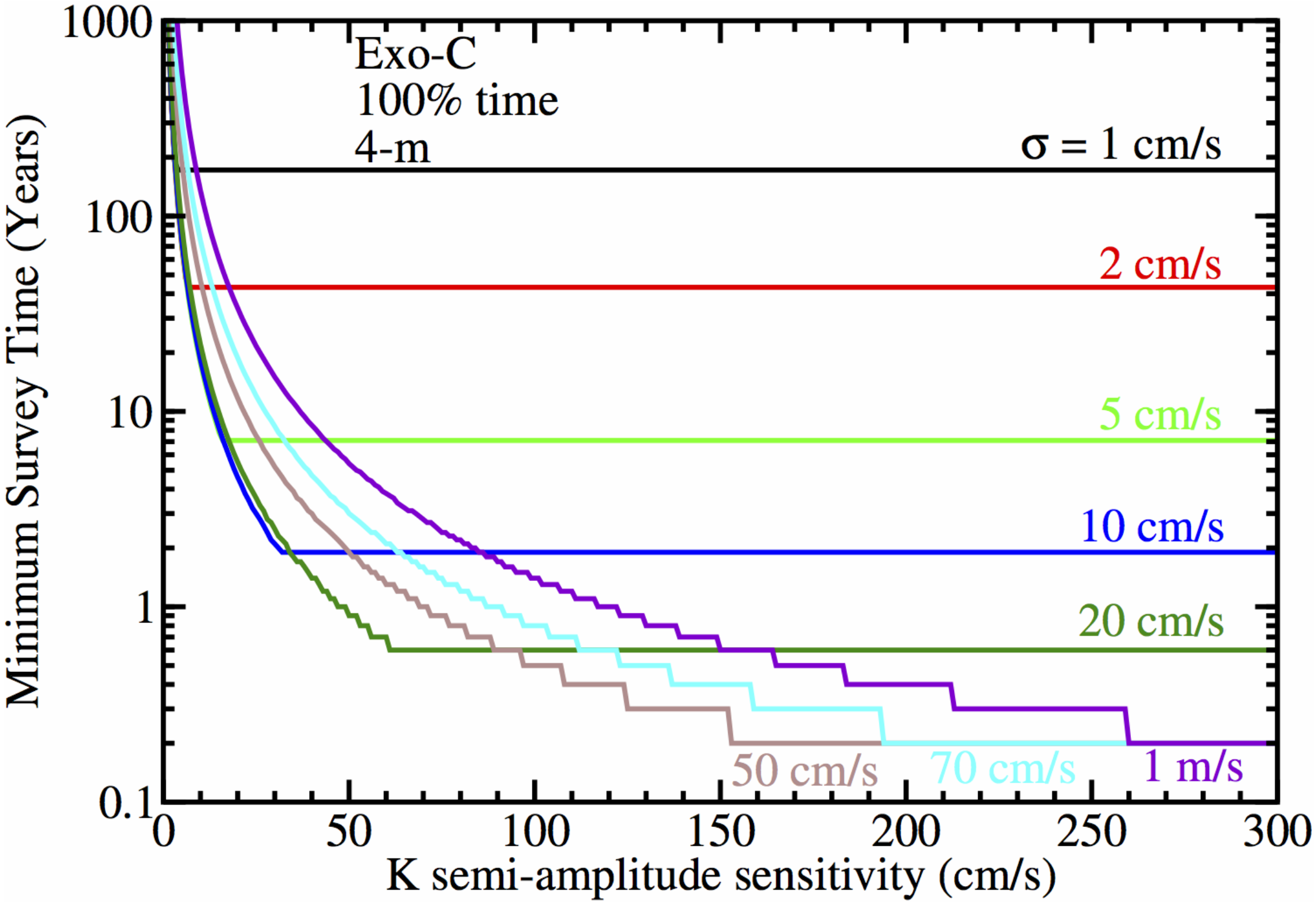

**Figure 18.** *Slices of Figure 14 at fixed values of the single measurement precision* $\sigma$ .

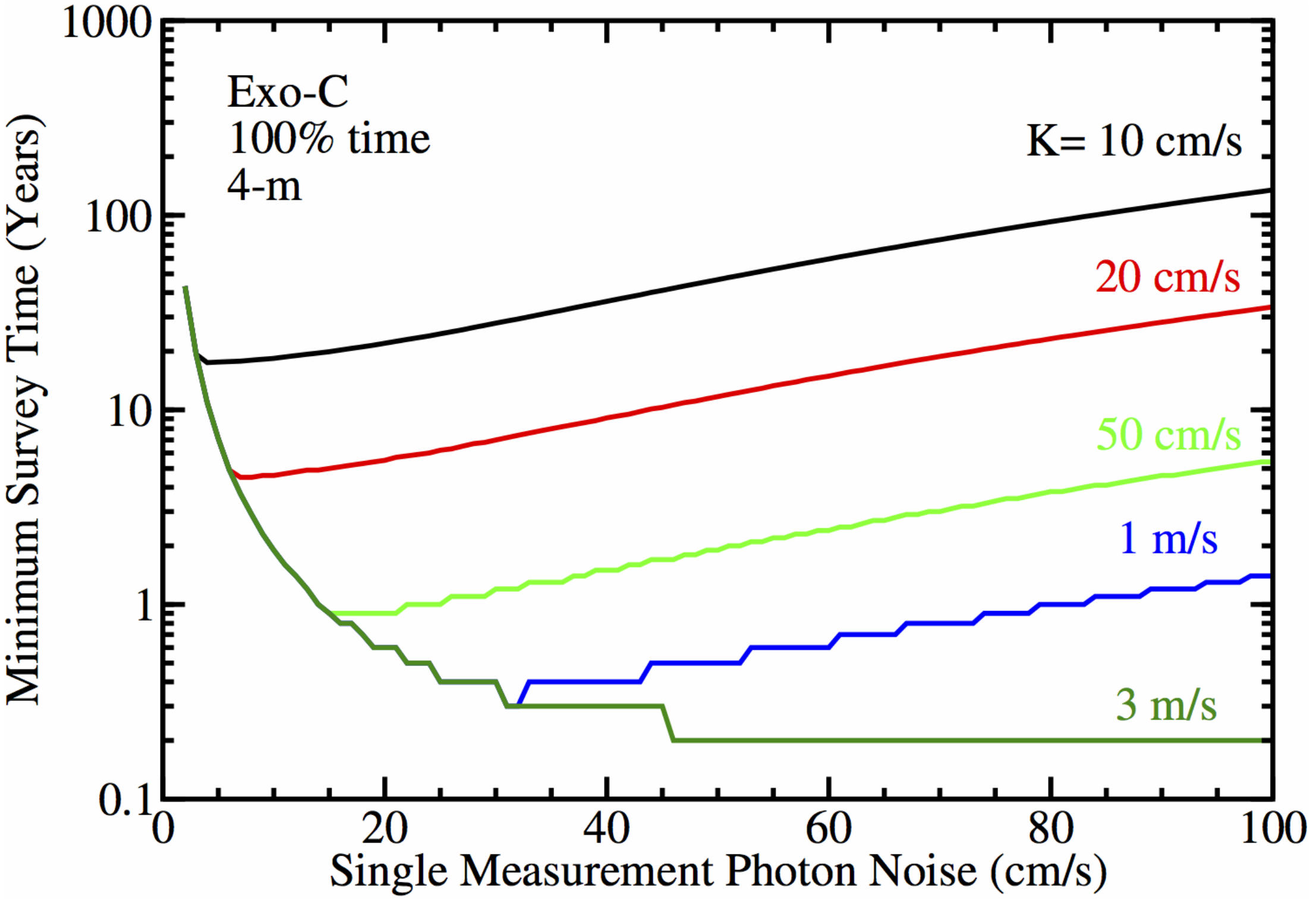

**Figure 19.** *Slices of Figure 14 at fixed values of the limiting semi-amplitude* K.

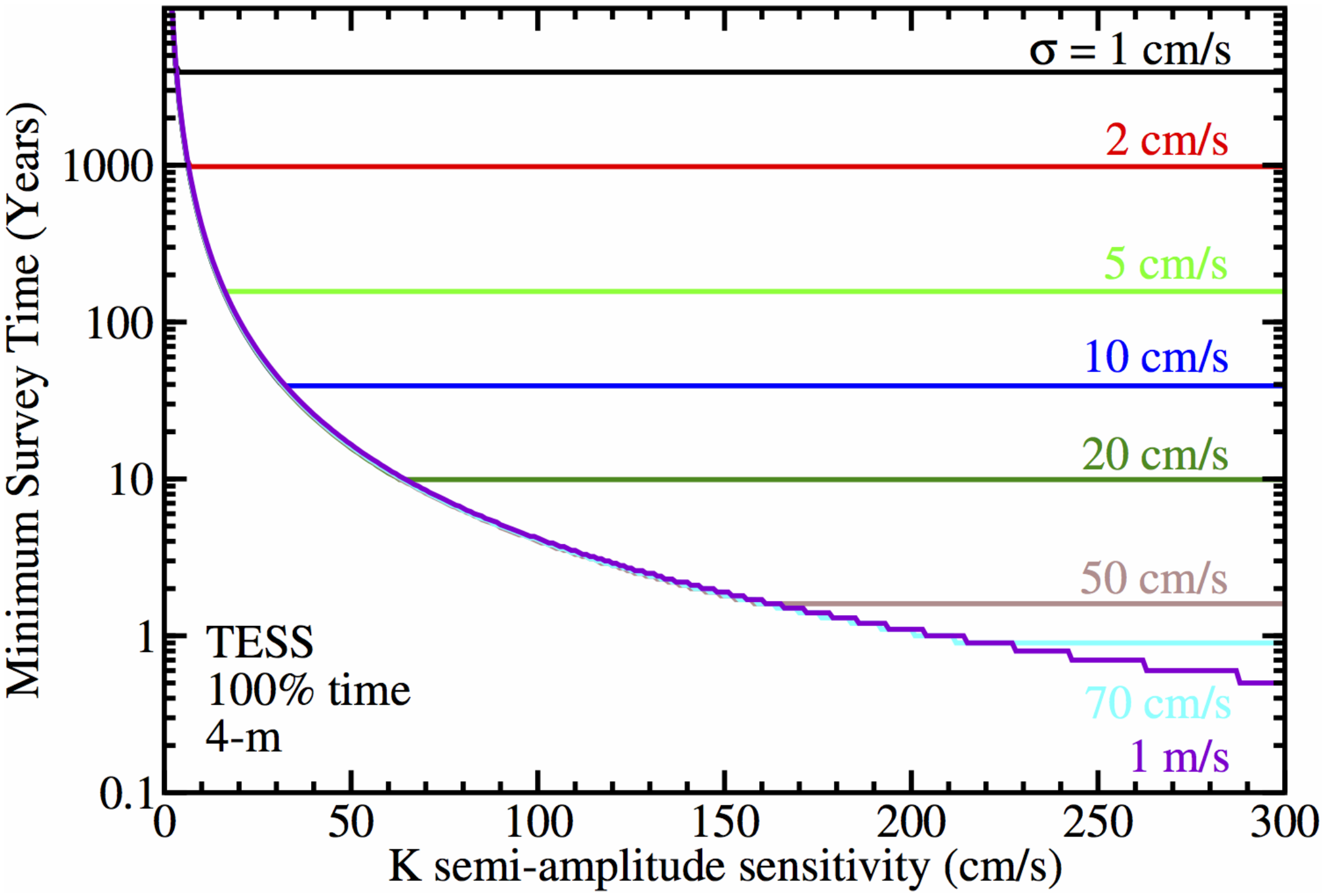

**Figure 20.** *Slices of Figure 15 at fixed values of the single measurement precision* σ *.*

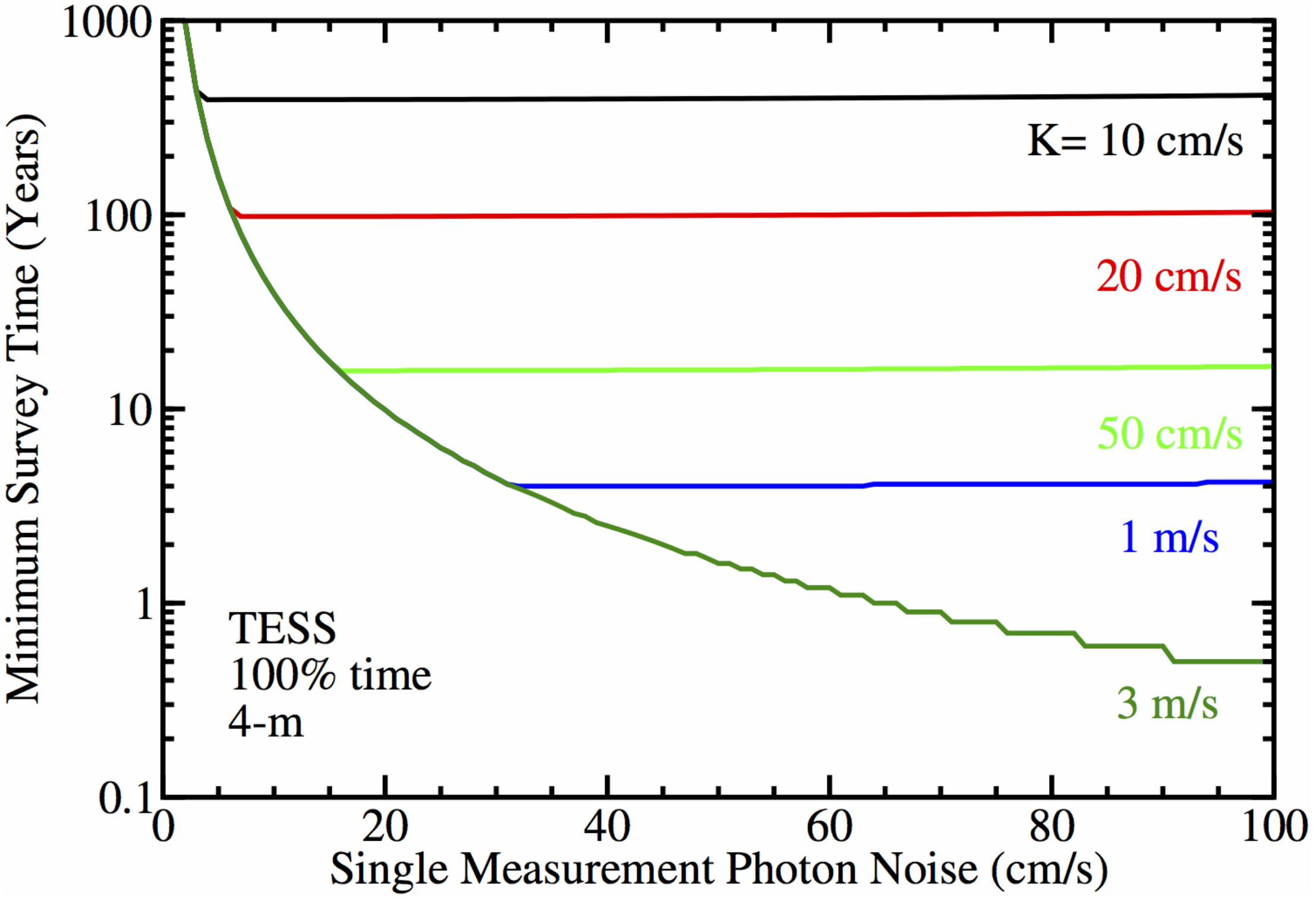

**Figure 21.** *Slices of Figure 15 at fixed values of the limiting semi-amplitude* K.

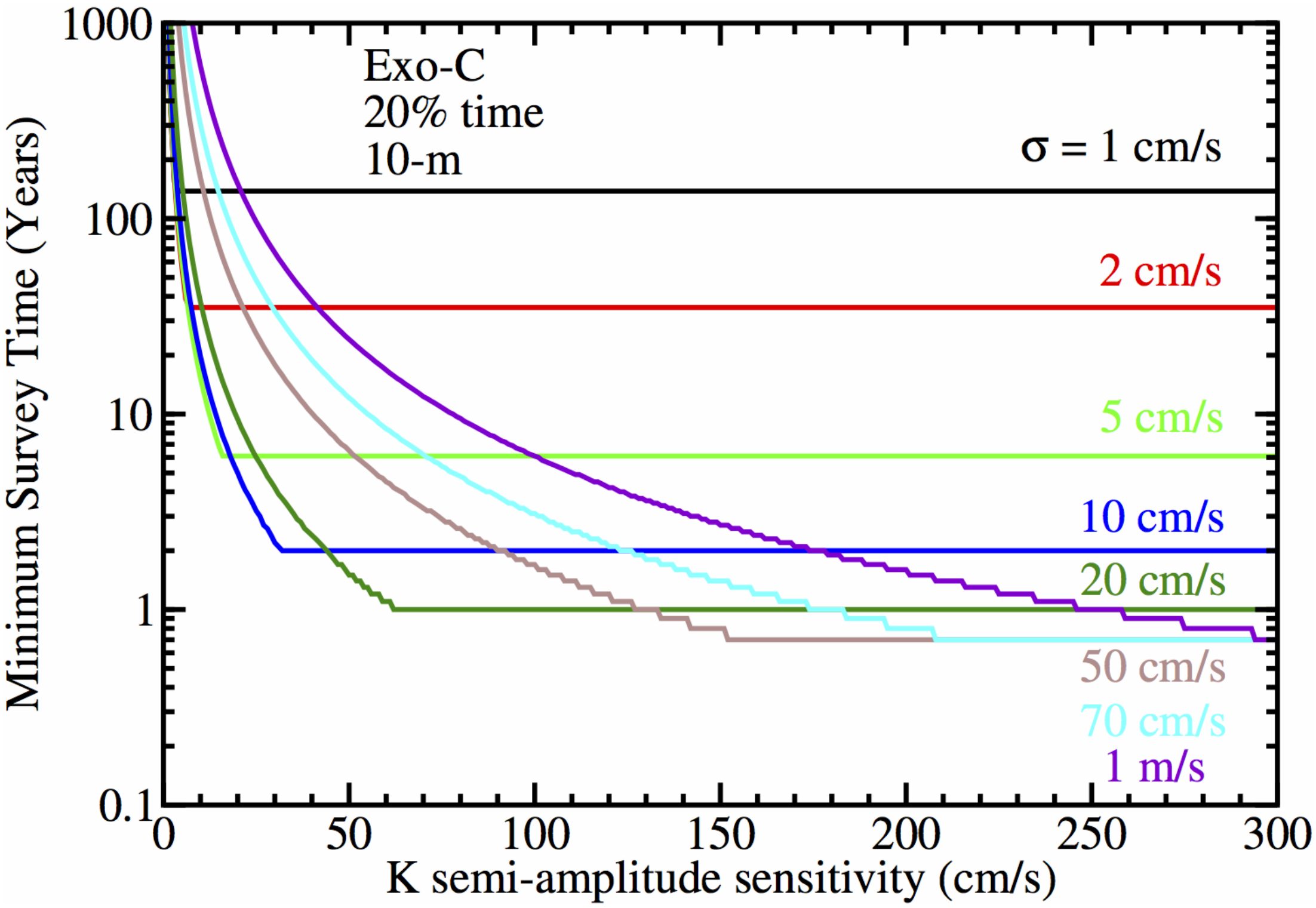

**Figure 22.** *Slices of Figure 16 at fixed values of the single measurement precision* σ *.*

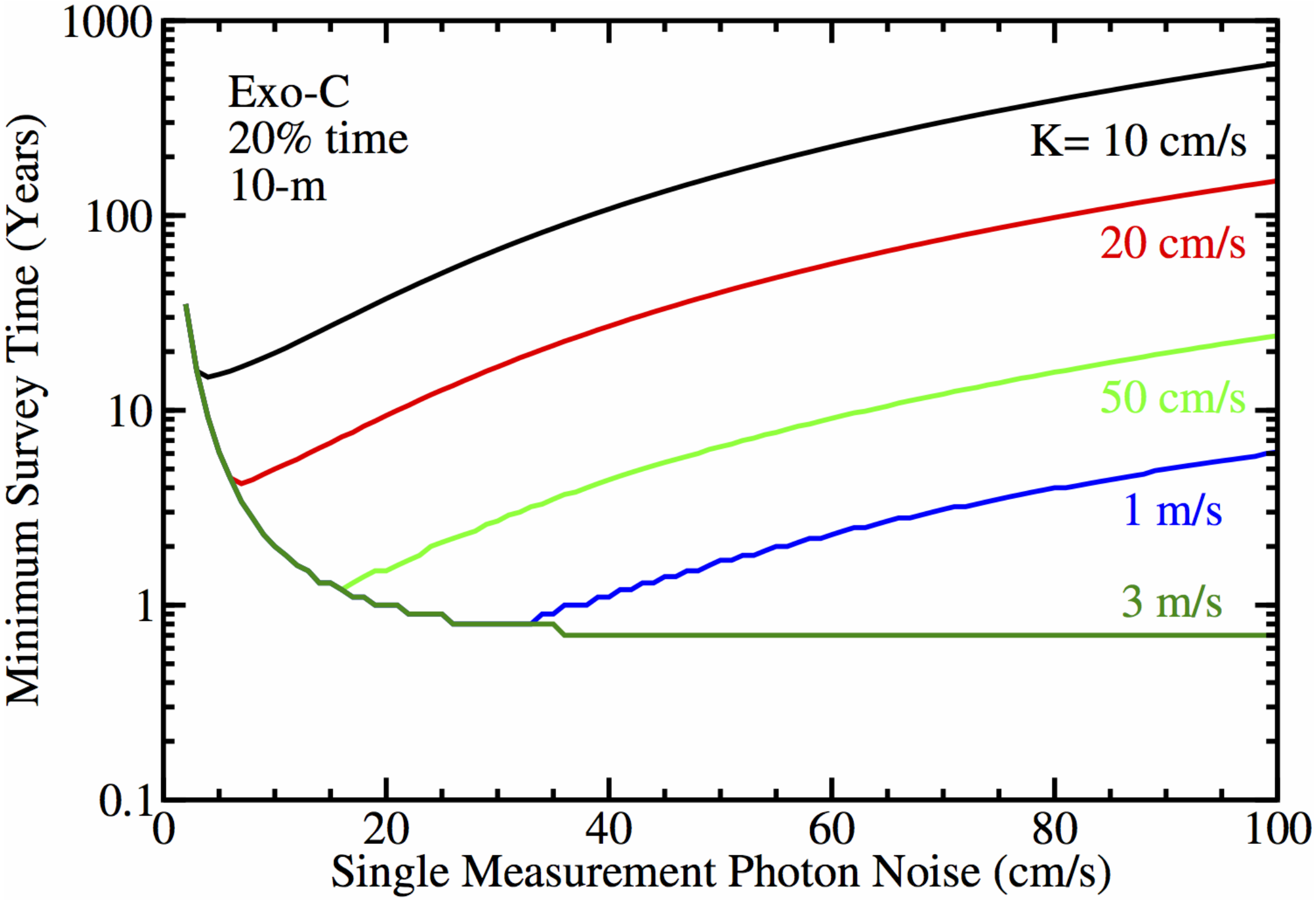

**Figure 23.** *Slices of Figure 16 at fixed values of the limiting semi-amplitude* K.

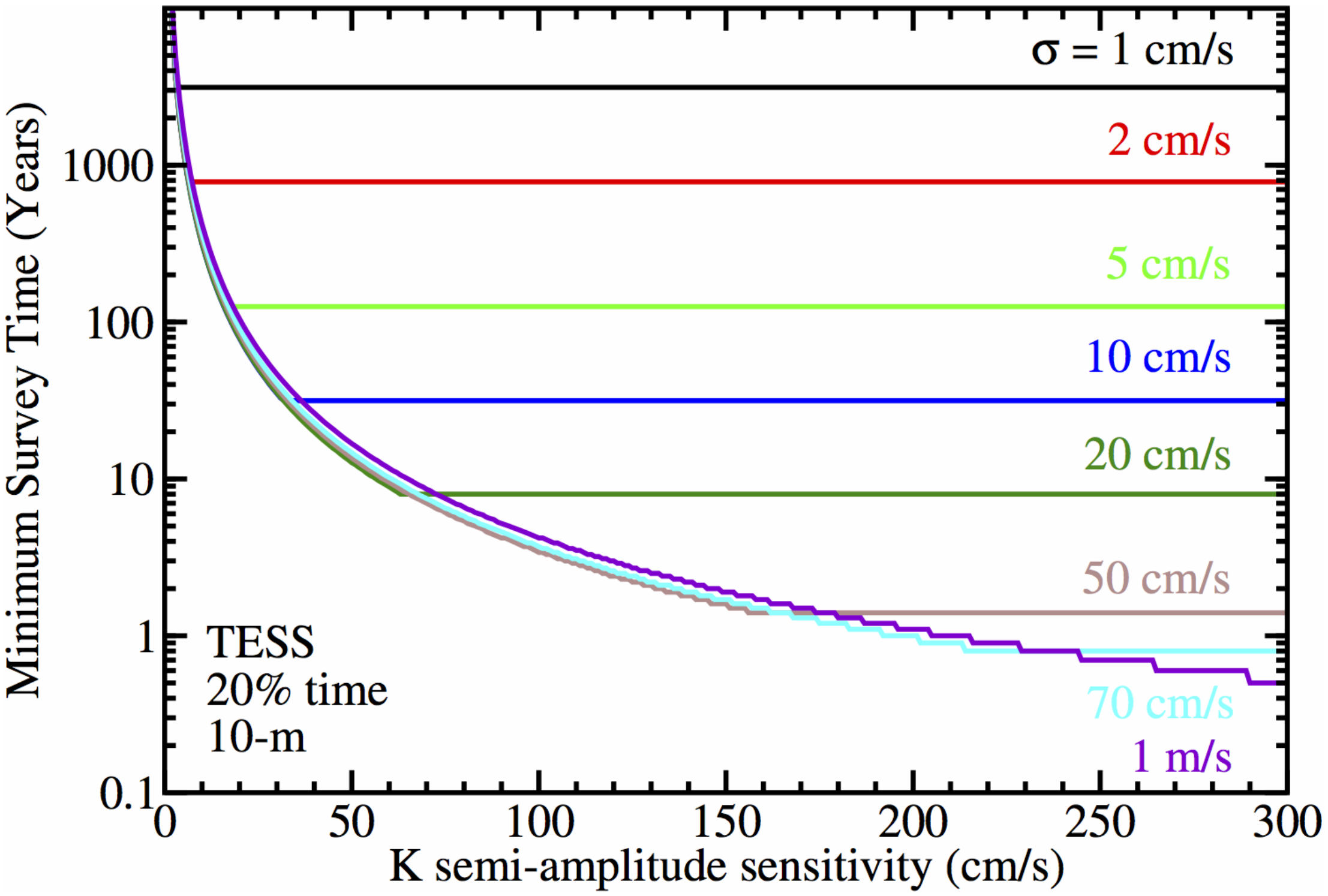

**Figure 24.** *Slices of Figure 17 at fixed values of the single measurement precision* σ .

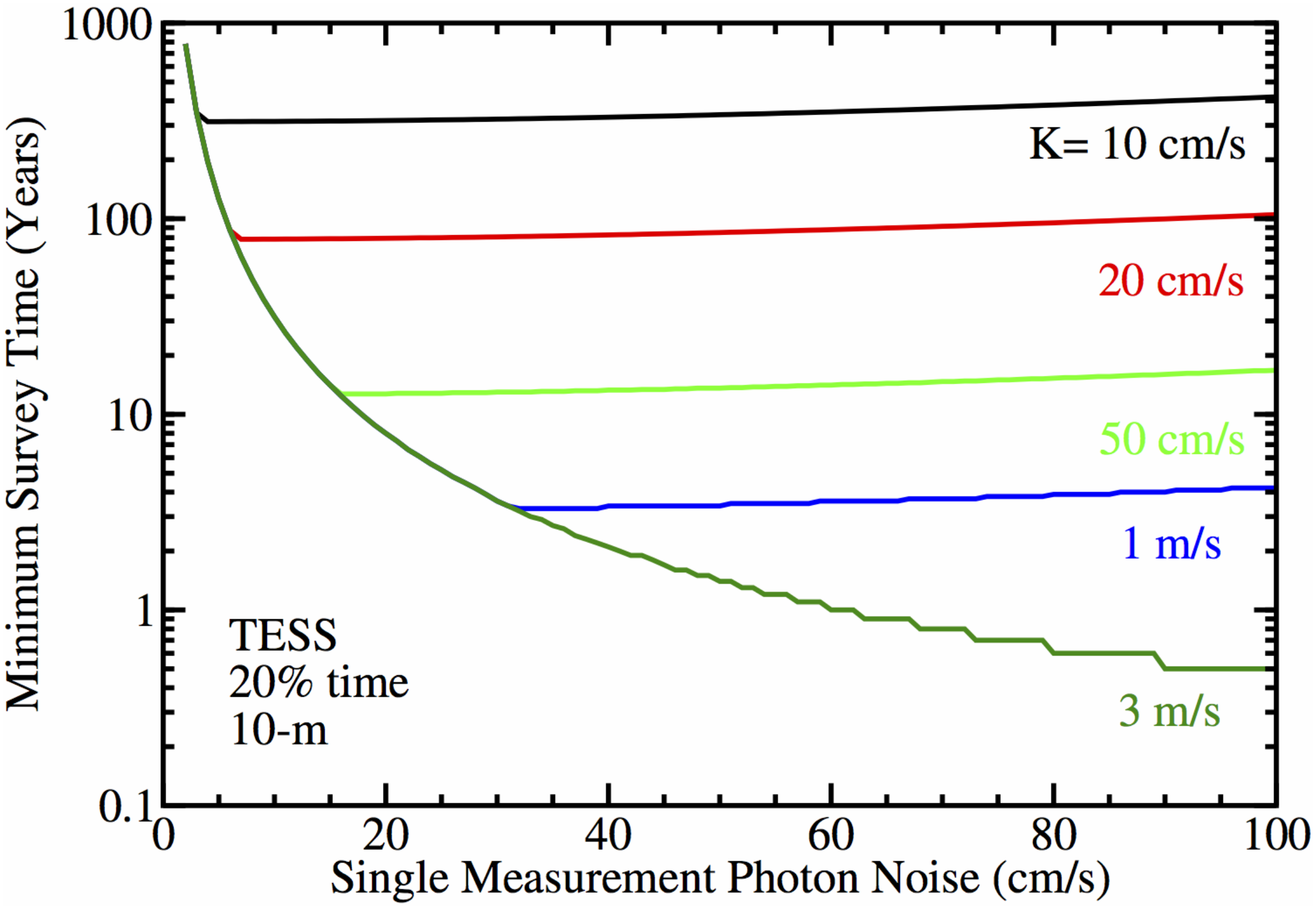

**Figure 25.** *Slices of Figure 17 at fixed values of the limiting semi-amplitude* K.

From these simulation results in Figures 14-25, we draw the following conclusions:

1) A spectrometer capable of 5 cm/s single measurement precision (e.g., an instrumental noise floor of ~1-2 cm/s) can survey 172 Exo-C targets for Earth-mass planets in 1 AU orbits (*K*~10 cm/s) with a minimum survey duration of ~15 years with 20% time allocation on a 10-m class telescope.
2) A survey of exoplanets down to the mass of Neptune at 5 AU (*K*~50 m/s) can be accomplished for Exo-C targets with a cumulative telescope time of ~1 year on either a 4-m at 100% time allocation or a 10-m telescope at ~20% time allocation. However, since the orbital period of such an exoplanet at 5 AU would be ~12 years, the survey duration would be necessarily longer, and could be designed to go down to smaller semi-amplitude sensitivity as a result (e.g. *K*=10-20 cm/s in bullet points #1 above).
3) A survey of 62 TESS targets down to K~50 cm/s would also take ~12 years.
4) It is prohibitive to survey 62 TESS targets for Earth-mass planets at 1 AU (*K*~10 cm/s), with a minimum survey duration of >300 years on a 10-m class telescope with 20% time allocation.

### 4.2.7 Summary Conclusions

To summarize, both of the first two spectrometer options presented at the beginning of Section 4 offer the capability for the identification of Neptune-mass and larger exoplanets for Exo-C direct imaging:

1) A ~50 cm/s instrument stability spectrograph on a 4-m telescope with 100% time allocation, and
2) A ~5 cm/s instrument stability spectrograph on a 10-m telescope with ~20% time allocation

Additionally, a ~5 cm/s instrument capability on a 10-m class telescope could enable the detection of Earth analogs for a future flagship direct imaging mission. Minimum survey durations of ~15 years will be necessary, requiring a long-term survey commitment and investment for both the survey and follow-up of candidate exoplanets. For option #1, the minimum survey time is due to the orbital period of the Exo-C exoplanets, and for option #2 it is due to the SNR needed to reach this level of precision.

## 4.3 NIR and/or Visible

in Section 4.2, we discussed the merits of visible PRV spectrometers for support for the NASA Exo-C and TESS mission objectives. In this subsection we discuss the merits of NIR PRV spectrometers to meet NASA mission objectives. A more general background context can be found in Section 5.3.

The primary motivations to pursue NIR PRVs is two-fold:

1) Cool stars that emit most of their bolometric flux in the NIR (e.g. $T_* < 4000$ K). While the Exo-C target list features only two late dwarfs with V-K>4, many of the TESS candidates are thought to be found around low-mass M dwarfs.
2) Stars for which the limiting sensitivity is due to wavelength-dependent stellar jitter. As we will show, this is increasingly important PRVs push sensitivity to sub-m/s semi-amplitudes for all FGKM stars. A more thorough discussion of the relevance of stellar jitter is presented in Section 5.2.

The primary drawbacks to pursue NIR PRVs is three-fold:

1) Telluric line absorption is much more substantial in the NIR compared to the visible (see Section 6.2.1.1).
2) Demonstrated on-sky RV precision of ~5 m/s in the NIR lags behind the visible (e.g. Bean et al. 2010, Plavchan et al. 2013a,b).
3) Cryogenic enclosures and optics with high throughput at NIR wavelengths often add a factor of 2-4 in increased cost compared to their visible counterparts (e.g. NIRSPEC on the Keck telescope cost $8M in 1998 dollars).

We next quantify the primary motivations to pursue NIR PRVs by constructing a model for obtainable RV precision as a function of stellar temperature and wavelength in Section 4.3.1, and including the impact of stellar jitter in Sections 4.3.2 and 4.3.3. As we will show, the inclusion or exclusion of stellar jitter drastically alters the conclusions over the optimal wavelength of observation for a variety of stellar effective temperatures.

### 4.3.1. The spectral type and wavelength dependence of RV photon noise

Two recent and independent efforts - Bottom et al. (2013) and Beatty & Gaudi (2015) - investigate the dependence of photon noise on spectral type and wavelength. Herein we summarize and compare their results, and use them as the starting point in our analysis. Bottom et al. (2013) investigates the optimization of an RV survey for HZ exoplanets with the MINERVA telescope array. In Figure 26, we reproduce Figure 1 from Bottom et al. (2013) showing the achievable measurement precision in 60-second integration times for a main sequence star at 10 pc. In Beatty & Gaudi (2015), the authors derive a similar quantity, but in units of precision per photon per velocity element. In Figure 27, we reproduce Table 1 from Beatty & Gaudi (2015).

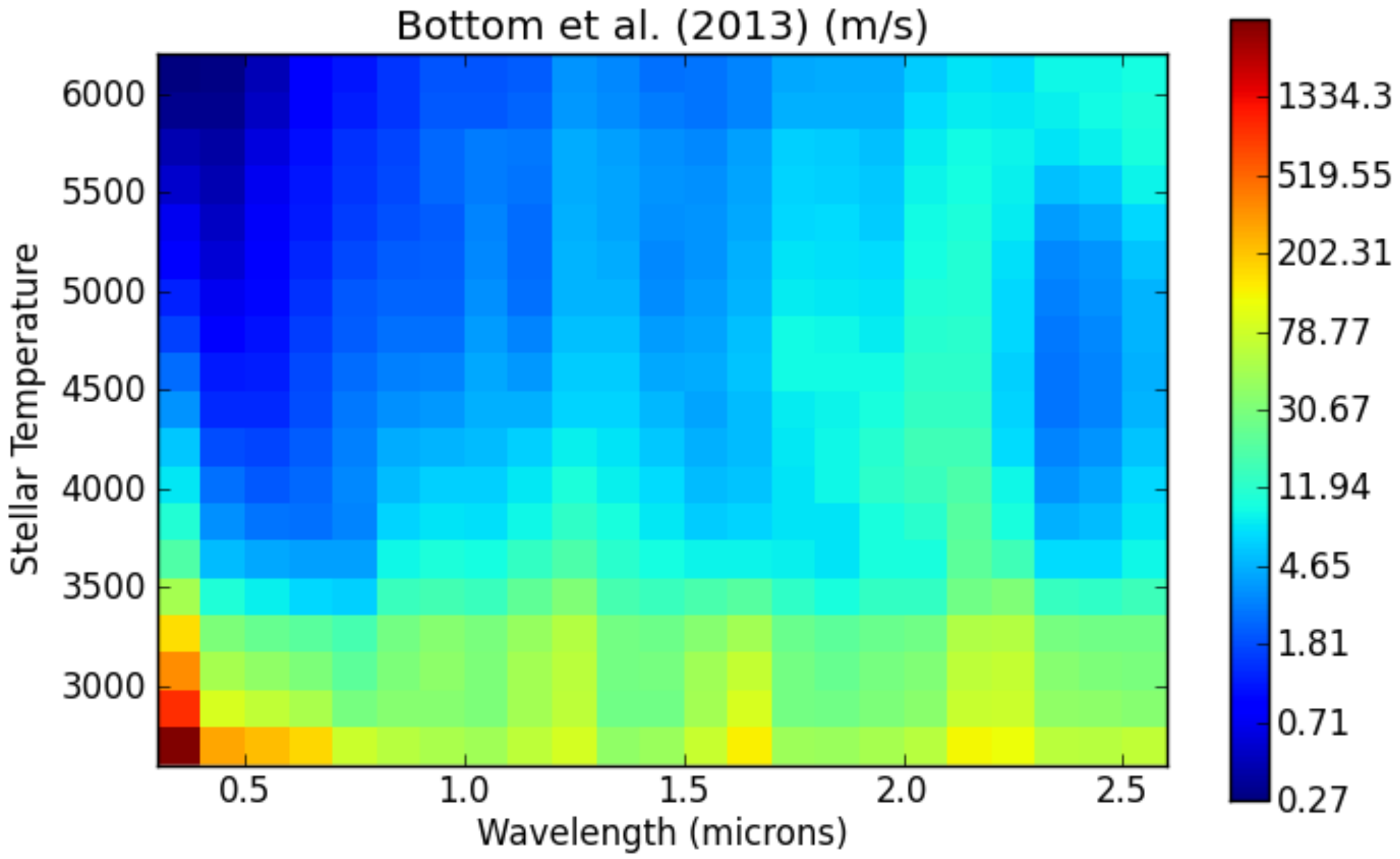

**Figure 26.** *A reproduction of Figure 1 in Bottom et al. (2013): "Doppler precision as a function of wavelength range and star temperature for a fixed amount of observing time (60 s). The stellar spectra are derived from rotationally-broadened main-sequence templates from 2600-6200K, stepped in 200K increments, and the wavelength range is stepped in 100 nm increments. … From the perspective of velocity precision, the best result is achieved in the range of 400-600 nm. ... This simulation assumes a 1.28 m$^2$ telescope dish, a spectrograph with R=75,000, and sky-to-detector throughput of 10%."*

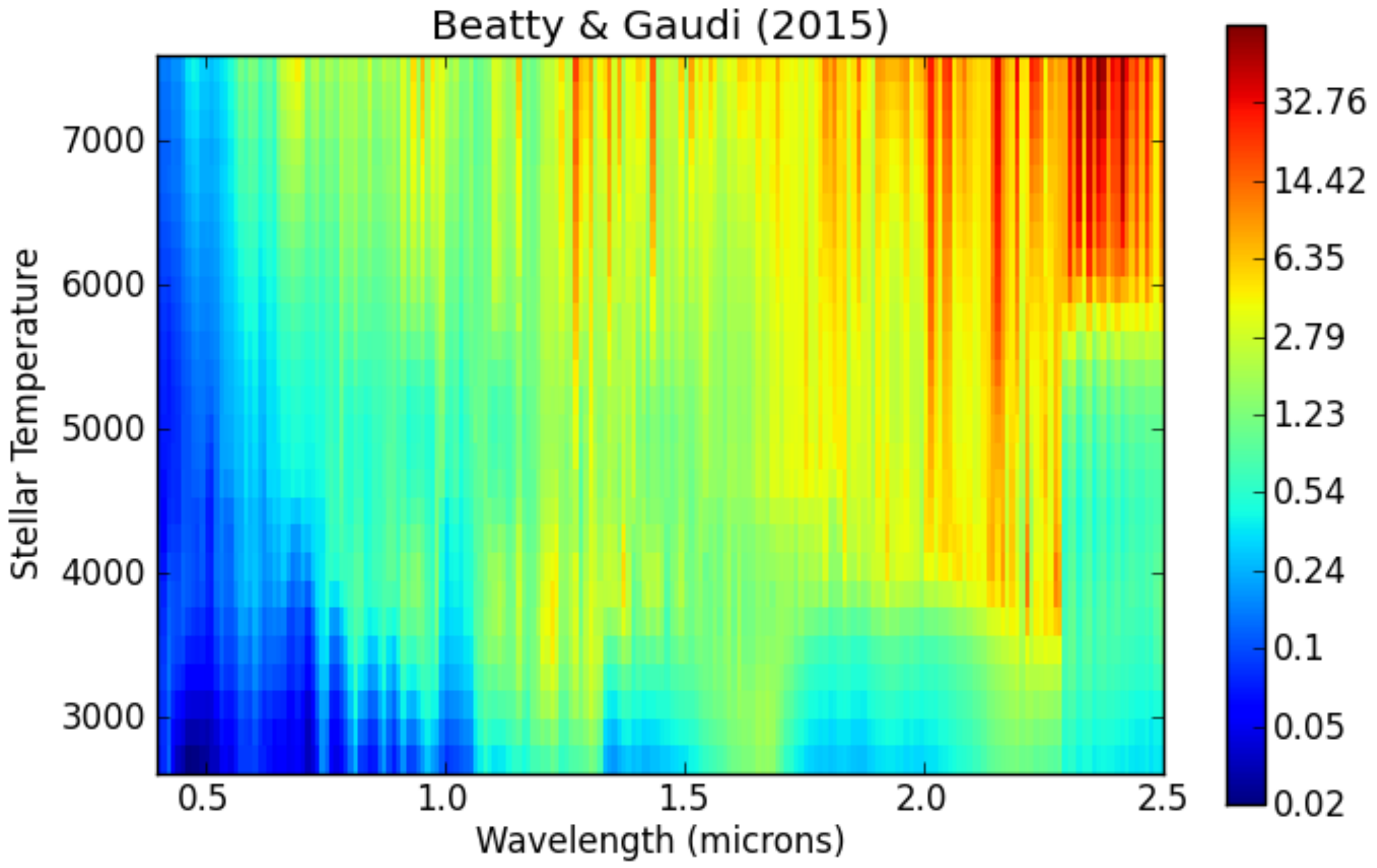

**Figure 27.** *A 2D surface plot visualization of Table 1 from Gaudi & Beatty (2015). The intensity units are in km/s per velocity element at constant flux over all wavelengths and stellar effective temperatures, and the color scale is logarithmic.*

In order to compare Figures 26 and 27 from Bottom et al. (2013) and Beatty & Gaudi (2015) respectively, we compute the brightness of a main sequence star at 10 pc (in Jy) as a function of wavelength in 10 nm bins, and as a function of stellar effective temperature in 200 K increments. For this computation, we use synthetic SEDs from the BT- Settl models at log g = 4.5 with solar metallicity (Allard et al. 2012), and the results are shown in Figure 28. In Figure 29, we divide Table 1 from Beatty & Gaudi (2015) and apply a normalization to scale the results from Beatty & Gaudi (2015) to match the parameters in Bottom et al. (2013). The figures are qualitatively similar, and in Figure 30 we plot the ratio of the two figures, interpolated on the Beatty & Gaudi (2015) wavelength and temperature grid. Figure 30 shows that the predicted RV precision from the independent works of Bottom et al. (2013) and Beatty & Gaudi (2015) agree to within ~40% over all wavelengths and stellar effective temperatures which is remarkable given the independent approaches taken by the authors. There are subtle differences between the two approaches:

1) Bottom et al. (2013) predicts a better RV precision than Beatty & Gaudi (2015) for stellar effective temperatures of >3500 K and wavelengths < 0.7 microns.
2) Beatty et al. (2015) predicts a better RV precision than Bottom et al. (2015) for a range of stellar effective temperatures and wavelengths of 1.0-2.3 microns, but particularly for M dwarfs with $T$ < 3000 K.

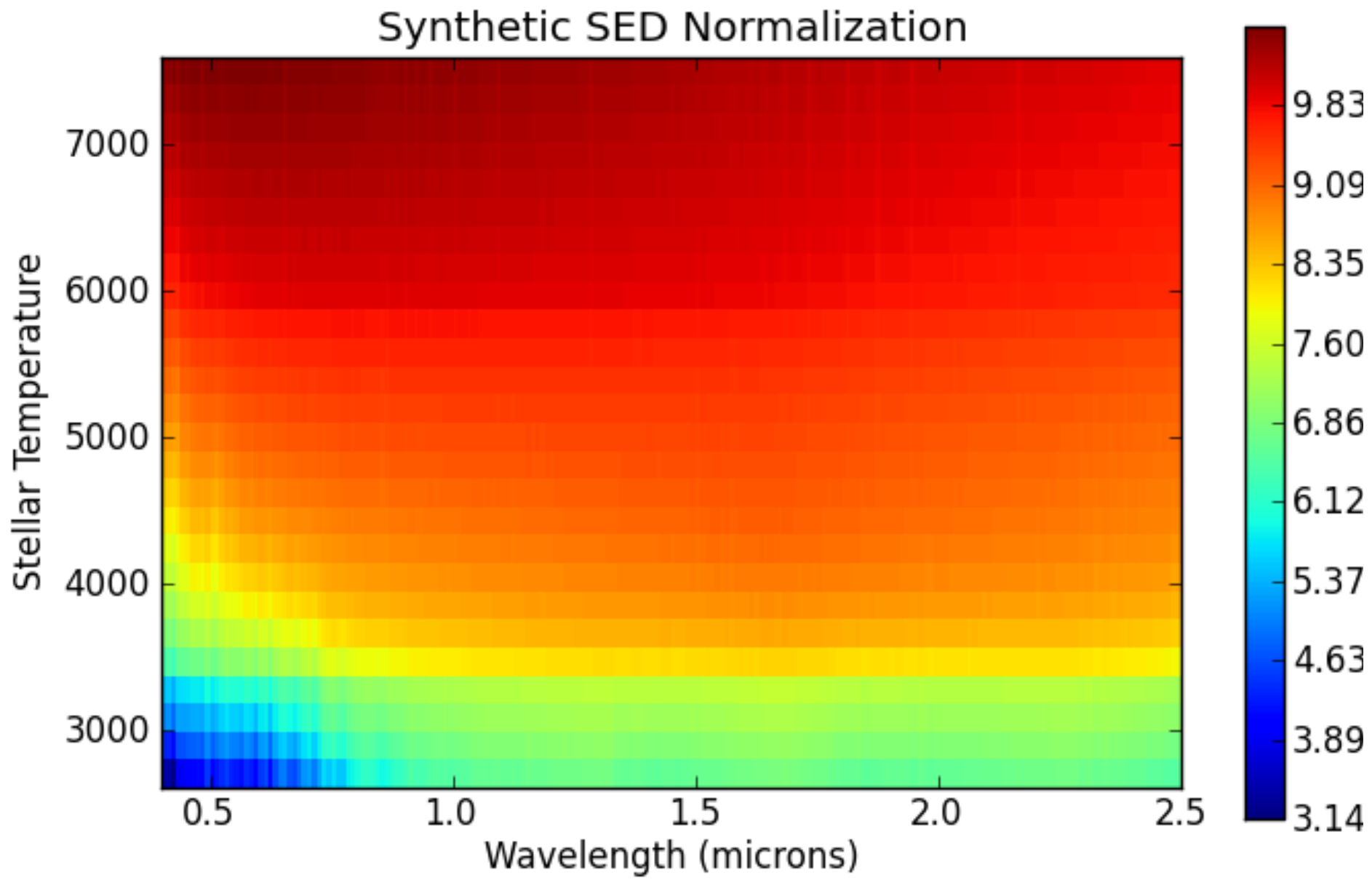

**Figure 28.** *A 2D surface plot of the log of the flux in log(Jy) from BT-Settl synthetic SEDs integrated over 10 nm wavelength bins and in increments of 200K in stellar effective temperature (Allard et al. 2012), to match the grid of wavelengths and temperatures in Beatty & Gaudi (2015). The color scale is linear.*

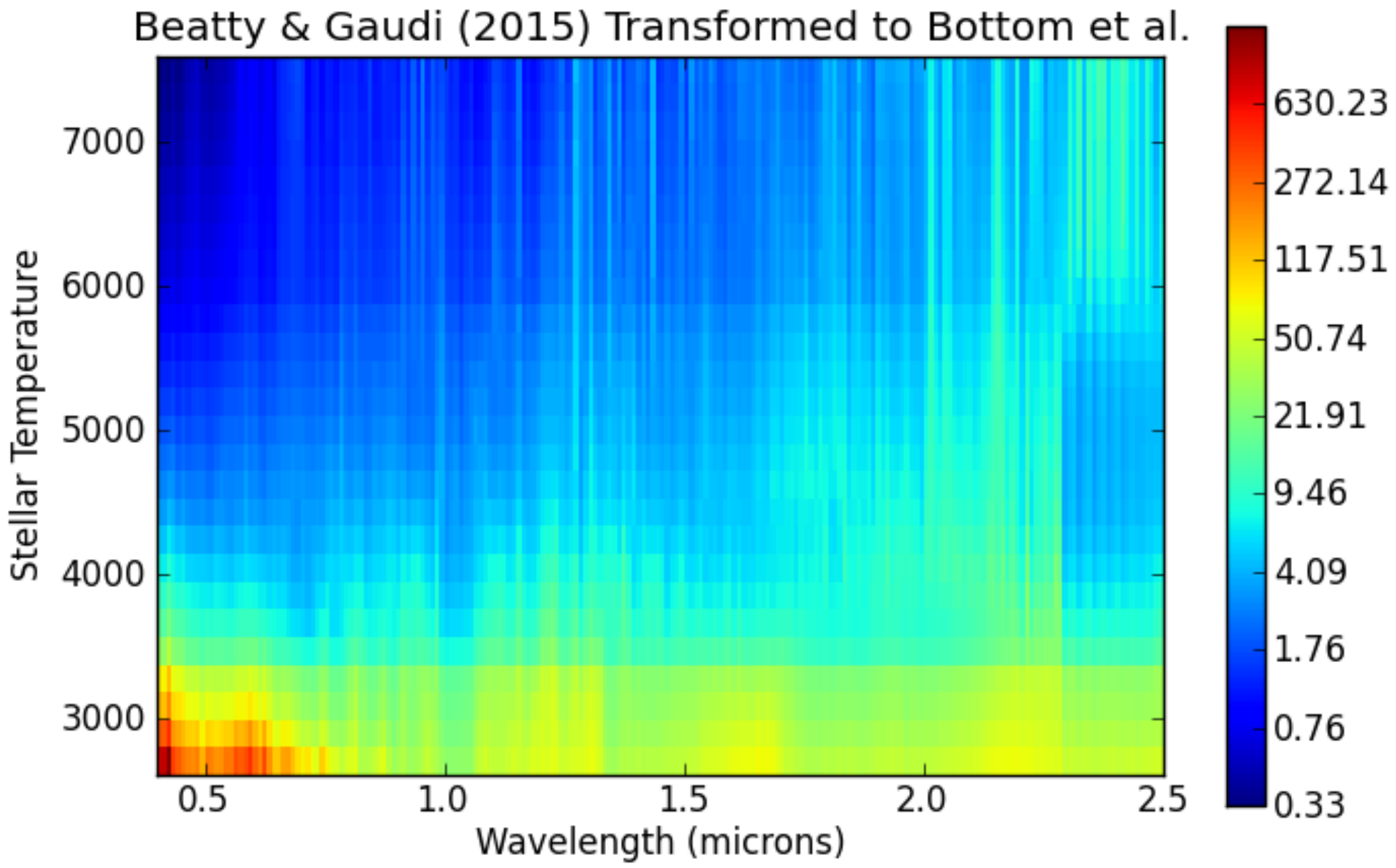

**Figure 29.** *A 2D surface plot of Table 1 from Beatty & Gaudi (2015) normalized and divided by Figure 28, to match the simulation parameter in Bottom et al. (2013) and units of m/s. The qualitative similarity with the independent analysis from Bottom et al. (2013) shown in Figure 26 is remarkable.*

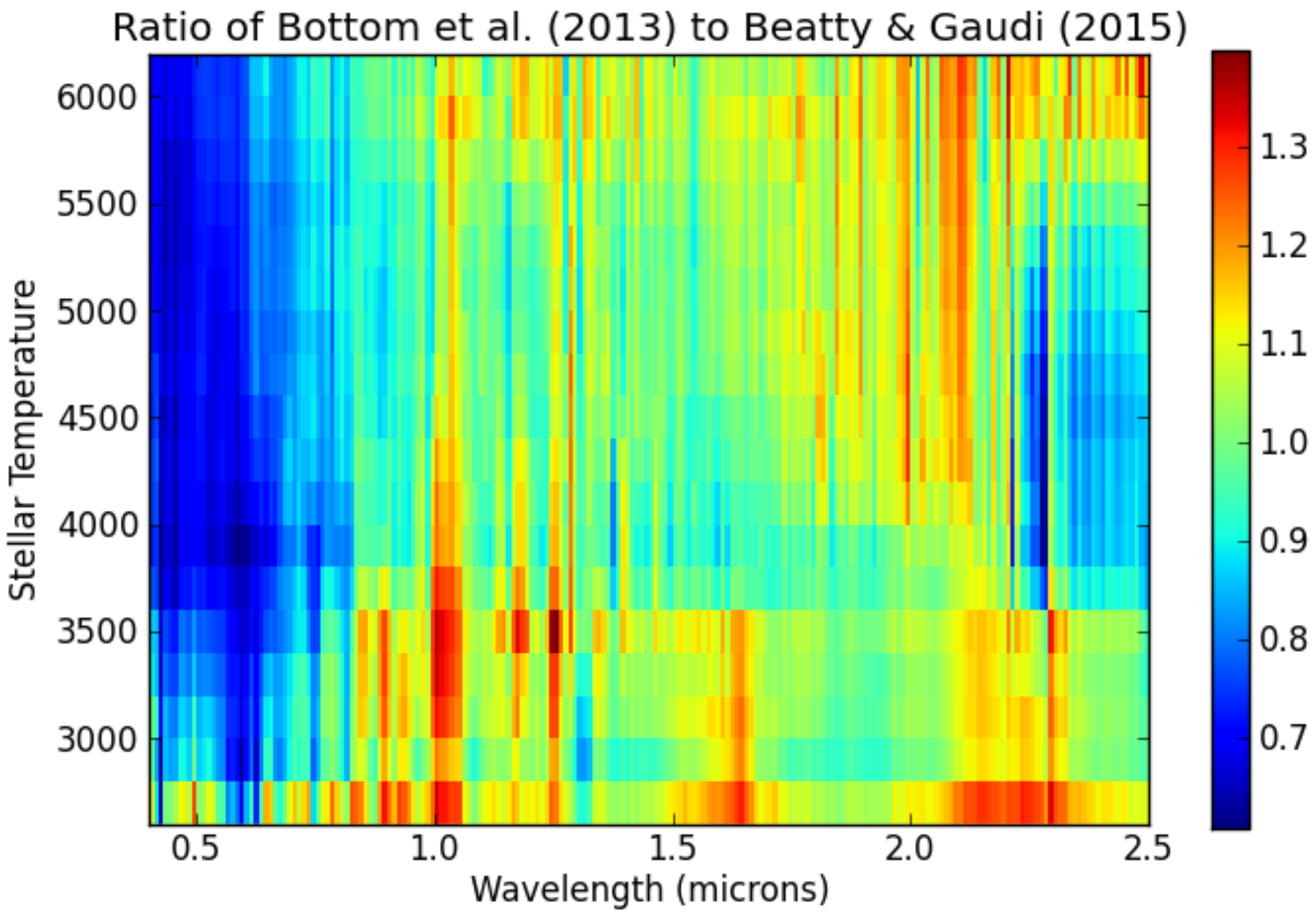

**Figure 29.** *A 2D surface plot the ratio of Figure 29 to Figure 27 of the simulated RV precision from Bottom et al. (2013) and Beatty & Gaudi (2015) as a function of wavelength and stellar effective temperature.*

Both studies conclude that, in the absence of stellar jitter:

1) RV precision for M dwarfs is comparably equal across a broad wavelength range of ~0.8 - 2.5 microns, and thus optimal in the visible given the lower costs and demonstrated instrumental RV stability, and
2) FGK stars are best studied at wavelengths of ~0.3-0.7 microns.

We now extend the analysis of Bottom et al. (2013) and Beatty & Gaudi (2015) to include a model for stellar jitter in Section 4.3.2.

### 4.3.2. The spectral type and wavelength dependence of stellar jitter

#### 4.3.2.1 Spectral type dependence

Vanderberg et al. (2015) assesses the impact on RV surveys from the rotational modulation of starspots and magnetic activity cycles as a function of spectral type and time-scale (see Section 5.2.4). By using the light curves of Kepler stars, and the relationship between RV stellar jitter and photometric variability from Aigrain et al. (2012), Vanderberg et al. (2015) derives the low, median and high values for the RV amplitude of stellar jitter from the rotational modulation of starspots as a function of spectral type. We reproduce Figure 2 from Vanderberg et al. (2015) as Figure 30. We adopt these low, median and high stellar jitter RV amplitude curves as a function of stellar effective temperature for our analysis (from 3200-6000 K), assigning a wavelength equal to 650 nm, the approximate center of the wide Kepler band-pass.

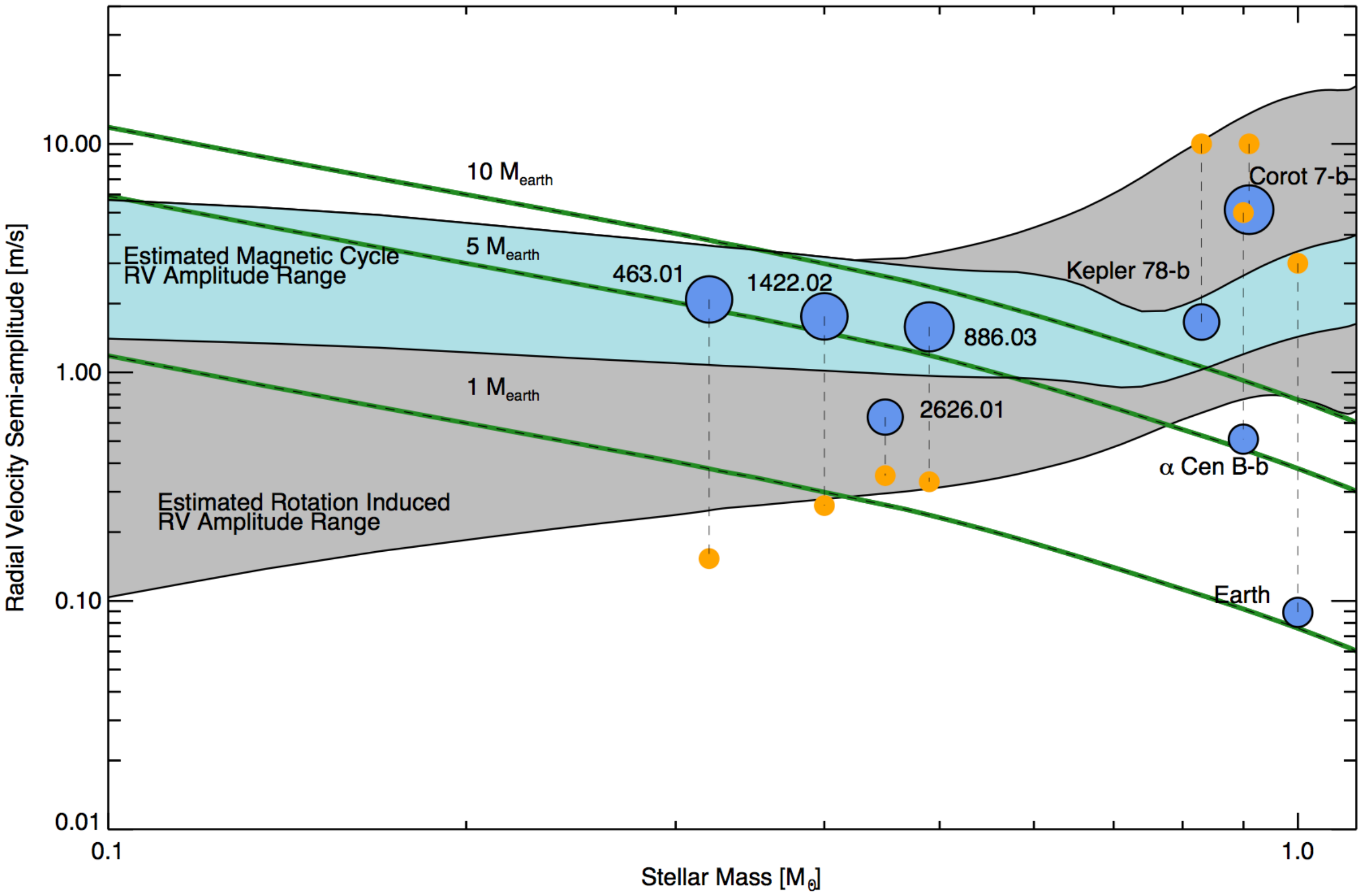

**FIgure 30.** *Reproduced from Figure 2 in Vanderberg et al. (2015): "Estimated characteristic semi-amplitudes of oscillatory RV signals. The RV semi-amplitude of habitable zone exoplanets of various masses are plotted in green lines. The semi-amplitude of rotational modulations are shown in grey, and the semi-amplitude of magnetic cycle RV modulations are shown in light blue. The rotational modulations are calculated assuming a uniform distribution of stellar inclinations. Characteristic exoplanets are plotted as blue dots, with the size of the dot corresponding to the planetary radius. Each exoplanet plotted is connected by a thin-dashed vertical line to an orange dot, corresponding to the predicted or observed RV semi-amplitude induced by stellar rotation. The four Kepler candidates shown have lower photometric amplitudes than typical M-dwarfs, hinting at observational biases in Kepler data against detecting smaller planets in long period orbits around photometrically noisy stars."*

#### 4.3.2.2 Wavelength Dependence

Reiners et al. (2010) presents an assessment of the utility of NIR PRVs, in particular finding that the RV photon noise is worse in the visible compared to the NIR only for spectral types later than ~M3, similar to the analysis presented in Section 4.3.1 from Bottom et al. (2013) and Beatty & Gaudi (2015). Reiners et al. (2010) take their analysis further, and presents an initial assessment of the wavelength dependence of stellar jitter. Reiners et al. (2010) develop a "toy" model for the rotational modulation of starspots that shows the amplitude of stellar jitter from starspot rotational modulation decreases at longer wavelengths.

The wavelength dependence of stellar jitter from the rotational modulation of starspots is proportional to *1/λ* to first order for temperature contrasts of 200 K between the starspots and photosphere. This follows directly from the Taylor series expansion for the ratio of black-body functions. To convert the flux ratio to a RV jitter amplitude, we reproduce Figure 10 in Reiners et al. (2010) with a $v \sin i$ = 2 km/s for the stellar rotation, a starspot contrast of 200 K, and a starspot filling factor of $f$=0.0125. At 650 nm, the "toy" model in Reiners et al. (2010) predicts a stellar jitter amplitude of 3.4, 7.3, and 11.4 m/s for stellar effective temperatures of 5700, 3700 and 2800 K respectively. Vanderburg et al. (2015) derive from Kepler light curves a RV jitter median amplitude of 2.7 m/s for a 5700 K main sequence star, comparable to the Reiners et al. (2010) model. However, the cooler stars RV jitter amplitudes in Reiners et al. (2010) are in disagreement with Vanderburg et al. (2015).

We adopt the general result from Reiners et al. (2010) that the RV jitter amplitude scales with wavelength as the ratio of blackbody functions, which is *1/λ* to first order for all stellar effective temperatures. In our analysis, we retain the full blackbody function flux ratio, and not the 1st order approximation. In Section 5.3 we present literature results on the wavelength dependence of stellar jitter that support our adoption of this wavelength dependence.

#### 4.3.2.3 Combining the wavelength and spectral type dependence of stellar jitter

By combining the stellar effective temperature dependence of RV jitter amplitudes from Vanderberg et al. (2015) with the wavelength dependence of RV jitter amplitudes from Reiners et al. (2010), we present in Figures 31-33 RV jitter models as a function of stellar effective temperature and wavelength for low, median and high stellar jitter levels.

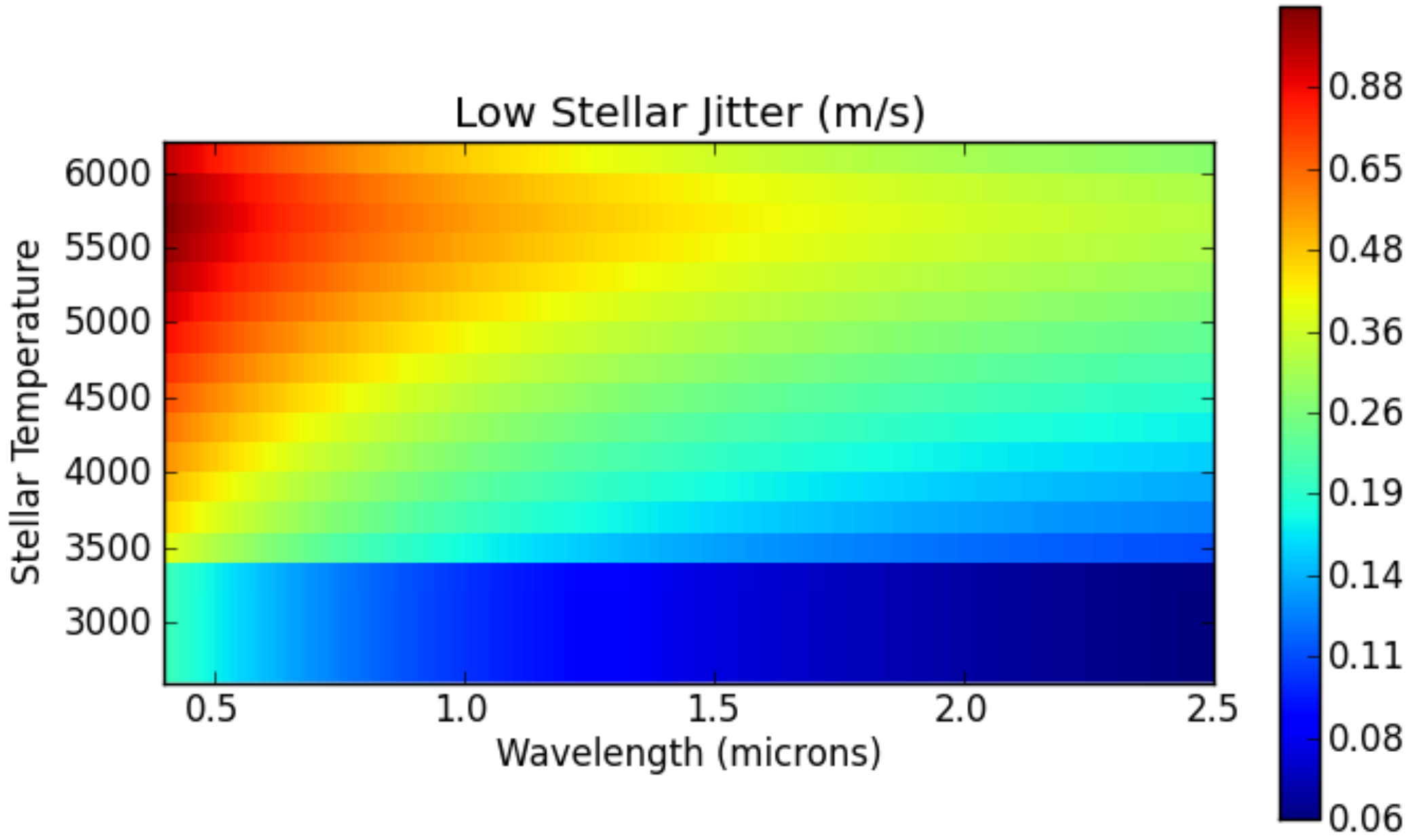

**FIgure 31.** *Our model for RV stellar jitter as a function of stellar effective temperature and wavelength. Values for temperatures < 3200 K are set equal to that for T=3200 K to match the Beatty & Gaudi (2015) temperature minimum. The color scale is logarithmic. Units are m/s.*

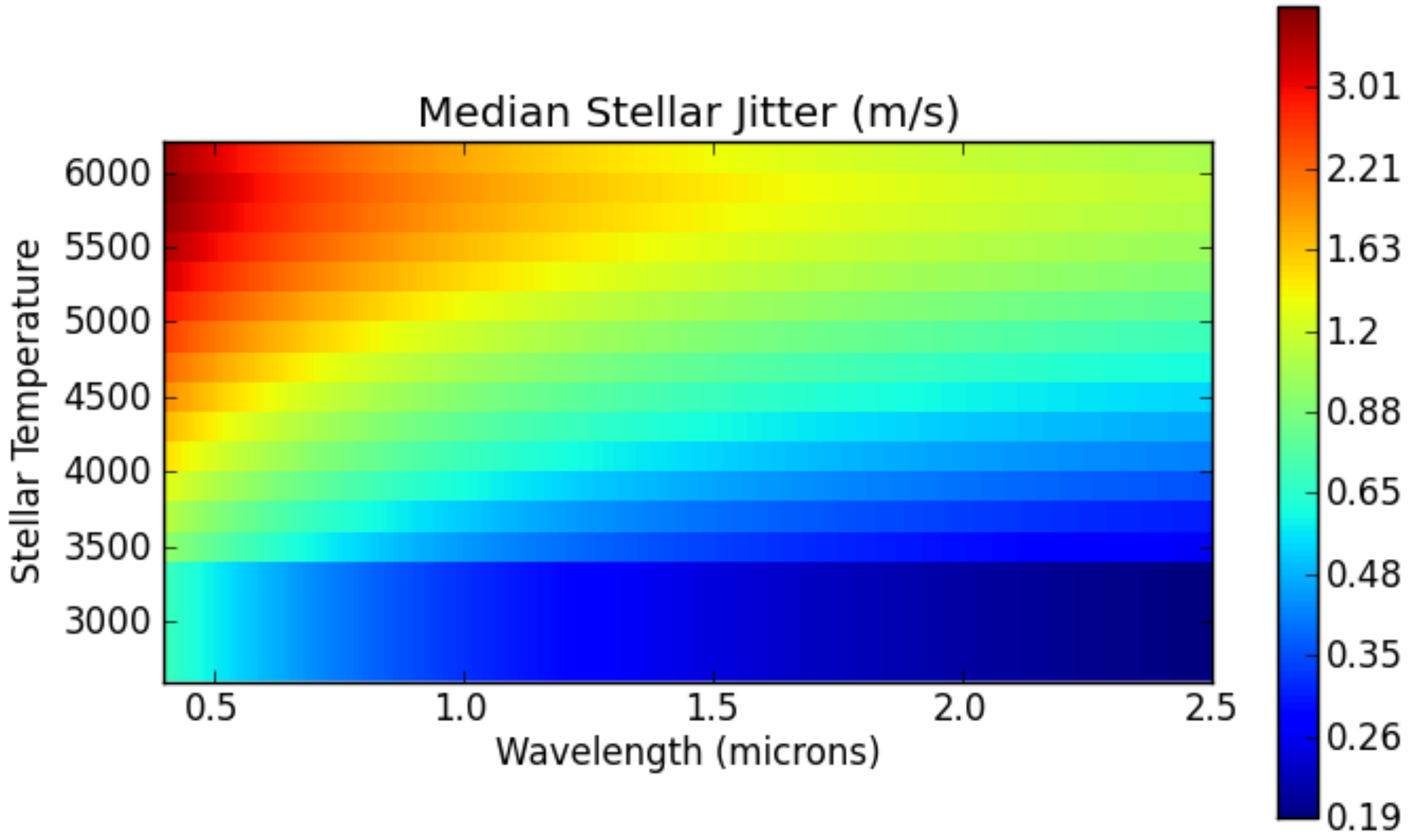

**FIgure 32.** *Same as Figure 31, but for median stellar jitter.*

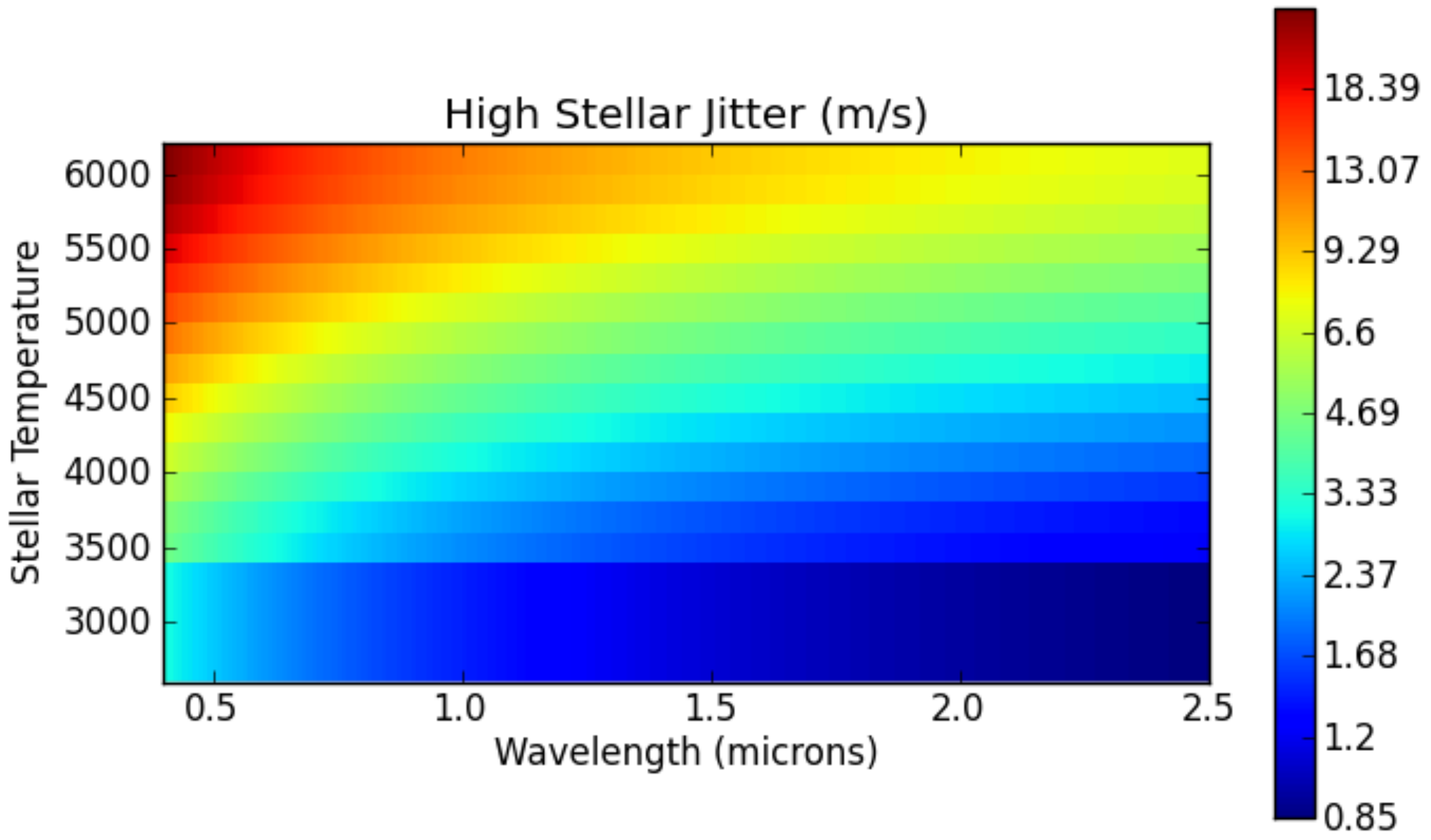

**FIgure 33.** *Same as Figure 31, but for high stellar jitter.*

### 4.3.3. Combining photon noise and stellar jitter as a function of spectral type and wavelength

With a model for stellar jitter, we can now combine it with the analysis in Bottom et al. (2013) to evaluate the importance of stellar jitter and photon noise on PRV measurements as a function of stellar effective temperature and wavelength. We add the noise terms in quadrature, for exposure times of 1 hr, four levels of stellar jitter (none, low, medium and high), and for 4-m and 10-m class telescopes. We present the model results in Figures 34-41.

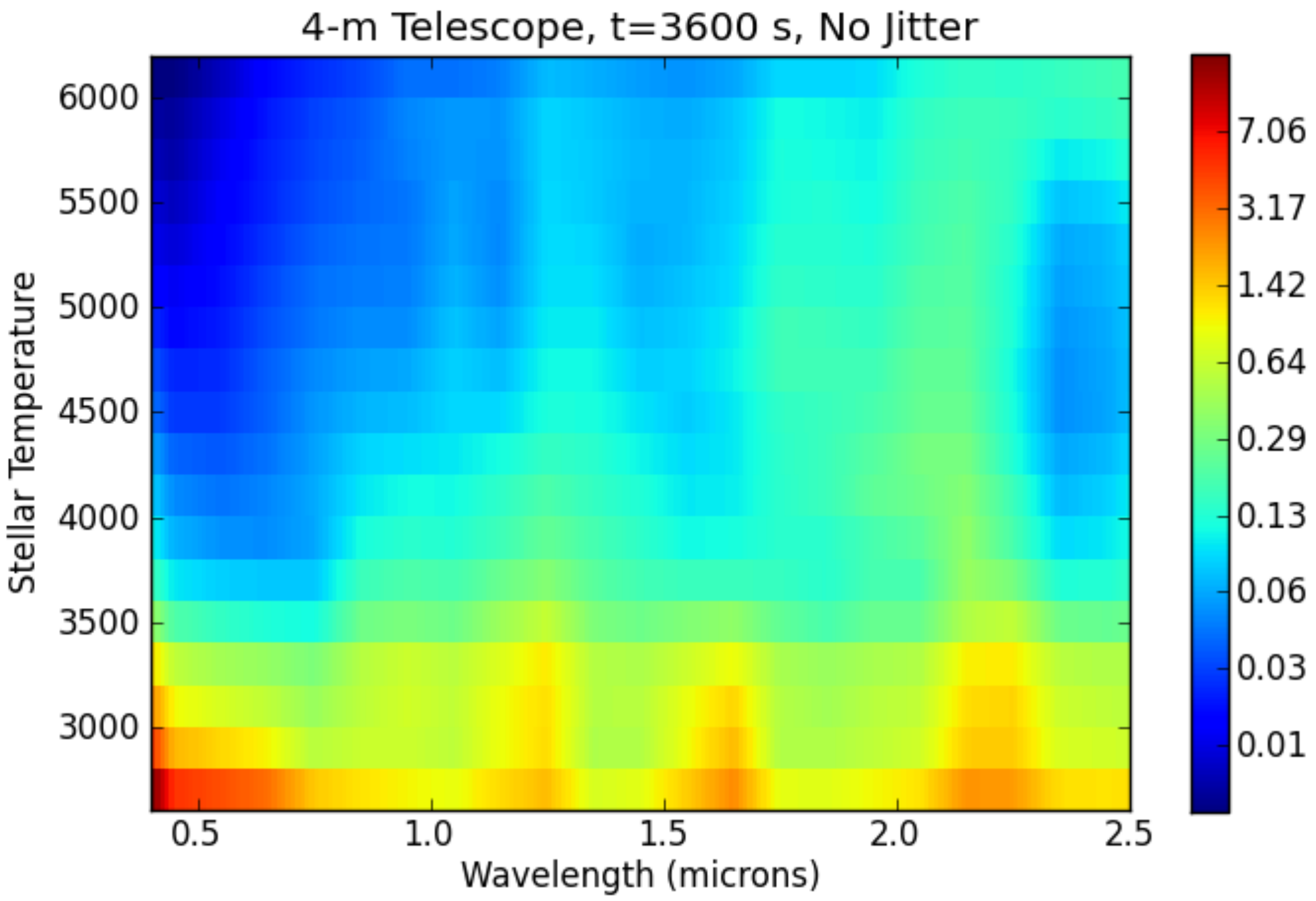

**FIgure 34.** *For a main sequence star with a distance of 10 pc, we plot the measurement precision from photon noise as a function of wavelength and stellar temperature for a 4-m telescope with an exposure time of 1 hour, scaled from Figure 1 in Bottom et al. (2013). The units of the 2-D surface plot are m/s. The color intensity has a logarithmic scale. In this plot, no contribution from stellar jitter is included.*

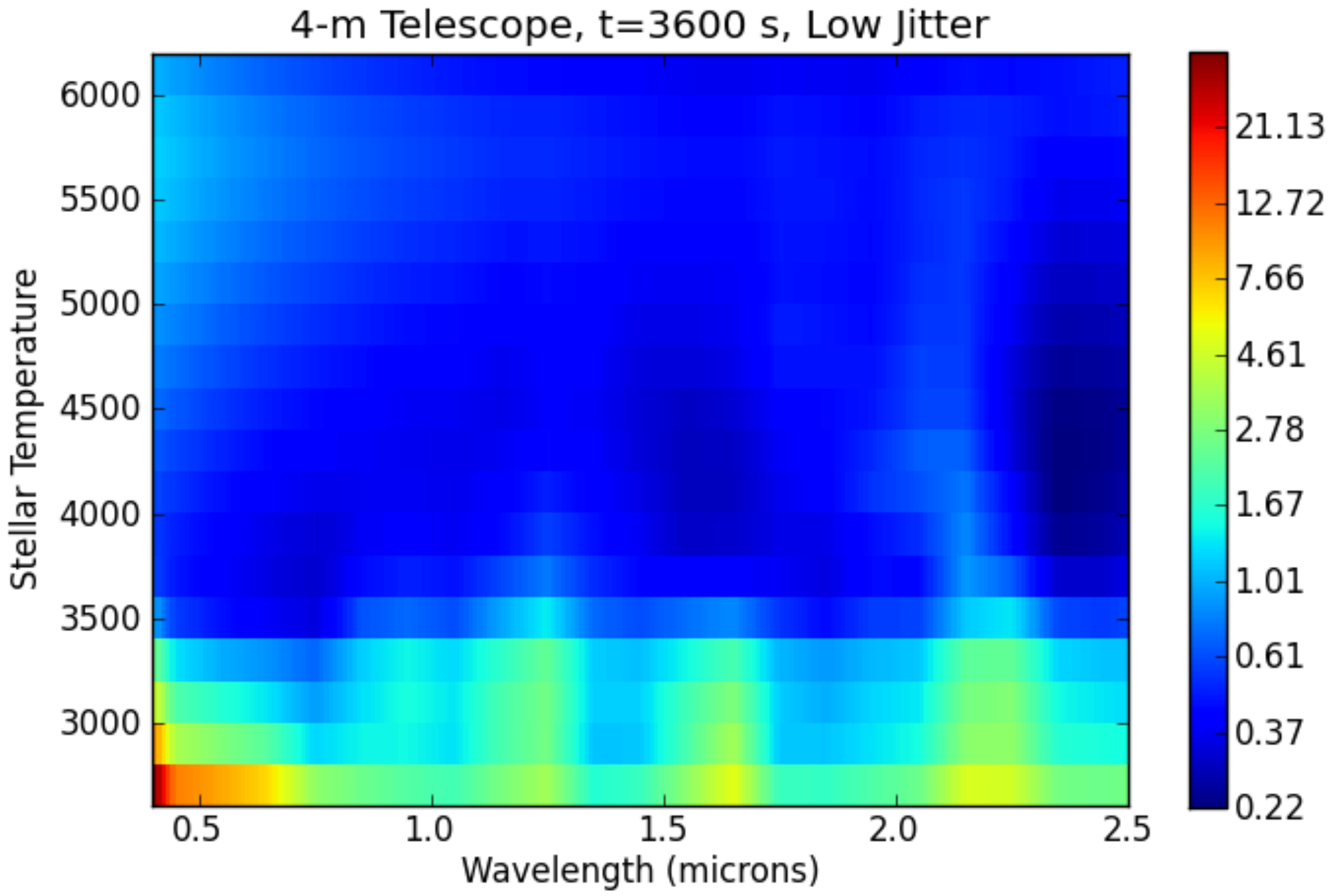

**FIgure 35.** *The same as Figure 34, but for low stellar RV jitter added in quadrature.*

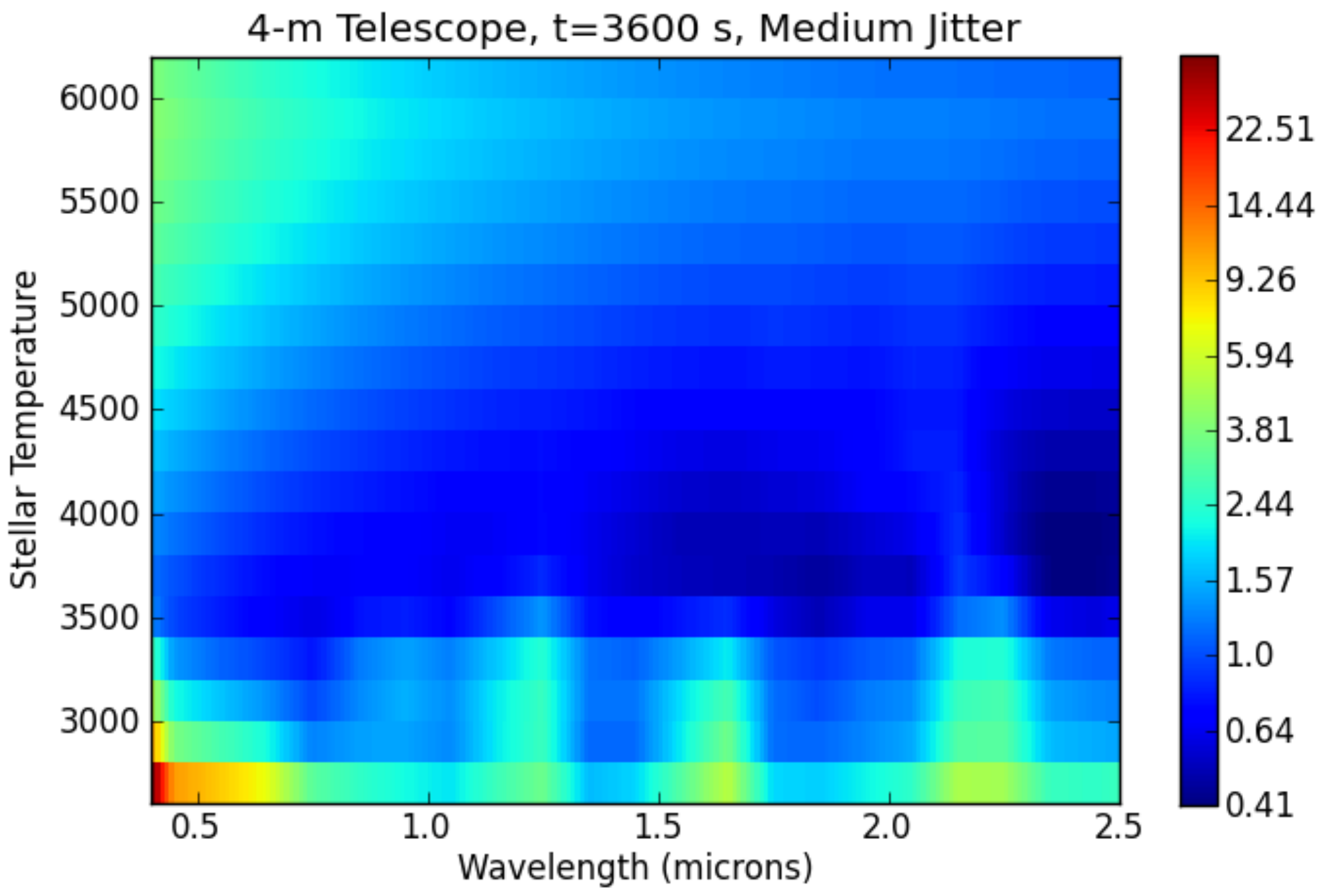

**FIgure 36.** *The same as Figure 34, but for low medium RV jitter added in quadrature*

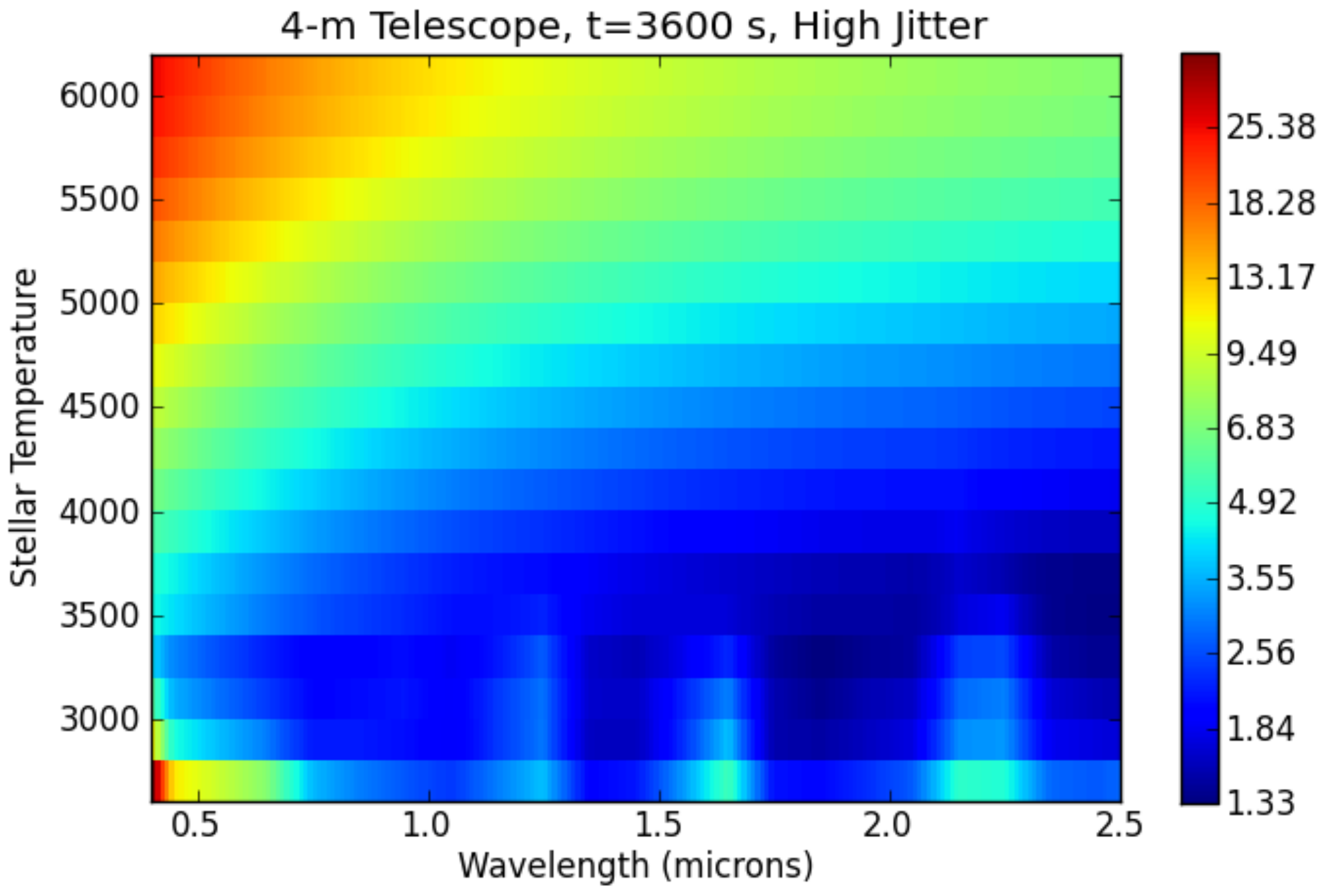

**FIgure 37.** *The same as Figure 34, but for low high RV jitter added in quadrature.*

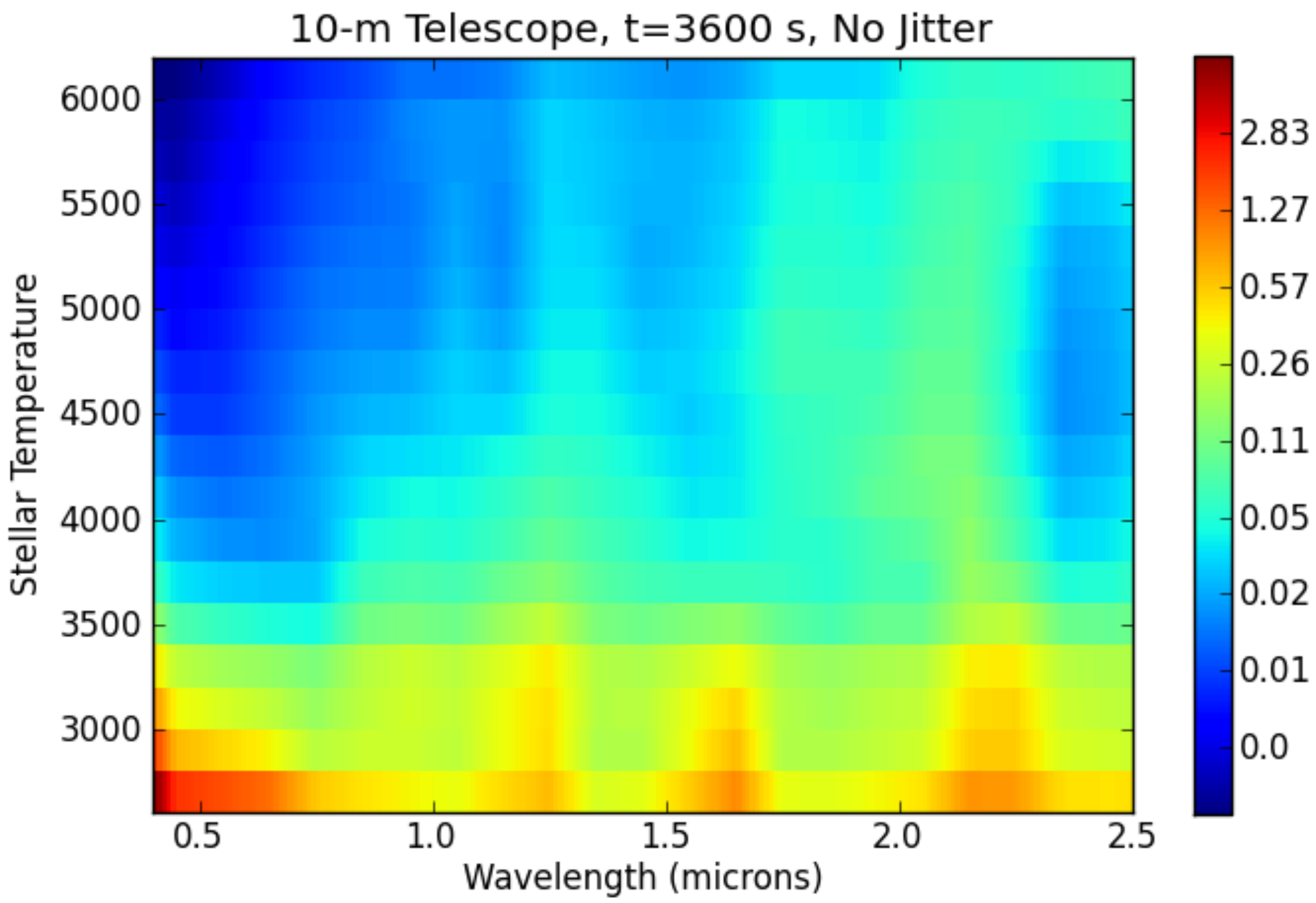

**FIgure 38.** *The same as Figure 34, but for a 10-m telescope.*

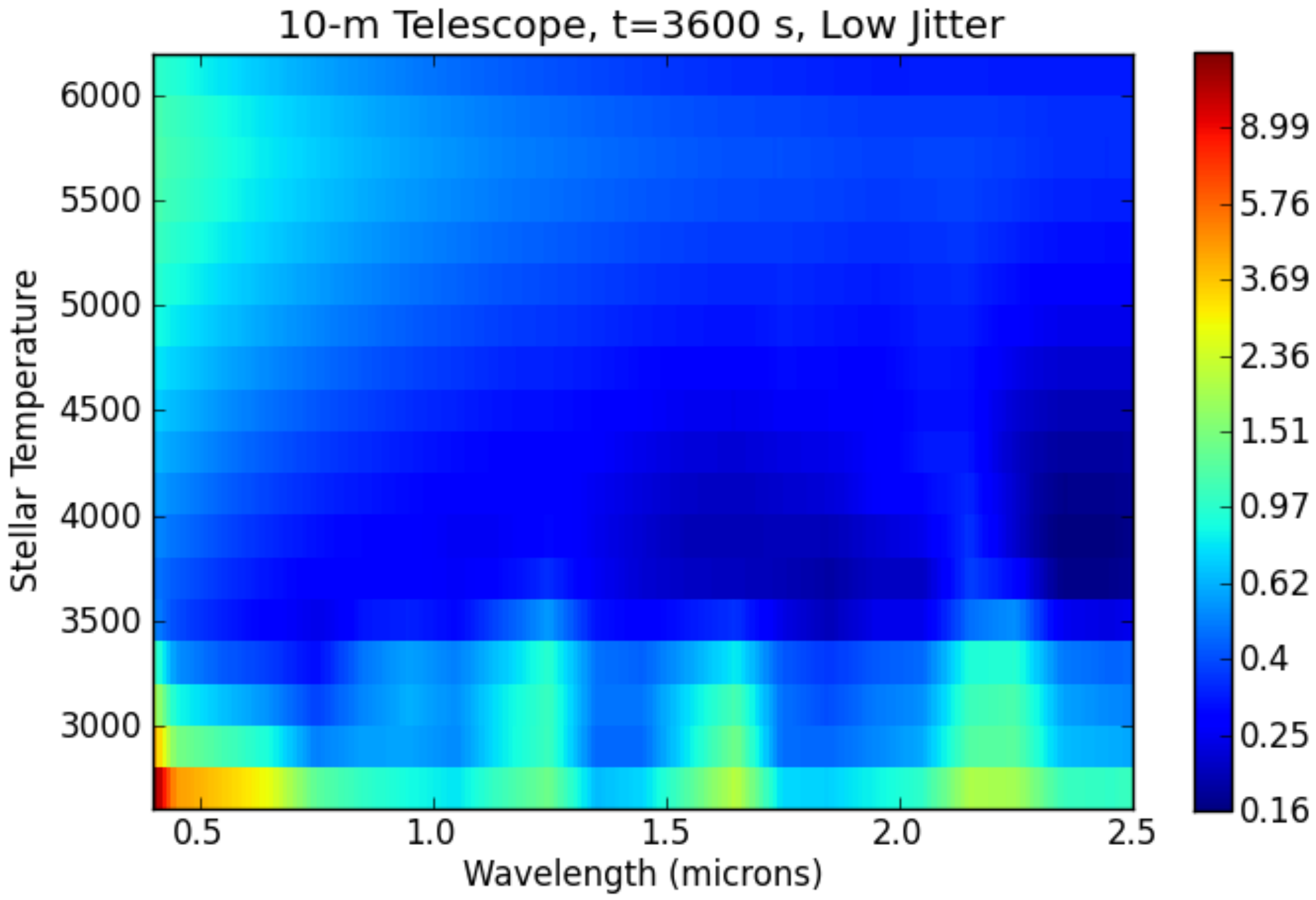

**FIgure 39.** *The same as Figure 35, but for a 10-m telescope.*

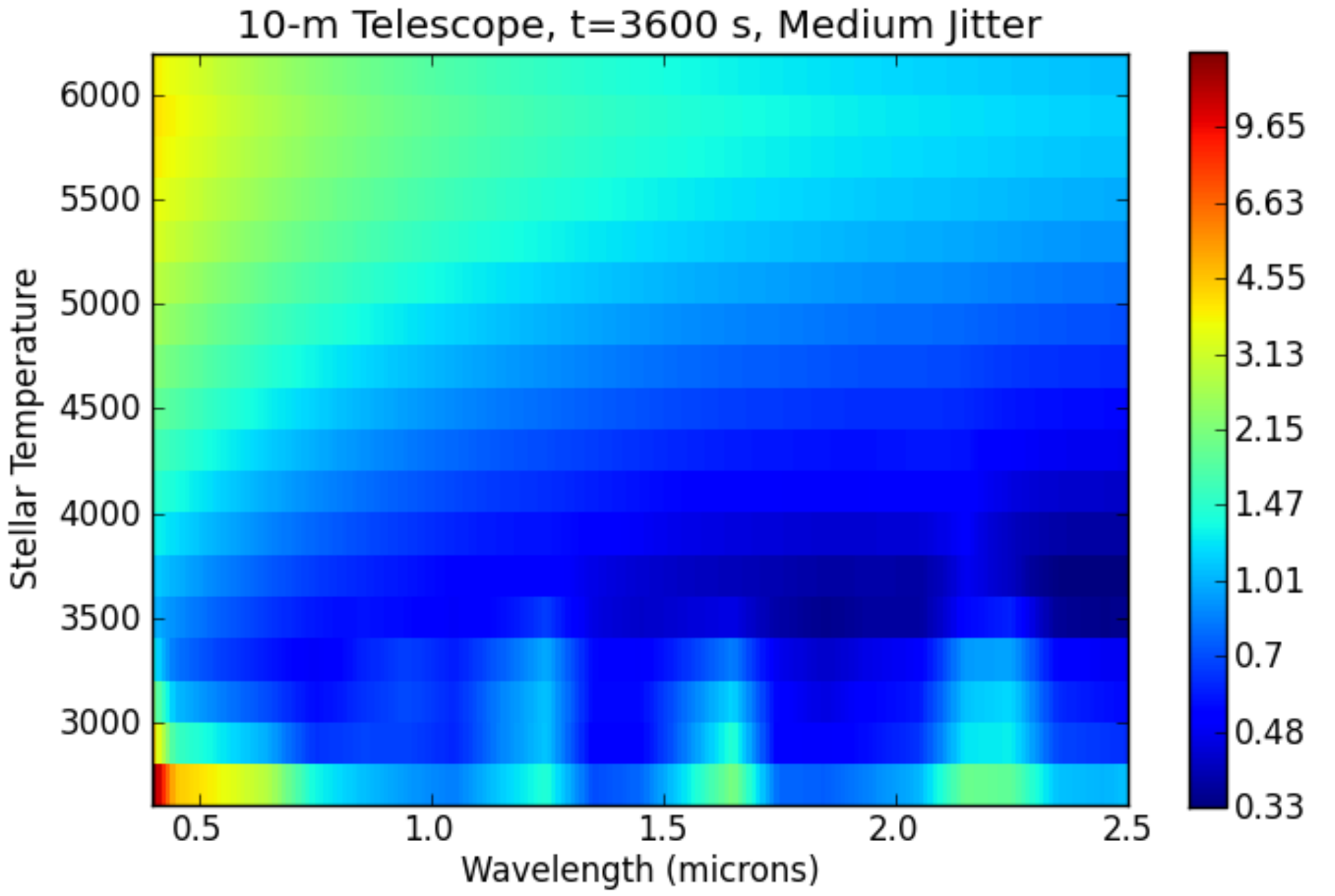

**FIgure 40.** *The same as Figure 36, but for a 10-m telescope.*

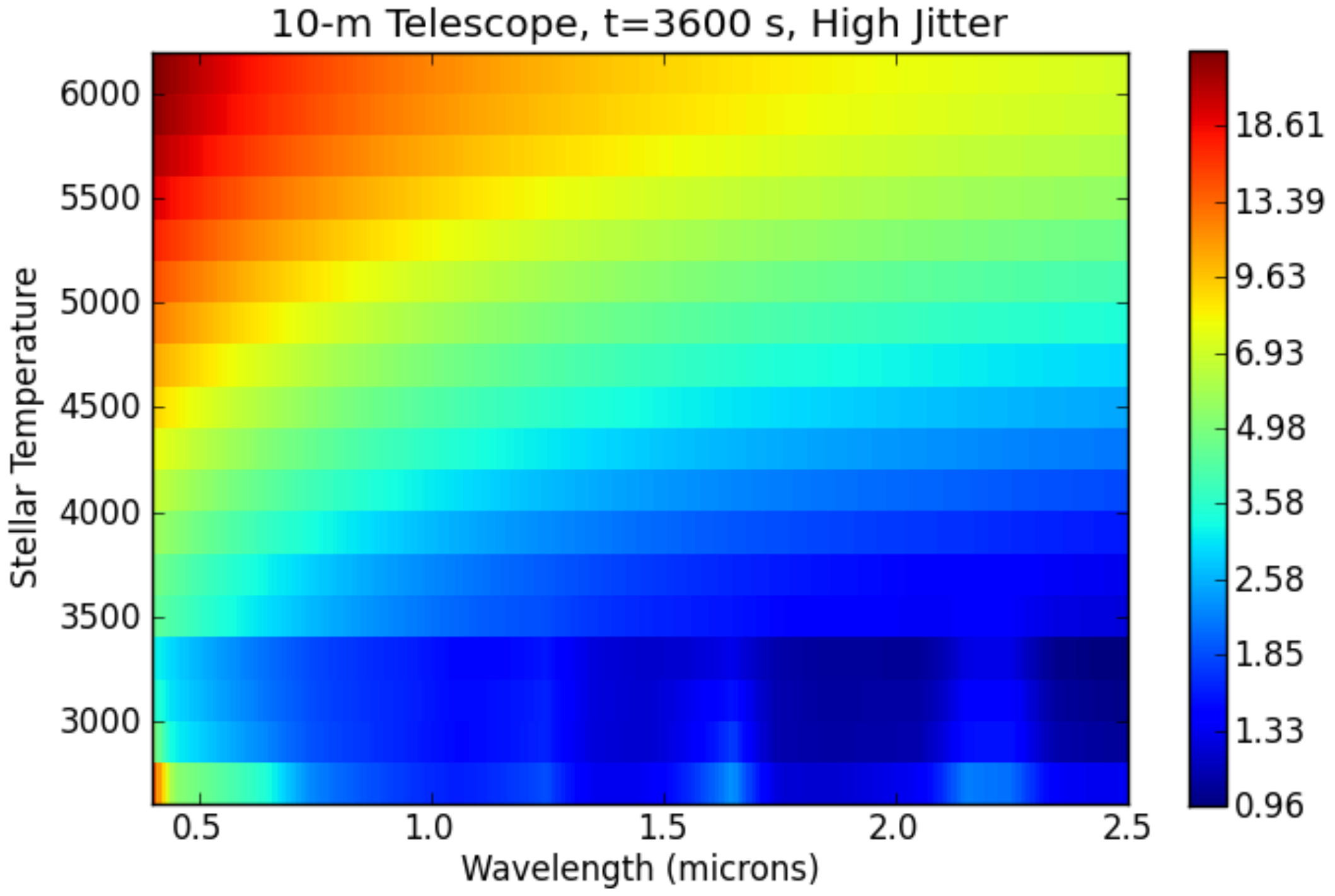

**FIgure 41.** *The same as Figure 37, but for a 10-m telescope.*

### 4.3.4. Summary Conclusions

We draw the following conclusions from Figures 34-41 in Section 4.3.3:

1) Stellar jitter, when treated as a systematic noise term, dominates the limiting RV precision for stars with T>4000 K, even for relatively quiescent stars, with integration times of 1 hour on 4-m and 10-m telescopes.
2) For F dwarfs and early G dwarfs, stellar jitter is less pronounced for $\lambda$ > 1 micron. However, for relatively quiescent stars, the NIR offers only marginal advantage over the visible. This advantage is more prominent for more active stars.
3) For mid-to-late G, K and early M dwarfs, in the presence of any stellar jitter, the optimal wavelength regime of observation appears to be $\lambda$ > 2.3 microns with the CO absorption bandhead. Thus, NIR PRVs are critical to reach sub-m/s exoplanet signals.
4) For M dwarfs at 10 pc, with T<4000 K and median stellar jitter, photon noise dominates the stellar jitter for $\lambda$ > 0.8 microns, with some marginal advantage at $\lambda$ > 2.3 microns with the CO absorption bandhead.
5) Even low stellar jitter can triple the RV uncertainty compared to photon noise alone, for a given effective temperature and wavelength of observation.

In Section 5.2, we discuss the time-scales and properties of stellar jitter. In our preceeding analysis, we have treated stellar jitter as a noise term. However, stellar jitter is a signal. With proper PRV wavelength coverage (ie, spectral grasp) to take advantage of the wavelength dependence of stellar jitter, simultaneous photometric monitoring, and/or a more dense observation cadence on the time-scales of stellar rotation periods (e.g. one observation at least every week for Sun-like rotation period), it may be possible to model and remove stellar jitter from RV time-series. However, we expect the signal analysis will not be ideal, particular in the presence of multiple exoplanets (Robertson et al. 2014, Robertson & Mahadevan 2014).

These future challenges represent a risk that requires investment in instrument technology and data analysis techniques, especially given the lack of demonstrated NIR precision of < 5 m/s to date. The instrumentation risks may be retired in the next 1-3 years (e.g. by 2017) as planned NIR PRV spectrometers are commissioned by the PRV community independently of NASA efforts (Section 6.1). Assuming this risk is retired and irrespective of cost, NIR PRVs offer a promising wavelength regime to detect exoplanet signals of $K$ < 1 m/s. This will be particularly important for surveying Exo-C targets for terrestrial mass planets at 1 AU.

To conclude, the third spectrometer options presented at the beginning of Section 4 and in the Executive Summary offers the potential to mitigate the impact of stellar jitter on PRV sensitivity to exoplanets:

3) Access to an IR spectrograph or second arm of a visible spectrograph on a 4-m class telescope equipped with a CARMENES-like instrument stable at the level of <1 m/s dedicated to PRV observations, to be available to the TESS mission, to assess the removal of wavelength-dependent stellar-jitter, and to survey nearby M dwarfs for habitable planets.

Such a capability will both complement and enhance visible PRV instrumentation.

## 4.4 Seeing or Diffraction Limited

It is possible to correct for image-blurring effects introduced by Earth's turbulent atmosphere using AO to reach the diffraction limit of a telescope. For large-aperture telescopes, this corresponds to an improvement in spatial resolution by a factor of 10-30. Coupling a spectrometer to the AO-corrected starlight completely changes the optical, mechanical, and thermal design possibilities for the spectrometer.

The primary benefits of a diffraction limited spectrometer compared to a seeing limited spectrograph are driven by the smaller spectrograph size and optics:

1) Cost, without including the cost of equipping a telescope with an AO system.
2) The prospect for illumination stability via the use of a single mode fiber.
3) The potential for improved mechanical/thermal stability.
4) High spectral resolutions are possible.

The primary drawbacks of a diffraction limited spectrometer compared to a seeing limited spectrograph are:

1) Again cost, for telescopes not already equipped with an adaptive optics system.
2) The lack of on-sky demonstrated performance with diffraction limited PRV spectrographs to date.
3) Throughput efficiency coupling starlight into single-mode fibers, which is somewhat mitigated by other throughput savings (ie, from not needing optics to mitigate modal noise).

Fortunately, to address the drawbacks there are active areas of PRV research. There are several groups pursuing diffraction-limited or single mode fiber spectrometers, including iLocater (PI: Justin Crepp), and the Small Red Spectrograph (PI: Cullen Blake). Lab tests of spectrometers have shown promise in instrument illumination stability. However, coupling efficiency for PRV spectrometers remains a concern for matching diffraction limited PSFs to the spatial profiles of single mode fibers, particularly when some of the possible spectral grasp is being used for the adaptive optics system. We discuss diffraction limited high-resolution spectroscopy, and its science application strengths, in more detail in Section 6.4.

We conclude that the fourth spectrometer options presented at the beginning of Section 4 and in the Executive Summary offers the potential for long-term precision goals and cost savings, possibly requiring future investment in risk retirement with on-sky demonstrations:

> 4) Access to a NIR or visible diffraction-limited spectrograph for a large-aperture telescope with extreme AO to achieve unprecedented precision and compete with ESPRESSO.

# 5. Exoplanet Science Objectives

## 5.1 A Summary of the Historical Relevance of PRVs to the Exoplanet Field

Over 650 exoplanets have been discovered or confirmed over the past 23 years with the PRV method (Akeson et al. 2013, Wright et al. 2011), starting with the discoveries of HD 114762 b (Latham et al. 1989) and 51 Peg b (Mayor & Queloz 1995). The PRV method exploits the time domain and spectroscopy to measure subtle changes in Doppler shifts with time in Nyquist sampled spectra with resolutions on the order of 100,000 (R=lambda/delta_lambda). The Doppler shift is produced by the reflex motion of the star in response to a Keplerian orbital companion. The projected reflex velocity semi-amplitude for a circular orbit companion is given by:

$$K = 28.4\ M_p \sin i\ /\ (M_* + M_p)^{2/3}\ P^{-1/3}\ \mathrm{m/s}$$

where $M_p$ is in Jupiter masses, $(M_*+M_p)$ is in solar masses and P is in years, and the unknown inclination i is in radians (Ch 7, Exoplanet Community Report). Doppler velocities of a few m/s produce shifts in the observed spectra of ~1/1000th of a detector pixel. For comparison, the barycenter motion of the Earth can produce apparent radial velocity shifts of up to ±30 km/s, and instrumental changes in atmospheric pressure, temperature, illumination of the optics, and orientation in the Earth's gravity on a classical spectrograph can produce systematic shifts of ~5 km/s. Two different approaches have been pursued for measuring the much smaller changes in stellar RVs induced by orbiting exoplanets. One approach introduces a gas absorption cell into the beam of stellar light entering the spectrograph, allowing high-fidelity calibration and removal of the various instrumental shifts. The second approach focuses on stability, both by illuminating the spectrograph with stable images using fiber feeds and by careful design and construction to eliminate instrumental drifts. Particularly important is control of the atmospheric pressure at the dispersing elements, to eliminate changes in the index of refraction. This is the reason the HARPS spectrographs are in vacuum chambers. Continuous monitoring of instrumental drifts has also been implemented in some spectrographs. Precise determination of the barycenter motion of the Earth taking into account any changes in the rate of exposure during an observation can also be important. The capability of the community to correct for systematic sources of Doppler shifts has steadily improved with time, and consequently the exoplanet detection sensitivity has continually improved to smaller mass planets and longer orbital periods. (Figure 43).

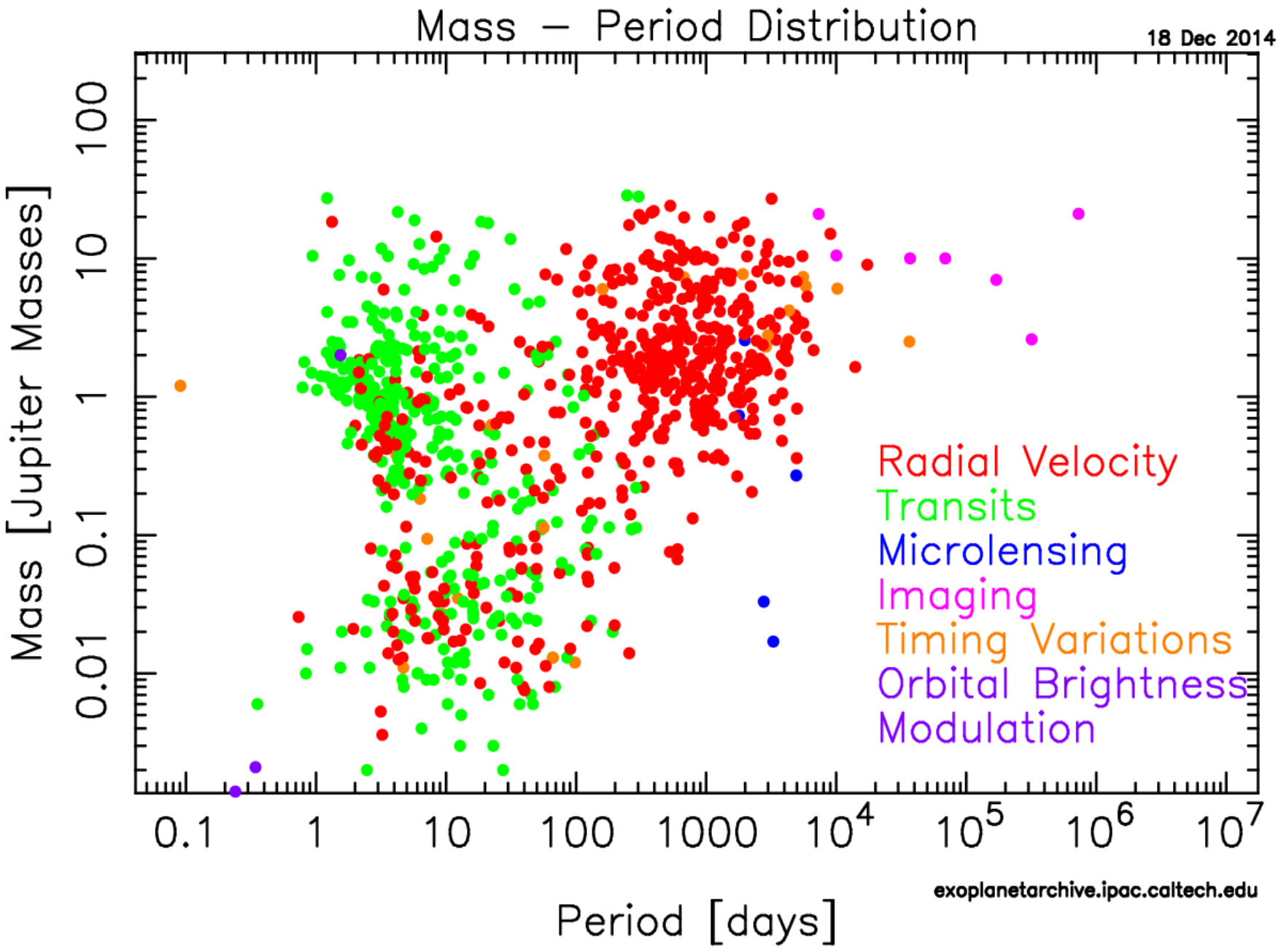

**Figure 42.** *Confirmed exoplanets as of 18 December 2014, from different exoplanet discovery techniques, including PRVs, transits, microlensing, pulsar timing and direct imaging, plotted as a function of orbital period and exoplanet mass (NASA Exoplanet Archive, Akeson et al. 2013).*

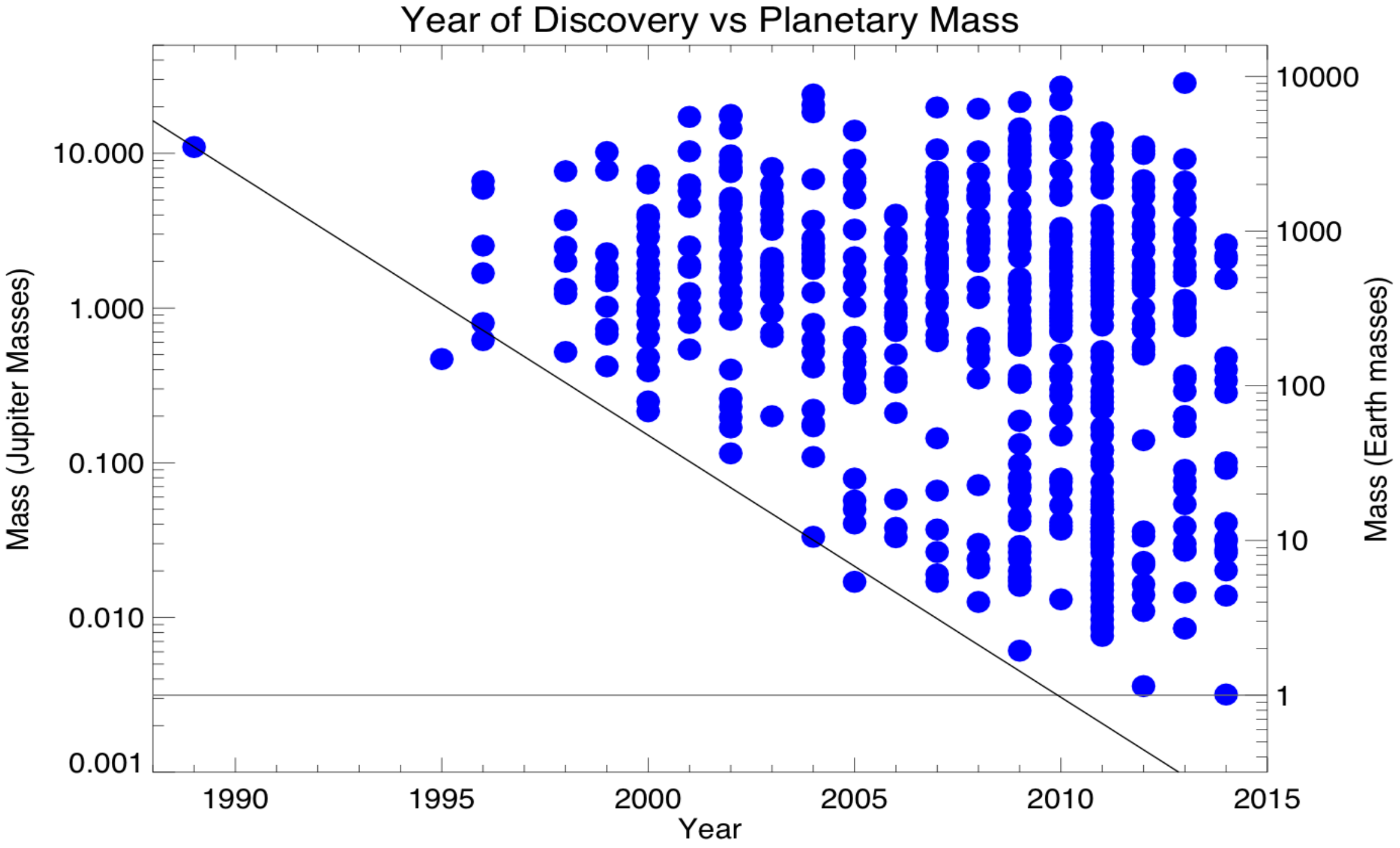

**Figure 43.** *Exoplanet masses determined using RVs as a function of discovery year. For transiting planets the actual mass is plotted. For RV systems with unknown orbital inclinations the minimum mass is plotted. The trend towards lower planet masses has not proceeded the past 5 years as extrapolated from the first 20 years of discovery.*

With these advances in calibration and/or stability, the PRV technique enabled the discovery of hot Jupiters (Mayor & Queloz 1995, Marcy & Butler 1996), systems with more than one planet (Butler et al. 1999), Jupiter analogs (Marcy et al. 2002), planets in orbital resonance (Laughlin et al. 2005), super-Earths and possible water-worlds (Mayor et al. 2011, Pepe et al. 2011), planets around evolved subgiant stars (Johnson et al. 2011), the Rossiter-McLaughlin effect and determination of spin/orbit (mis)alignment (Queloz et al. 2000, Winn et al. 2005, Triaud et al. 2009), and many other discoveries. Consequently, much of what we know today about exoplanets is motivated by the PRV discoveries of the past quarter-century.

PRVs complement the discovery space of other exoplanet detection methods, including the transit method, astrometry, direct imaging, and microlensing. When combined with PRVs, the transit method enables empirical constraints on the internal structure and composition of exoplanets (Seager et al. 2007, Bean et al. 2011). Combined with precise astrometry, full 3-D orbital solutions may be obtainable for non-transiting exoplanets orbiting FGK stars, and astrometry can detect exoplanets around more massive radiative-atmosphere main-sequence stars that lack significant RV information content in their visible spectra (Brown et al. 2009). Microlensing and direct imaging probe exoplanets at larger orbital separations (Gaudi et al. 2008, Gould et al. 2010, Marois et al. 2008, 2010), and are just starting to intersect in semi-major axis with the RV method with projects such as TRENDS (Crepp et al. 2012, 2013a,b). The ensemble of exoplanet discoveries with RVs and other detection methods place into context the future long-term goals of the exoplanet community to discover, directly image, and characterize habitable Earth twins.

## 5.2 Astrophysical constraints on the radial velocity method

Astrophysical sources of RV noise and their characteristic time-scales and wavelength dependence limit the predictable yield of future PRV surveys. Stellar jitter can include P-mode oscillations (minutes), granulation (hours), rotationally modulated spots and plages (days to months), magnetic activity cycles (years), and rotational line-broadening in younger stars (which lowers the RV information content). These astrophysical noise sources are discussed in greater detail in Dumusque et al. (2011a,b), Reiners et al. (2010), Frohlich & Lean (2004), Kjeldsen & Bedding (2011), Meunier et al. (2010), Basri et al. (2010, 2011), McQuillan et al. (2012), Barnes et al. (2010), and in particular for the Sun in Lockwood et al. (1992, 2007), Radick et al. (1998), Radick (2003), and Makarov et al. (2009).

The present state-of-the-art in correcting PRVs for stellar jitter is illustrated by the reported detection of an Earth-mass planet, α Cen B b, by Dumusque et al. (2012). After correcting the velocity of the V=1.3 mag K1 dwarf host star, α Cen B, for its 80-year orbital motion with its visual companion α Cen A; for the shifts of a few m/s due to a magnetic cycle of 8.8 years; for shifts connected with the star's rotation in 35 days; and after minimization of the shifts due to granulation over a few hours and acoustic oscillation over a few minutes by a combination of extended exposure times and repeated visits during the same night; then a 3.23-day orbit with semi-amplitude of K=51 +/- 4 cm/s emerged, corresponding to a minimum mass of $M_P \sin(i)$

=1.1 $M_{Earth}$ , and a maximum mass of 2.7 $M_{Earth}$ from formation scenarios and dynamical simulations (Plavchan et al. 2015). A total of 453 observations spanning four years were required for this RV detection. Current near-infrared PRV performance has not yet achieved an adequate precision (< 3 m/s) to assess the potential benefit for correcting stellar jitter in quiescent main sequence stars, but has been applied to younger, more active stars (e.g., Bailey et al. 2012).

### 5.2.1 Stellar Activity Features and Time-Scales

A number of authors have proposed that stellar RV jitter can be mitigated by adopting optimized observing cadences (Dumusque et al. 2011a,b; Lagrange et al. 2010, Meunier et al. 2010). Dumusque et al. (2012) utilized the time-scale and lack of phase coherence to identify, model, and remove the RV activity due to the rotational modulation of star-spots, and utilized the correlation of RVs with the Calcium H&K indicator to identify, model and remove the RV distortions due to the long-term magnetic activity cycle. Makarov et al. (2009) recommended taking advantage of the magnetic activity cycle of stars to focus exoplanet searches at periods of minimum activity. A one-year habitable zone orbit is temporally situated nicely between the rotational and magnetic activity time-scales, and thus the techniques presented in Dumusque et al. (2012) may be applicable at longer time scales in future surveys. One clear consequence of stellar jitter is the need to average over p-mode oscillations with minimum integration times of ~5 minutes for Solar-type stars. Thus, the number of targets that can be surveyed with a precision of ~10-50 cm/s in a given night is largely independent of telescope aperture for telescopes larger than 4m (Section 4.2).

In the sections that follow, we present the different types of stellar jitter signals that have been proposed, and give estimates of the induced RV amplitude that have been measured with the same spectrographs that are used for exoplanet RV work. We present these different signals in order of timescale. In Figure 44 we present an overview of the stellar signals that are understood and can be mitigated, and the stellar signals that still pose problems.

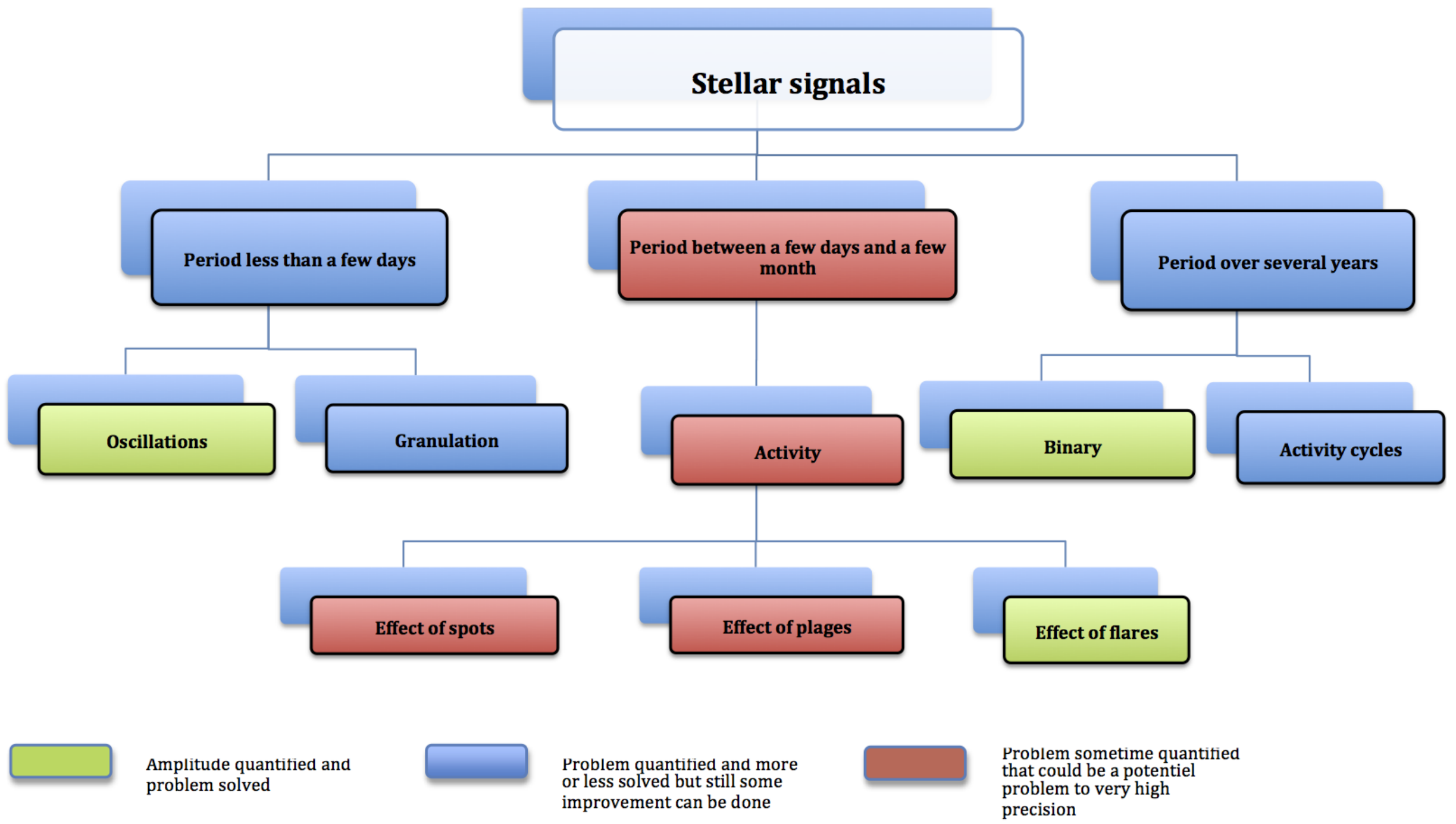

**Figure 44.** *The different types of stellar jitter affecting RV measurements.*

### 5.2.1.1 Stellar signal with period less than a few days

#### 5.2.1.1.1 Stellar oscillations

Pressure waves, also called P-modes, propagate at the surface of solar-type stars leading to a expansion and contraction of external envelopes over timescales less than 15 minutes for GKM dwarfs (5-15 minutes for the Sun; Broomhall et al. 2009, Arentoft et al. 2008, Schrijver & Zwann 2000). The individual radial-velocity amplitudes of P-modes are typically a few tens of cm/s, but the interference of tens of modes with close frequencies introduces RV variations on the order of 1 m/s for GKM dwarfs, depending on the spectral type and evolutionary stage of the star (Bedding et al. 2007, Bouchy & Carrier 2003, Bedding & Kjeldsen 2003, Schrijver & Zwann 2000, Kjeldsen & Bedding 1995).

Because of the short timescale of oscillations, it is possible to average out this signal by increasing the exposure time to at least the period of the signal. In addition, theory and observations show that the frequencies of the P-modes rise with the square root of the mean density of the star and that their amplitudes are proportional to the ratio of the stellar luminosity over the stellar mass (Arentoft et al. 2008, Christensen-Dalsgaard 2004, Kjeldsen & Bedding 1995). Therefore when going from G to M dwarfs, the period of these oscillations decreases, as well as their amplitudes. Observing later spectral-type stars therefore implies a smaller contribution of oscillations, and a shorter exposure time required to average them out.
As an example, for the solar analog α Cen A (spectral type G2V), the amplitude of the oscillation signal measured using HARPS is 1.2 m/s (Dumusque et al. 2011a). For the K0V dwarf α Cen B, the amplitude of the oscillations measured with the same instrument is 0.39 m/s.

#### 5.2.1.1.2 Granulation

At the surface of the Sun, a transition in the way energy is transported occurs. The surface becomes optically thin and energy can escape by radiation, carrying out the major part of the energy. This sudden cooling down of convection cells gives rise to a phenomenon called granulation. When reaching the stellar surface, the ascending hot gas of the convection cell flows horizontally, which form the granules, before descending in dark intergranular lanes once it has cooled down.

The average lifetime of a granule is approximatively ten to thirty minutes (Title et al. 1989), Bahng & Schwarzschild 1961). The velocity of convection inside these granules is a few km/s, but averaging the velocities of granules and intergranules over the entire stellar disc reduces the net signal to an amplitude of a few m/s.

At a larger scale, supergranulation resulting from convection at much larger scales, can also be observed on the Sun. The size of supergranulation cells is on average 30,000 km, with velocity fields of hundreds of m/s, and lifetimes up to 2 days (Del Moro 2004a). Like granulation, the average of all the supergranulation cells on the stellar disc reduces the net signal to an amplitude of a few m/s.

In addition to granulation at small scales and short periods, and supergranules at large scales and longer periods, a more or less continuous convection pattern in size and timescale exists. This induces a continuous RV signal from 15 minutes induced by granules, up to 48 hours for supergranules.

An approach for averaging out this signal is to observe a star several times during the same night, spanning as long a time interval as possible. Then, binning the data over the night helps to reduce the jitter induced by granulation phenomena. This strategy has been applied to a sample of ten stars observed several times per night with HARPS, resulting in the detection of seven low-mass planets, including the Earth-mass exoplanet orbiting α Cen B (Pepe et al. 2011). A possible way to account for the uncertainties introduced by the signal of granulation phenomena is to add a red noise model when fitting exoplanet orbits to the data.

### 5.2.1.2 Stellar signal with period between a few days and a few months

#### 5.2.1.2.1 Effects of starspots and plages

A starspot, emitting less light than the average stellar surface, will break the symmetry between the blue-shifted approaching limb and the red-shifted receding limb of a rotating star, and will induce an apparent RV variation as it passes across the visible stellar disc.

A plage at the disc center is only slightly hotter than the average stellar surface temperature and will induce a small flux effect. A plage on the limbs will be brighter due to a limb-brightness dependence (Meunier 2010a, Unruh 1999, Frazier 1971), while the star emits less light at this location due to limb-darkening. Plages induce a smaller flux effect than spots, even though they tend to be an order of magnitude more extended than starspots (Chapman 2001).

Plages and starspots are both affected by strong local magnetic fields. These magnetic fields suppress the convection locally, thus suppressing the convective blueshift effect inside the active regions (~300 m/s for the Sun, e.g. Dravins 1981). These regions therefore appear redshifted in comparison to the quiet photosphere affected by convective blueshift, which can induce an RV variation as the star rotates. This RV effect of the inhibition of convective blueshift will be of the same order of magnitude for spots and plages, because both regions are affected by strong local magnetic fields.

Starspots dominate the stellar jitter for rapidly rotating stars, because the distortion of the RV increases with the projected rotational velocity, $v\sin(i)$. Above a $v\sin(i)$ of 3 or 4 km/s, starspots dominate the RV variation even if plages are present. For slowly rotating stars, stellar jitter induced by the inhibition of convection by plages can dominate.

A recent study by Dumusque et al. (2014) based on observations of the Sun and including the effect of the inhibition of convection inside active regions can be used to estimate the RV effect of spots and plages. The maximum peak-to-peak amplitude in m/s that a spot with a filling factor *f* can induce is given by

$$\Delta RV = f \times (1.04 + 4.65\, v\, sin(i) + 0.40\, v\, sin(i)^2 - 0.02\, v\, sin(i)^3)$$

Big spots on the Sun have filling factors of about 0.1%, which implies a peak-to-peak RV jitter of 1.2 m/s. For a plage, the maximum peak-to-peak RV amplitude is given by

$$\Delta RV = f \times (3.87 + 0.11\, v\, sin(i) + 0.10\, sin(i)^2)$$

Plages are ~10 times more extended than spots (Chapman et al. 2001). A big plage on the Sun will therefore have a filling factor of 1%, which induces a peak-to-peak RV jitter of 4.5 m/s.

For slowly rotating stars, the main activity-induced RV variation is caused by the inhibition of the convective blueshift inside active regions. The velocity of convection is smaller for later spectral type stars, so K to M dwarfs should be less affected by stellar activity than G dwarfs.

For GK dwarfs, photometric variations (e.g. Sanchis-Ojeda 2012), the calcium index (e.g. $R'_{HK}$), the bisector span (BIS SPAN), and the full-width half maximum (FWHM) of the cross-correlation function (Queloz 2009, Bonfils 2007, Queloz 2001) are observables sensitive to stellar activity. Variation of these indicators is a warning that observed RV variations may be due to plages and/or spots on the stellar surface. For M dwarfs the H-alpha emission and sodium activity index are useful activity indicators (Gomes da Silva 2012).

Stellar activity can induce a signal at the rotational period of the star and its harmonics. It is therefore possible to fit sine waves with periods fixed at the rotational period of the star and its harmonics to account for the RV activity signal. This approach has been applied successfully to several cases (e.g. Boisse 2011, Dumusque 2012).

It is also possible to go further and model the activity signal using the photometric, $R'_{HK}$, BIS SPAN, and FWHM amplitudes. If the activity of the star can be explained by a major active region on the stellar surface, this technique has been shown to be efficient (Dumusque et al. 2014). Gaussian processes can be trained on the photometric light curve alone, and can then be used to model the RV jitter (cf. the FF' method, Aigrain 2012; Haywood 2014).

Another small effect of spots is due to the fact that spots are depressed compared to the photosphere (the Wilson depression) because they are cooler. The depth of the Wilson depression has been estimated in many studies, ranging from 500 - 2500 km. This induces a modification of the gravitational redshift inside the spotted region, and thus an RV variation of a few cm/s (Cegla 2012).

##### 5.2.1.2.2 Effect of flares

Flares occur when accelerated charged particles, mainly electrons, interact with the plasma medium. Magnetic reconnection between loops of magnetic field lines releases a lot of energy, which accelerates the charged particles in the close environment and creates the flaring event.

The effect of flares on RV measurements can be a problem when searching for planets orbiting active M dwarfs. This effect is not well quantified, but could induce an amplitude of ~0.5 m/s for common flares (Saar 2009). These phenomena are not periodic and can be quite frequent (Audard 2000). Nevertheless, not all M dwarfs are flaring, and some of them are extremely quiet (e.g. Bonfils 2013a). The Balmer line H-alpha is very sensitive to these events and can be used as a proxy for flares.

#### 5.2.1.3 Stellar signals with a period over several years

##### 5.2.1.3.1 Orbital motion due to companion stars

Binary stars or multiple stellar systems can induce a strong RV variation. Close binaries induce variations of many km/s that are easily detected but must be removed very accurately to allow a solution for a planet orbiting one of the stars. Long-period binaries can induce a long-term trend, which may not mask the detection of planets with much shorter periods. In some cases the change in position on the sky due to orbital motion in a binary is large enough so that it must be taken into account when calculating the barycentric correction.

#### 5.2.1.3.2 Magnetic activity cycles

Most solar-type stars exhibit Sun-like magnetic activity cycles over a timescale of several years (Lovis 2011, Dumusque 2011c). A magnetic cycle is characterized by very small or no active regions at minimum, and an important number of active regions at maximum. A simple approach for avoiding stellar jitter due to magnetic activity is to observe stars only near the minimum of their activity cycle. For the Sun, the expected amplitude of the magnetic cycle effect is 16.4 m/s (Lovis 2011), and the minimum phase lasts for three to four years.

It has been observed that the long-term RV variation induced by a magnetic cycle correlates well with the variation in activity indices, and therefore this correlation can be used to correct the RV variation induced by a magnetic cycle (e.g. Dumusque et al 2012, Dumusque et al. 2011c, Meunier & Lagrange 2013).

### 5.2.2 Spectroscopic Properties of Stellar Jitter

The embedded signals from photospheric velocities do not appear as Doppler shifts. Photospheric velocities (or spots) affect the line profiles of spectral lines - they do not imprint a shift of $\lambda \times v/c$. However, current spectrometers do not have resolution to distinguish the photosphere from center of mass stellar velocities. The ability to identify and remove photospheric signals is at the very heart of whether the PRV community can ever make use of 10 cm/s instrumental precision on real stars.

With current spectral resolutions ($R$ up to 115,000) the photospheric velocities are indistinguishable from Doppler shifts. Attempts to decorrelate spot signals and granulation have had some success, but will be increasingly more difficult for velocity semi-amplitudes $K < 0.5$ m/s. Improper decorrelation can introduce false-positive signals. What is possible to do at 1 m/s precision may be difficult and ambiguous for long term RMS velocities at the level of 10 cm/s.

In addition to trying to continued efforts to decorrelate stellar noise, it is worth considering whether instruments can be built to detect the photospheric signals. Astrophysicists who model granulation compare line bisectors from their 2D and 3D convection models to line bisectors in the $R$~500,000 FTS solar spectrum (Asplund et al. 2000). A spectrum with $R$=500,000 and greater than Nyquist sampling can detect time correlated changes in stellar granulation. Spectrometers with sufficient resolution to detect and model the stellar noise may succeed in the search for 10 cm/s signals. The Doppler shifts will trivially come for free. Along these lines, ESPRESSO is designed for a second resolution of $R$=230,000 in addition to the $R$=150,000 mode (Section 6.1).

### 5.2.3 Wavelength Dependence

Reiners et al. (2010) concludes that near-infrared (NIR) PRVs are only advantageous for stars later than M4 in spectral type, based upon estimates of signal-to-noise and RV information content. Bottom et al. (2013) and Beatty & Gaudi (2015) also conclude that red (~0.8-1.0 micron) PRVs are optimal for early M dwarfs. However, these conclusions do not take into account the effects of the reduced stellar jitter in the NIR. In Section 4.3, we present a model for and analysis of the wavelength dependence of stellar jitter. We show that red and NIR PRVs may provide higher precision due to the reduced stellar jitter amplitudes at longer wavelengths (instrumentation and telluric absorption challenges aside).

These theoretical arguments in favor of red and NIR PRVs have also recently been supported by PRV observations. For example, Anglada-Escude et al. (2012a) demonstrate that Barnard's Star (M4V) is stable down to 80 cm/s over 4 years when using the red-most part of the HARPS wavelength range (between 650-680 nm). Including the bluer part of the spectrum (380-650 nm) increased the long-term RMS to about 1.4 m/s, and using the Iodine wavelength range only (450-620 nm) the RMS was as high as 1.6 m/s. Moreover, the authors argue that the increase in noise at bluer wavelengths is not the result of lower SNR, but is due to wavelength-dependent sources of stellar noise acting more strongly in the blue. A similar explanation could account for the difference in the RMS of PRV measurements of Kepler-78b with HARPS-N compared to HIRES (Howard et al. 2013, Pepe et al. 2013). These results provide motivation to move to longer wavelengths in order to achieve sub-1 m/s long-term precision for stars of all spectral types.

### 5.2.4 Photometric Correlation

A recent re-analysis of the CoRoT-7 system by Haywood et al. (2014) has shown that activity-induced RV variations can be well-reproduced by simple models based on the off-transit photometric variations of the star. CoRoT-7 is an active G9 star displaying peak-to-peak activity-driven RV variations of the order of 20 m/s. It hosts a compact planetary system including at least one transiting super-Earth, CoRoT-7b (Léger et al. 2009) and one sub-Neptune mass planet (Queloz et al. 2009). The CoRoT-7 system was observed intensively with the HARPS spectrograph for 26 consecutive nights from 2012 January 12 to February 6, with multiple well-separated measurements for each night. These observations were analysed in conjunction with simultaneous CoRoT photometry, allowing the activity-driven RV signal to be reproduced using the FF' method of Aigrain et al. (2012) and a Gaussian process that has the covariance properties of the lightcurve (Haywood et al. 2014). The FF' method uses the off-transit variations in the star's lightcurve and its first time derivative to model the suppression of convective blueshift and the flux blocked by starspots on a rotating star. The Gaussian process accounts for any remaining activity-induced signals that may result from other activity-related processes such as photospheric inflows towards active regions or limb-brightened facular emission.

Accounting for stellar activity in this way allowed the authors to estimate the masses of CoRoT-7b and CoRoT-7c with realistic allowance for the uncertainty introduced by the stellar activity, and to rule out the presence of an additional planetary companion at an orbital period of 9 days.

### 5.2.5 Time-scale Dependence on Stellar Parameters

Stellar activity affects RVs in a number of ways, over a wide range of different timescales. Here, we summarize how radial velocity jitter scales with stellar parameters like mass and age. Processes like stellar surface granulation and seismic oscillations introduce noise on timescales of minutes to hours. While the timescales for these oscillations are typically on the order of minutes for dwarf stars, for evolved stars they increase and can be as long as hours. Kjeldsen & Bedding (1995) found that asteroseismic oscillation amplitude varied close to linearly with the luminosity to mass ratio of stars, implying that asteroseismic radial velocity jitter is more important for stars more massive than the sun than it is for lower mass stars. There is, however, a large amount of scatter in oscillation amplitudes for a given spectral type, and it is hard to predict which stars will show large amplitude oscillations. Out of 2000 dwarf stars observed by Kepler for asteroseismology, only 500 exhibited detectable oscillations (Chaplin et al 2011).

Stellar surface granulation happens on slightly longer timescales than asteroseismic oscillations. Bastien et al (2014) showed that RV jitter correlated with 8 hour flicker, which Bastien et al (2013) and Cranmer et al (2014) have linked to stellar granulation processes. The inverse correlation between flicker derived stellar surface gravity and RV jitter implies that similarly to asteroseismic oscillations, stellar granulation induces lower amounts of RV jitter for lower mass stars (Dumusque 2011a). In particular, Bastien et al (2014) demonstrated a roughly linear relation between granulation induced RV jitter and stellar surface gravity, with typical levels of RV jitter dropping below a 2-3 m/s observational floor for stars with log(g) > 4.5.

RV jitter due to stellar rotational modulations takes place on timescales of days to months. Models (e.g. Aigrain et al. 2012) have been developed and validated to estimate RV variations from white light photometric measurements (assuming the photometric bandpass and spectroscopic bandpass are similar), so it is possible to estimate the ensemble properties of starspot modulations from photometric measurements of stars of various spectral types. The sample of Kepler stellar rotation periods and photometric amplitudes from McQuillan et al. (2014) suggests that while rotation periods increase towards lower stellar masses, both the fraction of stars showing photometric rotational modulations and the amplitude of those modulations typically increase towards lower mass. The net effect is that the typical RV amplitude of starspot modulations stays close to constant at the ~1 m/s level for FGK and M-type stars, albeit with a wide scatter in amplitudes (Vanderberg et al. 2015). In addition to the dependence on spectral type, the stellar rotation periods (and therefore ages) are closely related to the level of RV jitter. Young, fast rotating stars will exhibit larger radial velocity variations (see Section 5.4; Saar 1998). Similarly, other tracers of stellar age-like magnetic activity indicators (particularly $R'_{HK}$) can determine the level of RV jitter.

Due to the relatively recent development of PRV technology capable of measurements on the 1 m/s level, the jitter induced by long period magnetic activity cycles has not been thoroughly explored as a function of stellar properties. Lovis et al (2011) searched HARPS data for magnetic activity cycles and reported on their strengths and correlations with velocity for stars of various effective temperatures. In particular, the authors found that for lower stellar effective temperatures, correlations between RV and the $R'_{HK}$ activity indicator became less significant, and that the typical RV semi-amplitude (and its scatter) caused by magnetic activity cycles increase with stellar mass. Stars below 5000 K have essentially no observable correlation between RV and magnetic activity cycles, which hotter stars can exhibit variations of up to 10 m/s semi-amplitude over the course of a magnetic cycle. Gomes da Silva et al (2011) extended this work into M dwarfs, albeit with fewer target stars and shorter observational baselines, and found that a Sodium-based activity indicator was a better predictor of RV jitter in M dwarfs than other indicators like $R'_{HK}$. Unfortunately, the short observational baseline, small sample, and observational sampling technique limited the conclusions about activity indicator correlation with RV jitter for M dwarfs in Gomes da Silva et al (2011).

### 5.2.6 The Sun in Context

Gilliland et al. (2011) reported that the Sun was among the quietest 15% of stars based upon the photometric variability of main sequence stars in the Kepler data set. Indeed, radial velocities surveys with HARPS and HARPS-N are targeting the most quiescent nearby stars, since they will be the easiest to probe down to lower exoplanet masses. However, the Gilliland et al. (2011) result makes use of a noise estimate on a time-scale of hours (CDPP6 from the Kepler mission), which is of course relevant to transiting exoplanets with transit durations on the order of several hours. This time-scale is not relevant for the radial RV search of exoplanets in Habitable-Zone orbits around main-sequence stars. Instead, Basri et al. (2013) report that the Sun is a more "normal" star in terms of stellar activity, from analysis of Kepler photometric variability on time-scales of months and years. The median $\log R'_{HK}$ of the Sun is ~ -4.95, which is slightly quieter than α Cen B's value of ~ -4.90. Compared to other Solar-type stars (e.g. Mamajek & Hillenbrand 2008), the Sun is also a typical star in this regard. The rotational modulation of starspots for the Sun is modeled from solar data to be ~0.4 m/s (Makarov et al. 2009). For comparison, the rotational modulation of star-spots for α Cen B is reported in Dumusque et al. (2012) to produce a semi-amplitude of ~2 m/s – a factor of five higher that can likely be attributed to its later spectral type.

The integrated RV of the Sun as a star can be estimated in several ways. One can use the empirical relation $\sigma_{RV}$ versus photometric RMS flux found by Paulson et al. (2004) for the Hyades. Using the multi-year photometric variability of the Sun estimated from the PMOD TSI (total solar irradiance) data, the average solar RV variation comes up to 1.7 m/s. However, the Sun is a much older star than the Hyades, and rotates more slowly. The RV variation caused by star spots and other photometric features on the surface is proportional to the projected rotational velocity, $v \sin(i)$, as shown in Makarov (2009). Assuming an average $v \sin(i)$ = 5.6 km/s for the Hyades sample, the estimate should be scaled down to 0.6 m/s for the Sun.

Alternatively, the solar RV variation can be estimated from the observed surface brightness distributions using the known differential rotation profile. Makarov ey al. (2010) integrated high-resolution solar surface brightness maps derived from Mount Wilson observations, which were processed and calibrated to ensure a close match with the observational TSI data. As a side product of that analysis, the daily variation of the integrated solar RV can be estimated. The average multi-year standard deviation was estimated at 0.21 m/s. This estimate should be used as a lower bound, because it accounts only for the part of variation induced by the photometric irregularities and asymmetries on the rotating surface. The physical motion in the photosphere caused by, for example, supergranulation or surface flows, is not included. A strong correlation of RV variation with the solar activity cycle was found, with the running standard deviation varying between 0.02 m/s on the quietest months and 0.5 m/s at the maximum of activity in 2000. Exceptionally active intervals of time, when a large group of spots and plages appears on one side of the visible hemisphere, are characterized by RV dispersions of up to 1.4 m/s.

Most of the long-term photometric variability of the Sun (442 ppm) comes from the solar cycle. Short-term variations of TSI (and, presumably, of integrated RV) are most of the time significantly smaller. If the PMOD TSI data are divided into consecutive 10-day intervals, the irradiance standard deviation is estimated for each interval, and the median value of the standard deviation is computed for 960-day intervals (a typical duration for exoplanet RV data sets), the range of the TSI standard deviation is between 70 ppm and 260 ppm depending on the magnetic activity level. This still underestimates the short-period variability, which is more relevant for exoplanet detection, The mean over the 10-day standard deviation values is 182 ppm, the median is 139 ppm, and the 15th and 85th percentiles are 53 and 320 ppm, respectively. Using the scaling relation from Makarov et al. (2009), the brightness-related RV standard deviation over several years is:

$$\sigma_{RV} = 0.442 \times 7 \times 10^5 \times 2 \text{ km/s} = 0.062 \text{ m/s}$$

but the short-term variation can be much smaller. Formally, the Earth's signal (0.089 m/s) can be confidently detected in a 2-3 years observing run for a Sun seen equator-on even at the height of magnetic activity. Yet, these numbers may strongly underestimate the overall RV jitter of the Sun.

Supposedly the most direct way of estimating the overall $\sigma_{RV}$ of the Sun is to use the multi-year series of the SOHO MDI data. The RV measurements taken with the Doppler spectrograph are pixellized over the visible disk, and it appears straightforward to integrate the measurements, convolving them with a weight function corresponding to the limb darkening function and subtracting the differential rotation pattern. Unfortunately, the MDI data are burdened by large systematic errors of unknown origin and instrumental resets. Still, the data can be divided into a number of "smooth" intervals, typically, 150-200 days long. High-order polynomials or Fourier series can be fitted to remove the systematic and instrumental effects. The robust standard deviations (estimated from the 85th and 15th percentiles) of the daily residuals range from 0.9 to 1.0 m/s. These values may represent more realistic estimates of solar RV jitter, because they include the physical motion of the surface.

## 5.3 Radial-Velocity Surveys in the Red and Near-Infrared

The precision of RVs in the NIR has historically lagged behind that achieved in the visible. As of 2010, the state of the art consisted of using telluric lines for wavelength calibration, resulting in achieved precisions of ~50-100 m/s (Tanner et al. 2012, Crockett et al. 2012, Bailey et al. 2012). Instruments that require cryogenic cooling can often be a factor of a few times more expensive than their corresponding visible equivalents, and as a result there are currently very few operational high resolution NIR spectrographs, none of which are optimized for radial velocities – CRIRES at the VLT, CSHELL at IRTF, ICRS at Subaru, Phoenix at Gemini and NIRSPEC on Keck. Fortunately, a number of facilities are planned to become operational in the next decade, including CARMENES, iSHELL (equipped with a gas cell), SPIRou, and the Habitable Planet Finder (Mahadevan et al. 2012, Quirrenbach et al. 2012, Yuk et al. 2010, Rayner et al. 2012). Additionally, recent advances have pushed the achievable RV precision in the NIR, with current facilities now closer to parity with the visible, using ammonia and isotopic methane absorption gas cells (Bean et al. 2010, Anglada-Escude et al. 2012b, Plavchan et al. 2013a), and the recent commissioning of a NIR fiber scrambler prototype with non-circular core fibers on CSHELL at the IRTF (Plavchan et al. 2013b). The next generation of NIR facilities hold the potential to achieve ~1 m/s single measurement precision, but much work will be needed over the next several years, including improvements in the analysis and removal of telluric lines and the generation of stellar templates (Seifahrt et al. 2010).

Several studies, including Reiners et al. (2010), Bottom et al. (2013), and Beatty & Gaudi (2015), have looked into the benefits of red and NIR PRVs compared to visible PRVs. Such studies assume telluric absorption can be perfectly modeled without sacrificing precision, which is one of the major NIR PRV challenges (Seifahrt et al. 2010). In particular, Bottom et al. (2013) looked at the optimal wavelength to observe main sequence stars of different effective temperatures for a HZ survey, and what spectral types such a survey would optimally contain (Figures 45,46). These studies did not consider the wavelength dependence of stellar jitter, which is a primary scientific driver for NIR PRVs, and which we discuss in detail in Sections 4.2 and 5.2.3.

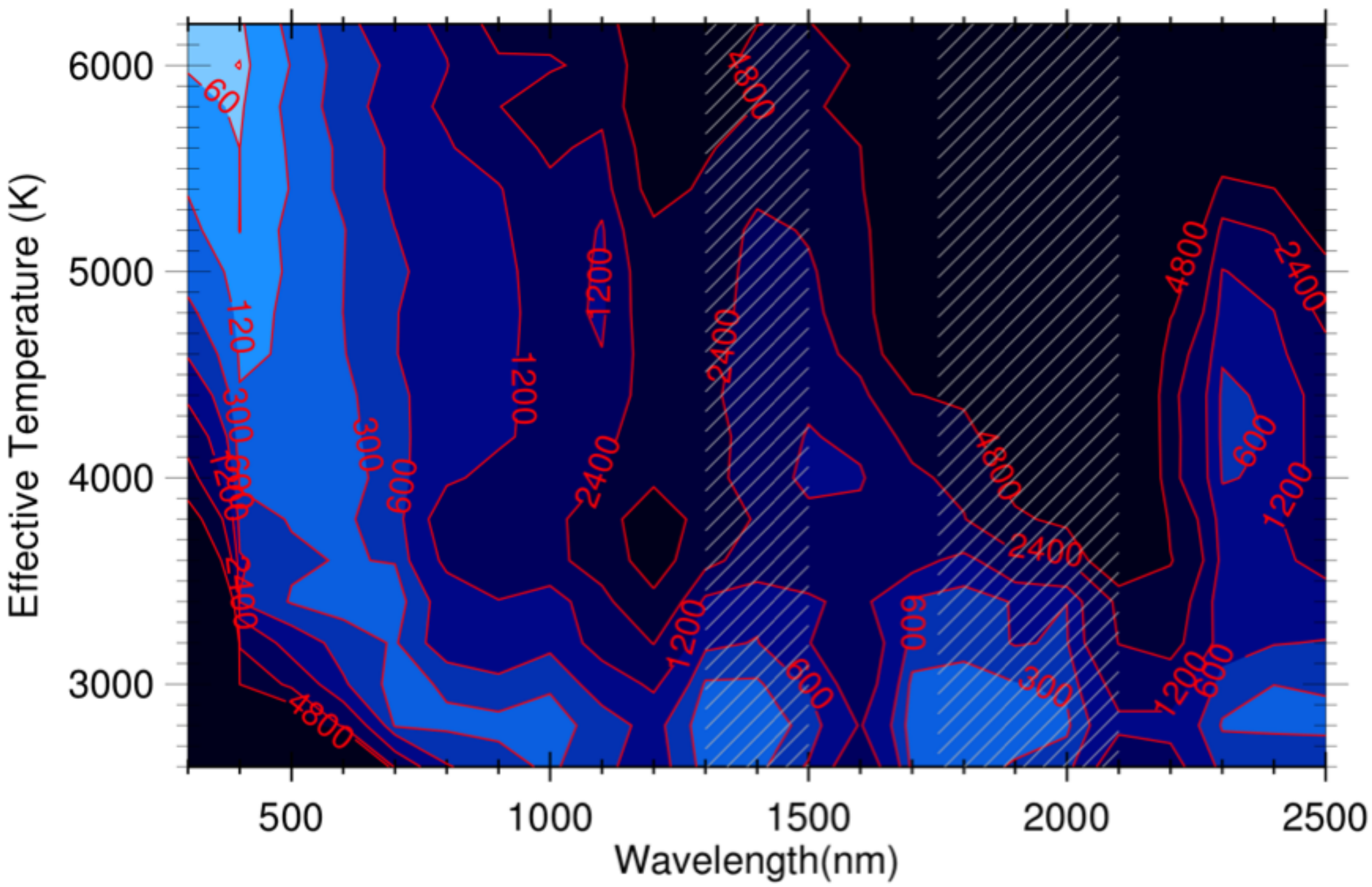

**Figure 45.** *Figure reproduced from Bottom et al. (2013). Relative time to detect a Habitable-Zone planet of a fixed mass as a function of stellar effective temperature and wavelength of observation, for a spectral grasp of 0.1 microns. Hashed regions indicate atmospheric telluric absorption bands. This figure demonstrates the advantage of going to longer wavelengths for lower-mass targets, and that there is little difference in survey time for M dwarfs from 0.7 to 2.5 microns. The analysis is based on photon plus read-noise estimates only, and does not include instrumental nor stellar activity noise models, the latter of which is likely to be wavelength-dependent (See Section 4.3).*

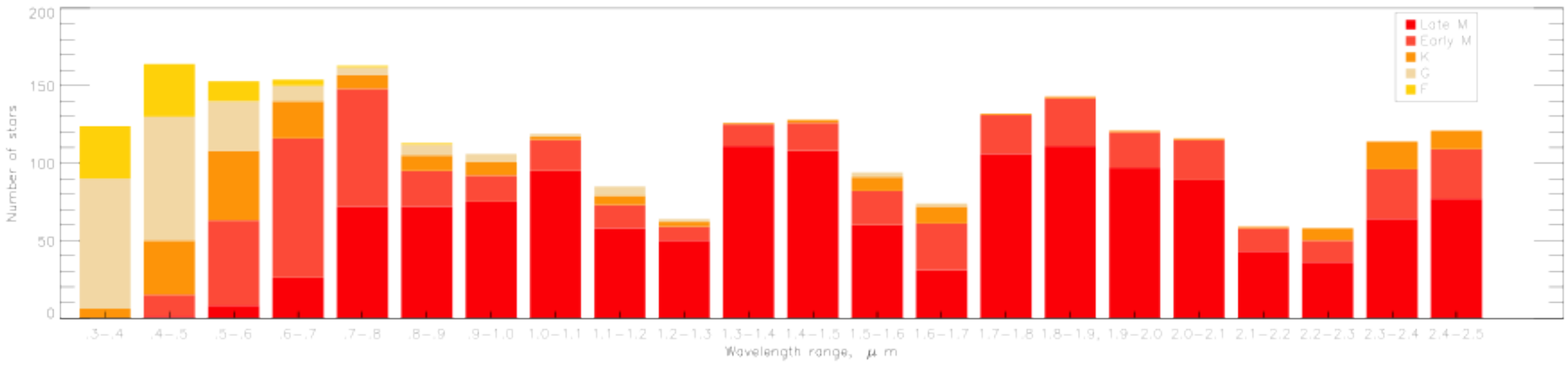

**Figure 46.** *Figure reproduced from Bottom et al. (2013) Survey sample spectral types for a Habitable-Zone complete survey as a function of wavelength with fixed survey duration, using the RECONS list of nearby stars. This figure demonstrates that an optimal survey largely consists of M dwarfs at wavelengths longward of ~0.7 microns. This is due to the closer habitable zones of M dwarfs and lower stellar masses, resulting in larger Habitable-Zone RV amplitudes and thus requiring a shorter observation time per target than earlier type stars. This analysis does not include a model of stellar activity RV noise, which will affect the total number of targets, both as a function of wavelength (favoring longer wavelengths), and in accounting for the minimum integration time necessary to average over p-mode observations, the latter of which will limit the number of targets to ~4 per hour in a given night. This analysis demonstrates that visible and NIR efforts can be optimally designed for HZ surveys and are complementary in target spectral types.*

Finally, perhaps the most exciting recent discovery with NIR RVs is the direct detection of RV variation from Jovian planets with significant CO absorption lines at 2.3 microns such as τ Boo b (Rodler et al. 2012a, Snellen et al. 2010). This method may be applicable to the direct RV detection of Earths with a future generation of NIR RV instrumentation and observatories (Snellen et al., 2013). Here the precision in the radial velocity is not as important, with semi-amplitudes on the order of ~10 km/s. Rather sensitivity with large apertures and low-read noise detectors combined with advanced modeling and removal of telluric and stellar signals becomes necessary, beyond current capabilities to detect such a faint signal.

## 5.4 Young Stars, Active Stars

The discovery of planetary systems that are drastically different from our own, including those with short orbital period gas giants and those with large obliquities, has led to significant revisions in planet formation and migration theory. While the ensemble properties of planetary systems and their stellar hosts can provide some constraints on formation models (e.g., Helled et al. 2013), a much more direct test of the new ideas would be to find planets soon after formation. Competing formation scenarios (e.g., cold start vs. hot start; Marley et al. 2007, Fortney et al. 2008, Galvagni et al. 2012) and migration scenarios (planet-disk vs. planet-planet; Lin et al. 1996, Lin & Ida 1997) predict observationally distinct differences that may persist for up to ~1 Gyr.

However, only a few of the nearly 1000 RV-detected exoplanets orbiting field stars have ages that are likely to be less than 1 Gyr, and none have ages that are precisely known. This is a consequence of RV surveys preferentially targeting older, less active field stars (e.g., Wright 2005). In contrast, the presence of starspots on a rotating star can periodically distort stellar absorption lines, and, if unaccounted for, can give the appearance of RV shifts of many hundreds of m/s (e.g., Queloz et al. 2001; Desort et al. 2007). The decline in stellar rotation and activity with age is both well established and quantified (e.g., Skumanich 1972, Barnes et al. 2010). Observations accumulated over the most recent decade are starting to quantify how stellar jitter declines with age and what the implications are for the detection of planets at young ages.

Observations at visible wavelengths of adolescent age stars in nearby open clusters have found mean stellar jitter measurements of 16 m/s in the Hyades (625±50 Myr; Paulson et al. 2004) and 13 m/s in Praesepe (580±50 Myr; Quinn et al. 2012). Nevertheless, with sufficient monitoring, short-period sub-Jupiter mass planets can and have been identified (e.g., Quinn et al. 2012, 2014). Observations of younger stars in nearby moving groups (10-150 Myr) find roughly an order of magnitude increase in stellar jitter (131 m/s for FGK stars; Paulson & Yelda 2006; Lagrange et al. 2013, Hillenbrand et al. 2014). This increased RV variability, which can be pseudo-periodic, has led to the premature announcements of two young planets in these groups (TW Hya, Setiawan et al. 2007; BD +20 1790, Hernán-Obispo et al. 2010), both of which were subsequently shown to be caused by starspots (e.g., Huélamo et al. 2008, Figueira et al. 2010a). The increased stellar jitter translates to typical detection limits of several Jupiter masses at short orbital periods; this is corroborated by the detection of one of the youngest planets known, a 2.8 $M_{Jupiter}$ giant planet with a 320 day period orbiting a weakly active ~150 Myr star (Borgniet et al. 2014). Several pioneering studies of the youngest T Tauri age stars have revealed stellar jitter values as large as several km/s (e.g., Huerta et al. 2008; Prato et al. 2008; Crockett et al. 2012). In these cases, exoplanet detection via visible RVs is may not be feasible, athough van Eyken et al. (2012) identified an enigmatic young candidate planet in Orion.

Given the underlying motivation for finding young planets, observers have recently developed techniques to mitigate the effects of stellar jitter. One direct way to accomplish this is to obtain precise RV measurements at NIR wavelengths; the brightness contrast between cool starspots and the photosphere is diminished in the near-infrared, which translates into smaller stellar jitter. Bailey et al. (2012) and Crockett et al. (2012) report reductions in stellar jitter by factors of 2-5 for infrared measurements versus visible measurements. We note though that the overall effectiveness of this is still being developed, since most infrared RV survey work has been conducted with 1990s generation infrared arrays (e.g., CSHELL, NIRSPEC) in contrast to more modern facilities (e.g., CRIRES, iSHELL, IGRINS). Nevertheless, with resulting jitter values of ~100 m/s, the detectable young planets are limited to gas giants with short orbital periods.

Finally, we highlight that some success is being achieved at overcoming the effects of stellar jitter by observing active stars at multiple wavelengths and/or with multiple techniques (photometry, spectropolarimetry). In one promising case, the observed stellar jitter was filtered by nearly two orders of magnitude (4300 m/s to 55 m/s; Donati et al. 2014), which would enable gas giants to be identified orbiting the youngest planets known.

# 6 Instrumentation Objectives

Current PRV detection sensitivity is primarily limited by two categories of systematic effects: astrophysical sources of radial velocity "jitter" and long-term instrumental stability, i.e. the extent to which wavelength shifts due to the instrument can be calibrated and corrected. In this section, we look towards future efforts to advance instrumentation stability.

There are significant questions from the astronomy community as a whole whether the PRV technique can be extended towards the goal of detecting the 9 cm/s K semi-amplitude signal of a 1 Earth-mass planet in a one year habitable orbit around a G type star. The uncertainty of the semi-amplitude must be reduced to ~1-2 cm/s on a time-scale two orders of magnitude longer than the 3.2-day planet orbiting α Cen B (Dumusque et al. 2012).

Invoking simple photon statistics to extrapolate from the current state of the art in PRVs towards future capabilities is dangerously naïve, since systematic errors due to the instrumentation will set the ultimate error floor. To achieve precisions better than 1m/s will require careful stabilization of the spectrograph pressure, temperature, and optical/mechanical structure, together with better wavelength calibration.

## 6.1 Current or Planned Spectrographs

Members of the PRV community have confidence that visible instruments will be developed with stability at the level of 10 cm/s over long periods of time (e.g. Pepe & Lovis 2008). Three examples are now under development: ESPRESSO on the VLT (first light goal 2016; Spanò et al. 2012, 2008) G-CLEF on the GMT (first light goal 2021; Szentgyorgyi et al. 2012) and CODEX on the E-ELT (Delabre & Manescau 2010; Pasquini et al. 2010a,b, 2008). Similarly, members of the PRV community are confident that near-infrared instruments can be developed to be stable at the level of 1-3 m/s, approaching the current performance of visible spectrographs. Three examples are now under development: CARMENES at Calar-Alto (~2015; Quirrenbach et al. 2012), the Habitable Planet Finder on the HET (~2016; Mahadevan et al. 2012), and SPIRou on the CFHT (~2017; Artigau 2014). There is a large number of planned or in development PRV spectrometers, both in the visible and near-infrared. Herein we provide a list of these spectrometers. Another excellent list is maintained by the CARMENES spectrograph team (http://carmenes.caha.es/ext/spectrographs/index.html).

### 6.1.1 Visible Spectrometers

**Table 2.** *Planned or current visible PRV spectrometers*

| **Instrument** | **Telescope** | **Measurement precision, Spectral Grasp, Resolution** | **PI; (relevant publications) / First Light** |
|---|---|---|---|
| APF | Lick 2.4 m | 1 m/s, 374-970 nm, R=120k / 490-600 with iodine cell | Vogt; (Vogt et al. 2014, Radovan et al. 2010) / 2013 |
| CHIRON | Chile | 0.5 m/s over 10 days, 2 m/s over 2 years, R~90k,130k | Debra Fischer; Commissioned 2012; Tokovinin et al (2013) |
| CODEX | E-ELT | 2 cm/s, 370-710 nm, R=120k | Pasquini; (Delabre & Manescau 2010; Pasquini et al. 2010a,b, 2008) / ~2025 |
| Coralie | Euler Swiss Telescope | 2 m/s, 391-681 nm, R=50k | (Queloz et al. 1999) / 1998 |
| ESPRESSO | VLT | 10 cm/s (5 cm/s), 380-686 nm, R=120k (220k) | Pepe; (Spanò et al. 2012, 2008; Pepe et al. 2010) / 2016 |
| EXPRES | DCT | 10 cm/s, 380-700 nm, R~200k | Fischer; 2016-2017 |
| G-CLEF | GMT | 20 cm/s, 350-950 nm, R=120k / also MOS mode | Szentgyorgyi; (Szentgyorgyi et al. 2012) / 2021 |
| Hamilton Echelle | Lick: Shane 3m CAT 0.6m | 3 m/s, 340-900 nm, R=60-100k, 490-600 with iodine cell | Vogt; (Vogt 1987) / 1986 |
| HARPS-N | TNG 3.6 m | 1 m/s, 380-680 nm, R=110,000k | Pepe; (Cosentino et al., 2012, 2014; Langellier et al. 2014) / 2012 |
| HARPS | ESO 3.6 m | 1 m/s , 380-680 nm, R=110,000k | Pepe; (Pepe et al. 2000, 2003; Rupprecht et al. 2004, Lovis et al. 2006) / 2002 |
| HIRES | Keck 10 m | 2 m/s, 360-1000 nm, R=85k / 490-600 with iodine cell | Vogt; (Vogt et al. 1994) / 1996 |
| HRS | HET | 2.5 m/s, 390-1100 nm, R=120k | MacQueen; (Tull et al. 1998) / 2001 |
| LCOGT NRES | Global network of 6 spectrometers | ~1-3 m/s, 390-860 nm; R~53k | (Eastman et al. 2014) / 2015-2016 |
| MINERVA | Mt Hopkins 4x0.7 m | ~1 m/s, 500-650 nm, R~50k | John Johnson (Swift et al. 2015) / 2015 |
| SHREK | Keck 10 m | 1 m/s, 440-590 nm, R=85k / red channel later | Howard & Marcy; (http://nexsci.caltech.edu/keck_strategic_planning_Sep2014.pdf) |
| Sophie | 1.93 m Haute-Provence | 3 m/s, 387-694 nm, R=75k | (Perruchot et al. 2008) / 2006 |
| TRES | Whipple Obs 1.5 m | 15 m/s, 380-900 nm, R=44k | Szentgyorgyi; (Szentgyorgyi & Furesz 2007) / 2007 |
| Tull Echelle | 2.7 m Harlan J. Smith | 340-1090 nm, R=60k, 240k | Phillip MacQueen; |

### 6.1.2 Red/Near-Infrared Spectrographs

**Table 3.** *Planned or current Red and NIR PRV spectrometers*

| Instrument | Telescope | Measurement precision, Spectral Grasp, Resolution | PI or relevant publication, First Light |
|---|---|---|---|
| APOGEE | 2.5-m Sloan Foundation Telescope | ~10 m/s, MOS, 1.51-1.70 microns, R=22.5k | Deshpande et al. (2013) |
| CARMENES | Calar Alto | ~3 m/s; 0.5-1.8, microns, R~80k | Quirrenbach et al. (2012), 2016 |
| CRIRES | VLT | 5 m/s, K-band, R~100k | Bean et al. (2010) |
| CSHELL | IRTF | 5 m/s short term, 35 m/s long term, K-band R=46k | Anglada-Escude et al. (2012b), Plavchan et al. (2013a,b) |
| ESPaDOnS | CFHT | 0.3-1 microns, R~70k | Jean-Francois Donati |
| HPF | HET | ~3 m/s, YJ bands R~50k | Mahadevan et al. (2012) |
| iSHELL | IRTF | ~2-3 m/s, HK bands R~75k | Rayner et al. (2012), 2016 |
| iGRINS | Harlan Smith @ McDonald | HK bands, R~40k | Dan Jaffe, (Yuk et al. 2010) |
| iLocater | LBT | 20 cm/s, 0.95-1.10 microns, R=150k | Justin R. Crepp, in design study phase |
| MINERVA-RED | Mt Hopkins 2x0.7 m | < 1 m/s, 0.8-1.0 microns | Cullen Blake, spectrometer in lab testing phase |
| NIRSPEC2 | Keck | J,H,K,L or M band, R~50k | Ian McLean, in design study phase |
| SPIRou | CFHT | 0.98-2.35 microns, R~70k | Thibault et al. (2012), 2017 |

## 6.2 Wavelength Calibration

There are two primary approaches to wavelength calibration for PRVs, represented by the mature HARPS-S and HIRES spectrometers on the ESO 3.6-m and Keck 10-m telescopes respectively. PRVs are obtained with HIRES via an iodine gas cell for a common optical path relative wavelength calibration in the 500-650 nm range to achieve a single measurement precision of ~1-3 m/s (Butler et al. 1996, Vogt et al. 1994). This approach is necessitated by the thermal, pressure and illumination instability of the spectrograph. HARPS-S relies on stabilizing the thermal, pressure and illumination of the spectrograph to use a non-common optical path wavelength calibration source, currently Thorium-Argon gas, over a broader wavelength range to achieve a single measurement precision of ~0.7 m/s (Mayor et al. 2003).

### 6.2.1 Common Optical Path

Herein we review current and planned approaches to spectrograph wavelength calibration in the visible and NIR.

#### 6.2.1.1. Telluric Lines

At wavelengths longer than about 600 nm, ground-based photometric and spectroscopic measurements must contend with the effects of absorption by molecules in Earth's atmosphere. These telluric absorption features result from rotational and vibrational transitions of molecules including H2O, O2, O3, CO2, CH4, and N2O. While there are atmospheric windows with comparatively little telluric absorption, low-level molecular absorption is present at nearly all wavelengths in the infrared. The average properties of the telluric absorption lines vary considerably between species, and in some cases, on short timescales for a given species. For example, the numerous H2O absorption features in the visible and infrared may undergo substantial variations in visible depth on short timescales as the concentration of water vapor in the troposphere changes. Variations in telluric absorption lines can impact broadband photometry when filters overlap with atmospheric features (e.g. Blake & Shaw 2011, Berta et al. 2012), and also have the potential to confound ground-based transit spectroscopy observations (Mandell et al. 2011). Variable telluric absorption lines also pose a problem for high-precision Doppler measurements at red and infrared wavelengths. These telluric lines must be precisely modeled, or completely avoided, to achieve the ultimate in Doppler precision (Cunha et al. 2014).

Telluric absorption lines do offer the tantalizing possibility of calibrating the wavelength scale of any high-resolution spectrum they contaminate. When Griffin and Griffin (1973) first proposed the simultaneous absorption reference as a means for obtaining PRV measurements with slit spectrographs, it was telluric O2 lines they were suggesting as a wavelength reference. However, Earth's atmosphere is no substitute for stable, controlled gas absorption cell. In Earth's atmosphere, the absorbing molecules are moving with the wind, mainly horizontally, at velocities of 10 to 20 m/s. This means that the telluric absorption spectrum is not a *zero velocity* wavelength reference, rather one with a small intrinsic velocity that changes with time and telescope altitude and azimuth (Balthasar et al. 1982). At the same time, molecular transitions are pressure sensitive, so variations in the pressure profile of Earth's atmosphere lead to subtle shifts in the centers and shapes of telluric lines. Since contributions to the optical depth for a given line come from a range of altitudes, and therefore pressures, telluric lines are also fundamentally asymmetric, and their shapes and positions change with time as barometric pressure changes (Caccin et al. 1985). Wind and pressure can conspire to shift the effective centers of telluric lines by up to a few tens of femtometers. Telluric lines also have appreciable widths. In the visible and infrared, the numerous $H_2O$ lines are pressure broadened and have widths of about 0.01 nm (FWHM), meaning that they may be marginally resolved by high-resolution spectrometers (Livingston & Wallace 1991). This limits their usefulness for recovering information about the LSF of a slit-based instrument. Finally, the significant temporal variations in the optical depths of telluric H2O lines pose a different challenge. Some lines will alternate between the optically thin and optically thick regimes depending on atmospheric conditions. This can result in very complex blends between lines that are close in wavelength but have different nominal strengths.

Despite these challenges, telluric lines still hold great promise as a simultaneous wavelength calibrator for moderate-precision RV spectroscopy. If these lines, and the behavior of Earth's atmosphere, can be sufficiently well modeled, Doppler precision approaching that achieved with other techniques may be possible. Forty years ago, Griffin & Griffin (1973) estimated that Doppler measurements with a precision of 10 m/s should be possible for bright stars using telluric O2 lines as a wavelength reference. Since that time, numerous authors have demonstrated that the Griffins' prediction was essentially correct. Recently, long-term precision better than 5 m/s has been demonstrated. In the visible, measurements of stellar radial velocities using telluric lines with a precision of 5 to 20 m/s have been demonstrated over a wide range of timescales using many different analysis techniques and instruments (e.g. Smith 1982; Cochran 1998; Snellen 2004; Barnes et al. 2012). Figueira et al. (2010b, 2011) carried out an extensive analysis of six years of archival HARPS data, and found that an overall Doppler precision of 10 m/s using telluric O2 lines could be improved to about 2 m/s by using a simple, but physically motivated, model of wind patterns above the observatory.

There has also been substantial progress at infrared wavelengths, where telluric lines are prevalent and there is increased interest in Doppler measurements of cool, red stars. Until recently, the relative dearth of wavelength calibration sources available in the infrared made the use of telluric lines even more appealing. Many authors have explored the use of telluric CO2, CH4, and N2O lines as a wavelength reference in the infrared, with some results approaching the level of precision achieved at visible wavelengths (e.g. Blake et al. 2007, 2010; Huelamo et al. 2008; Seifahrt and Kaufl 2008; Figueira et al. 2010b; Bean et al. 2010; Bailey et al. 2012; Tanner et al. 2012; Rodler et al. 2012b; Deshpande et al. 2013). Overall, the Doppler precision using telluric lines that has been demonstrated in the infrared is somewhat worse than in the visible, at the level of 5-10 m/s over periods of weeks using the CRIRES instrument on the VLT. This limitation is, at some level, instrumental, since there is no high-resolution, multi-order infrared instrument like HARPS or HIRES.

As a new generation of high-resolution, infrared spectrographs employing broad-band calibration sources, such as frequency combs and stabilized Fabry-Perot cavities, is about to come online (including IRD, SPIRou, CARMENES, and iSHELL; see Section 6.1.2), it is crucial that we extract the maximum Doppler information from these data by employing the best possible models of telluric absorption. These models can be based on radiative transfer calculations using molecular transition data and a model of Earth's atmosphere (e.g. Seifhart et al. 2010; Blake & Shaw 2011; Bender et al. 2012; Lockwood et al. 2014; Gullikson et al. 2014; Bertaux et al. 2014; Kausch et al. 2014), or strictly empirical and derived from an ensemble of spectra (Artigau et al. 2014). In most cases, today these models provide excellent fits to observed telluric spectra, with residuals in the cores of lines of moderate optical depth of a few percent. However, precise Doppler measurements using the entire available stellar spectrum, including those portions subject to appreciable telluric absorption, will require improvements to these telluric models.

#### 6.2.1.2. Visible Gas Cells

Iodine cells provide an easily implemented and inexpensive wavelength calibration for high resolution visible spectrometers, enabling them to reach a Doppler precision of 1.5 m/s. The superimposed iodine spectrum has three great attributes and one great drawback. The superimposed iodine spectrum is obtained simultaneously with the stellar spectrum. During an exposure of a star, any changes that occur in the optical performance of the spectrometer (wavelength or LSF) or in the illumination pattern of its optics (near-field or far-field) are simultaneously revealed in the shapes and positions of the thousands of iodine lines from 5000 - 6200 Angstroms. Thus, the three great attributes of superimposed iodine spectrum are the conveyed wavelength scale, LSF, and simultaneity with the stellar spectrum.

The drawback of the iodine technique stems from the blending of all stellar absorption lines with iodine lines. The blending forces the practitioner to construct a forward-modeling spectrum synthesis formed by multiplying the underlying iodine and stellar spectra, convolved with the floating LSF and placed on a floating wavelength scale. The great drawback is that the underlying intrinsic stellar spectrum is not known (i.e. prior to convolution with the LSF) and therefore must be deduced by taking a spectrum of the star without the iodine cell (a "template spectrum") and deconvolving it. The deconvolution suffers two indignities. The LSF at each wavelength is not known perfectly, so the deconvolved spectrum carries a systemic error, including asymmetries in the lines, caused by the poorly known LSF. Also, the deconvolution process is inherently ill-defined mathematically, causing "noise" to be injected into the deconvolved spectrum. Thus, this needed ingredient in the forward modeling process induces systematic errors in the resulting RVs, typically at the level of ~1.5 m/s.
A final drawback of the iodine technique is that the LSF and wavelength scale must be determined anew for each spectrum, as the spectrometer itself is typically unstable at the level of ~1 km/s. The need to determine LSF and wavelength scale anew, at each wavelength in the spectrum, necessitates higher signal-to-noise ratios in the spectrum and hence longer exposure times.

Ultimately, the iodine technique is roughly a factor of 10 “slower” than RV techniques that rely on intrinsic stability of the spectrometer itself. For example, work on Kepler-78 b by both Keck-HIRES and HARPS-N achieved similar Doppler precision with exposures of ~1 hour, but the telescope aperture ratio is a factor of 7 (Howard et al. 2014, Pepe et al. 2014).

#### 6.2.1.3. NIR Gas Cells

In the red end of the visible spectrum (e.g. 650-900 nm), there are no known suitable gases with high enough optical depth with large dipole for ro-vibrational transitions, with water being the strongest, for use in absorption gas cells. In the NIR, however, a number of candidates have emerged and been deployed with NIR high-resolution spectrographs that provide comparable benefits to the iodine gas cell in the visible (see Section 6.2.1.2). In particular, both the room temperature ammonia and isotopic methane gas cells have yielded PRV measurements of ~5 m/s in the NIR in the H and K bands over time-scales of a couple years (Mahadevan & Ge 2009, Bean et al. 2010, Anglada-Escude et al. 2012b, Plavchan et al. 2013a). The benefit of using an isotopic methane cell ($^{13}CH_4$) over regular methane ($^{12}CH_4$) is that the reduced mass of the molecule is changed with the presence of the extra neutron. The change in the reduced mass produces an effective shift of ~10 nm of all the ro-vibrational absorption lines in the NIR relative to the variable telluric methane absorption lines. However, care must still be taken to model and remove the dense telluric lines where telluric and gas cell absorption bands overlap. Additionally, the line density of isotopic methane is so high that identifying the continuum in a spectrum is not possible in the K-band. As a result, Plavchan et al. (2013a) resorts to simultaneously solving for the blaze function correction and RV measurements. The resulting forward model has too many free parameters to be readily fit with a least-squares brute force approach, and an amoeba-simplex parameter search algorithm is employed instead (Nelder & Mead 1965).

### 6.2.2 Non-Common Optical Path

With illumination, pressure, temperature and gravity stabilized spectrographs, non-common optical path wavelength references can be used to derive spectroscopic wavelength solutions and PRV measurements. Emission line lamps have been used in spectroscopy for wavelength calibration for centuries, and their specialized use for PRVs has progressed rapidly over the past decade.

In the visible, Thorium-Argon lamps are used with the HARPS spectrometers to achieve ~70 cm/s single measurement precision. Thorium-Argon wavelength calibration can have limited simultaneous application for faint targets due to scattered light and longer integrations, but can still be obtained before and after a science observation. Additionally, emission line lamps have finite lifetimes. HARPS is able to achieve such precise long-term wavelength calibration by using a nested series of Thorium-Argon lamp, including one that is only turned on once a year to calibrate the other Thorium-Argon lamps that are used and replaced more frequently (Pepe et al. 2000).

In the near-infrared, Mahadevan et al. (2010,2012) have shown that Uranium-Neon emission line lamps also produce a dense set of emission lines across the YJH bands, and was successfully demonstrated with the PATHFINDER spectrograph, and will be employed for wavelength calibration with the HPF spectrograph.

To mitigate the non-homogeneous coverage of emission lamp lines across orders, particularly towards the red, PRV scientists have started to explore the stabilization of Fabry-Perot etalons for PRV wavelength calibration, with lab-based stability of 30 cm/s demonstrated in the near-infrared (Halverson et al. 2014), and less than 1 m/s in the visible (McCracken et al. 2014) with the potential for 1 cm/s stability. Fabry-Perot etalons involve the injection of white light or laser light into a mirrored resonant cavity. The light that escapes from a pinhole in the cavity alternatively destructively and constructively interferes with itself as a function of wavelength, producing a repeating pattern of emission lines for wavelength calibration, with the shape of the lines affected by the finesse of the cavity. Without stabilization of the cavity length, the interference pattern can shift in wavelength with a radial velocity equivalent of hundreds of meters per second.

Several groups are also pursuing the use a laser frequency comb (LFC) to monitor the instrumental line profile of stabilized spectrographs in the visible and near-infrared (Li et al. 2008, Osterman et al. 2011, Langellier et al. 2014 ). LFCs are ultra-stable wavelength references, with equivalent radial velocity stability of < 1 cm/s. This extreme precision is achieved by the comb spanning a factor of two in wavelength (an octave), allowing the use of a frequency double to self-reference the comb to itself (from one end of the spectrum to the other), locking both the line-spacing and the wavelength zero-point. Since the LFC lines are narrow, the recorded LSF of a calibration exposure is a true representation of the Optical Transfer Function (OTF), and thus instrumental changes can be detected and distinguished. However, laser frequency combs still have shortcomings. One is the limited wavelength range (100-150 nm). Another technical challenge is the intrinsically varying line profile of the LFC. While source comb lines are practically delta functions on the dispersion scale of the spectrograph, those source comb lines are so dense that filtering of every N-th line (N=10…40) is necessary to clearly resolve a comb like spectrum with an astronomical spectrograph. This filtering relies on Fabry-Perot cavities, which are sensitive to environmental changes, and if stabilization is insufficient then the line profiles of the filtered LFC can change over time, introducing instrumental errors. State of the art LFCs suppress these effects to a cm/s level, although it is still an issue that these systems are very complex and expensive ($1-2M). Development efforts focus on long-term stability and "turnkey" operation (with or without a laser physicist on call), with such a comb to be tested on sky at HARPS in 2015.

Finally, to get around the need for line-filtering Fabry-Perot cavities in LFCs, which complicates the optical components and electronics considerably, other teams are pursuing the use of “micro-combs” on a chip, with a recent on-sky demonstration in the NIR (Yi et al. 2015). These combs have native line-spacing that matches the spectral resolution of astronomical spectrographs (e.g. R=50,000), eliminating the need for line-filtering. However, these combs are not octave-spanning, and are less stable (but still at a few cm/s) and must be locked to an external wavelength reference.

## 6.3 Spectrograph Illumination and Stabilization

Performing any measurement on a recorded spectrum it is important to understand how the spectral lines are formed by the instrument, and if that has any effect on the results. The shape and *position* (the RV signal) of the 1-dimensional extracted line profile, or LSF, is affected by the following:

A. The wavelength dependent intensity of the light source, the actual spectrum we like to recover
B. The stability of the spectrograph
C. The shape, size and stability of the slit
D. The light intensity distribution (illumination function) over the spectrograph slit
E. The illumination function of the spectrograph optics
F. The optical transfer function of the spectrograph[1]

Of course ***A*** is affected by astrophysical processes we do not control. As we advance in controlling effects of ***B-F*** it becomes more and more obvious that we need a better understanding of the actual stellar spectrum, if we aim to push RV measurements to the cm/s level. Stellar noise (granulation, spots, plagues, active regions, etc.) is imprinted in the spectrum in a way that is very hard to reconstruct, even if we have full control over the instrumental effects (Section 5.2).

HARPS has paved the way to show how and what extent can we eliminate ***B*** by tightly controlling and stabilizing the environment of a spectrograph (gravity invariant location via fiber feed, to avoid time dependent mechanical distortions; vacuum enclosure to mitigate pressure and temperature variations that affect the optical properties; mK level temperature stabilization to avoid mechanical distortions due to non-zero/differential thermal expansion coefficients of the opto-mechanical elements).

The importance of ***C*** is obvious, since essentially every spectral line is a monochromatic image of the slit. This is a static component: once the shape and size of the spectrograph input is chosen, it does not change over time. It is important to note, though, that for exquisite stability the slit cannot be altered in any way, i.e. by adjusting the slit width or exchanging fiber feeds, since any error in repositioning directly translates to RV errors.

The size of the slit, or fiber, is constrained by light-coupling efficiency at the telescope focal plane. Usually the slit width is matched to the size of the median seeing disk. With recent advances in adaptive optics (e.g. the MagAO system on the Magellan telescope) a significant gain of throughput (or resolution, by applying a smaller slit) can be achieved even in the visible by feeding a spectrograph via an AO system. The table below lists estimated slit efficiencies for the GMT AO system in the visible, using different level of correction and assuming 0.7 arcsec circular slit, matching the median seeing condition. Using high order corrections the gain is significant, 70% at 550nm, almost halving the exposure time. For NIR spectrographs this is even more important.

| *Wavelength (microns)* | 0.440 | 0.550 | 0.640 | 0.790 |
|---|---|---|---|---|
| *No sensor* | 0.3949 | 0.4126 | 0.4254 | 0.4458 |
| *Tip-tilt sensor* | 0.4935 | 0.5215 | 0.5417 | 0.5723 |
| *GLAO* | 0.5163 | 0.5461 | 0.5678 | 0.6000 |
| *High-order sensor* | 0.6764 | 0.7086 | 0.7302 | 0.7527 |

**Table 4.** *0.7” encircled energy for GMT as a function of wavelength and wavefront correction strategy.*

Besides overall throughput gain AO feeds can also contribute a lot to ***D***, by stabilizing the slit illumination function. Tip-tilt or “first-order” AO systems have been employed at several spectrographs (TRES, HARPS, etc.), but operating at relatively low (0.1-10 Hz) frequencies the main objective is to quickly correct telescope guiding errors. For slit-fed spectrographs the star wandering within the jaws of the slit (see Figure 47a) can cause severe RV errors, since the spectrograph optics translates a non-evenly illuminated slit to a side-lobed LSF.

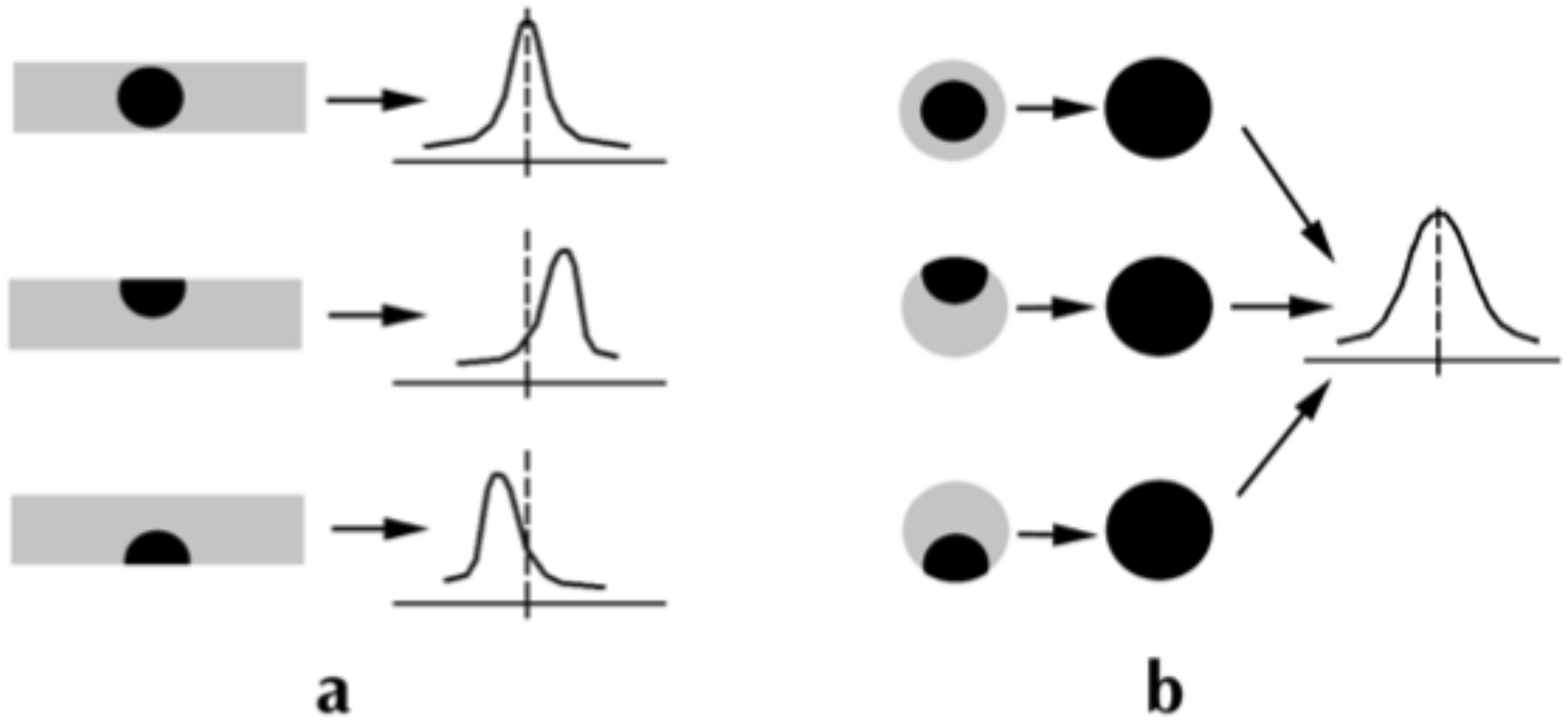

**Figure 47.** *Effects of varying slit illumination for a slit-fed and a fiber fed spectrograph.*

Thanks to the light-scrambling properties of fibers the output of an optical fiber, which serves as the “input slit” for the spectrograph, has much less of a resemblance how the light was fed into the fiber at the input (telescope) end. At the first order the star can be imaged a bit to the side of a fiber, the output end face is still evenly illuminated and thus results a stable LSF (see Figure 47b).

Circular cross section fibers, however, do not scramble light perfectly. The mixing of light rays in the azimuthal sense is nearly perfect, but in the radial direction it is not. (Shining a laser into a fiber at an angle, and projecting the light emerging from the fiber to a screen, can easily demonstrate this: the illumination pattern resembles a ring, not an expected Gaussian distribution.) Numerical simulations and laboratory tests clearly show (see Figure 48) that breaking the non-circular geometry of the fiber can significantly improve on the scrambling. It was first explored by industrial applications where laser beam delivery for metal cutting required an even and constant distribution of light/heat, and so these applications started using square fibers. But D-shaped or hexagonal fibers works just as well. Astronomy prefers octagonal shapes since that is very close to the overall circular seeing limited image of a star. Coupling at the telescope is most efficient this way, while the 1-dimensional projected image of an octagonal slit is the narrowest and thus it provides the highest possible spectral resolution. (Using square fibers core of the LSF becomes a top-hat function, which is “less sharp”.)

Since the LSF, as mentioned, is the monochromatic image of the slit, any change in the illumination distribution of the slit directly translates to changes in the line profile. These changes are often not 100% symmetric, resulting in the shift of the LSF barycenter, or an instrumental RV signal. Moreover, as apparent from Figure 48, the changes are wavelength dependent, so different spectral lines would exhibit different amount of instrumental drift, adding even further noise to a RV measurement.

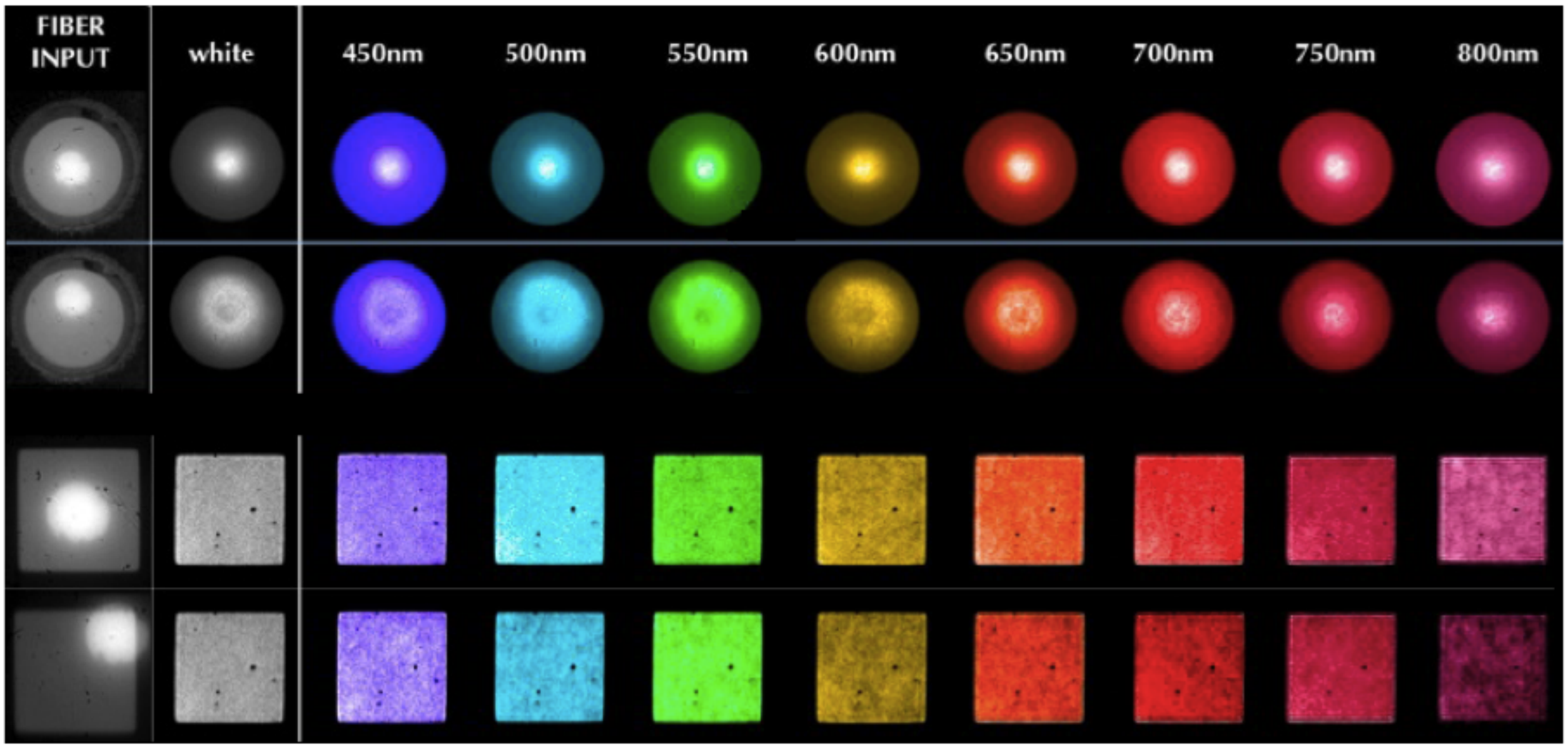

**Figure 48.** *Illumination patterns at the fiber output (near-field) for on-axis and off-axis input illumination, for circular (top) and square (bottom) fibers. The two leftmost columns shows the input and output in white light, while the colored columns give a sense of wavelength dependence in 50 nm increments from 450 nm to 800 nm.*

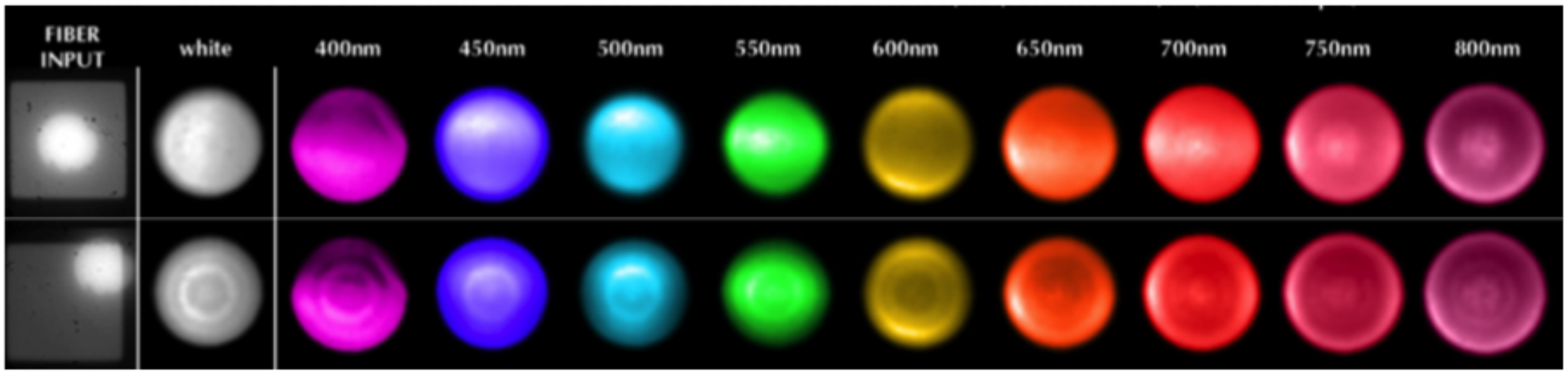

**Figure 49.** *Illumination patterns at the spectrograph optics (far-field, or pupil illumination) for on-axis and off-axis input illumination for a square fiber. Unlike the near-field of Figure 48, the far field is still very sensitive to the fiber input conditions, even for a non-circular fiber geometry.*

While non-circular fibers prevent changes in the slit illumination, or the so-called near-field, it only means that the spatial distribution of light rays, as those exit from the fiber, is even. But the angular distribution of light, so called the far-field (that one can observe projecting the light from a fiber to a distant screen) is also very important. The reason is that the optical transfer function (OTF) of the spectrograph highly depends on what part (and with what distribution of intensity) the optical elements are illuminated. Light passing through near the edges of lenses suffer more heavily of optical aberrations. Thus if a spectrograph is illuminated by a ring-like pattern it would produce heavily aberrated images of the slit, and the centroid of such LSF would be obviously different than the aberration free LSF of a more centrally illuminated optical system. Therefore the pupil illumination, ***E***, or far-field pattern within a spectrograph is very important. Numerical simulations or laboratory experiments have shown that using non-circular fibers does not solve this problem (see Figure 49). The way of minimizing this effect is therefore:

- Use spectrograph design in which the pupil illumination is very similar at all wavelengths, since the OTF is also wavelength dependent. Therefore the *white pupil optical arrangement is highly beneficial for PRV* spectroscopy.

- Use *non-circular fibers along with an optical double scrambler*. This is a simple (1-2 element) optical device that is inserted into a break of the fiber link. It injects the light emerging from the first section of the fiber run to the second section in such way, that the perfectly scrambled spatial distribution of rays (the near-field) is translated to an even angular distribution while the imperfectly scrambled angular distribution (the far-field) of the first fiber segment becomes the near-field for the second fiber segment. This way both the near field (the slit illumination) and the far-field (the spectrograph pupil illumination) gets evenly mixed and becomes uniform.

- Build *an optical system that is least sensitive to changes in pupil illumination, by design*. During the optimization of a spectrograph optical design one can illuminate the system on the edge and the center, and calculate how much the resulting centroid locations of simulated line profiles would shift. The HARPS instrument was not designed this way, although by chance the system had such a behavior that changing pupil illumination shifted lines several hundred m/s to the blue at the short wavelength end of spectral orders, while an almost exactly same amplitude but opposite shift occurred for the lines at the long wavelength end of a spectral order. Thus, doing cross correlation over an entire spectral order, these effects cancelled each other resulting a net zero RV error. (Note, that if the RV content of the spectrum is not evenly distributed within a spectral order the net effect is not zero.) But other spectrograph designs, like SOPHIE, were not as lucky. Realizing this the next generation of PRV instruments are being designed with this in mind. The G-CLEF optical system for the GMT and the ESPRESSO design for the VLT both incorporated pupil-illumination insensitivity to the optical design process.

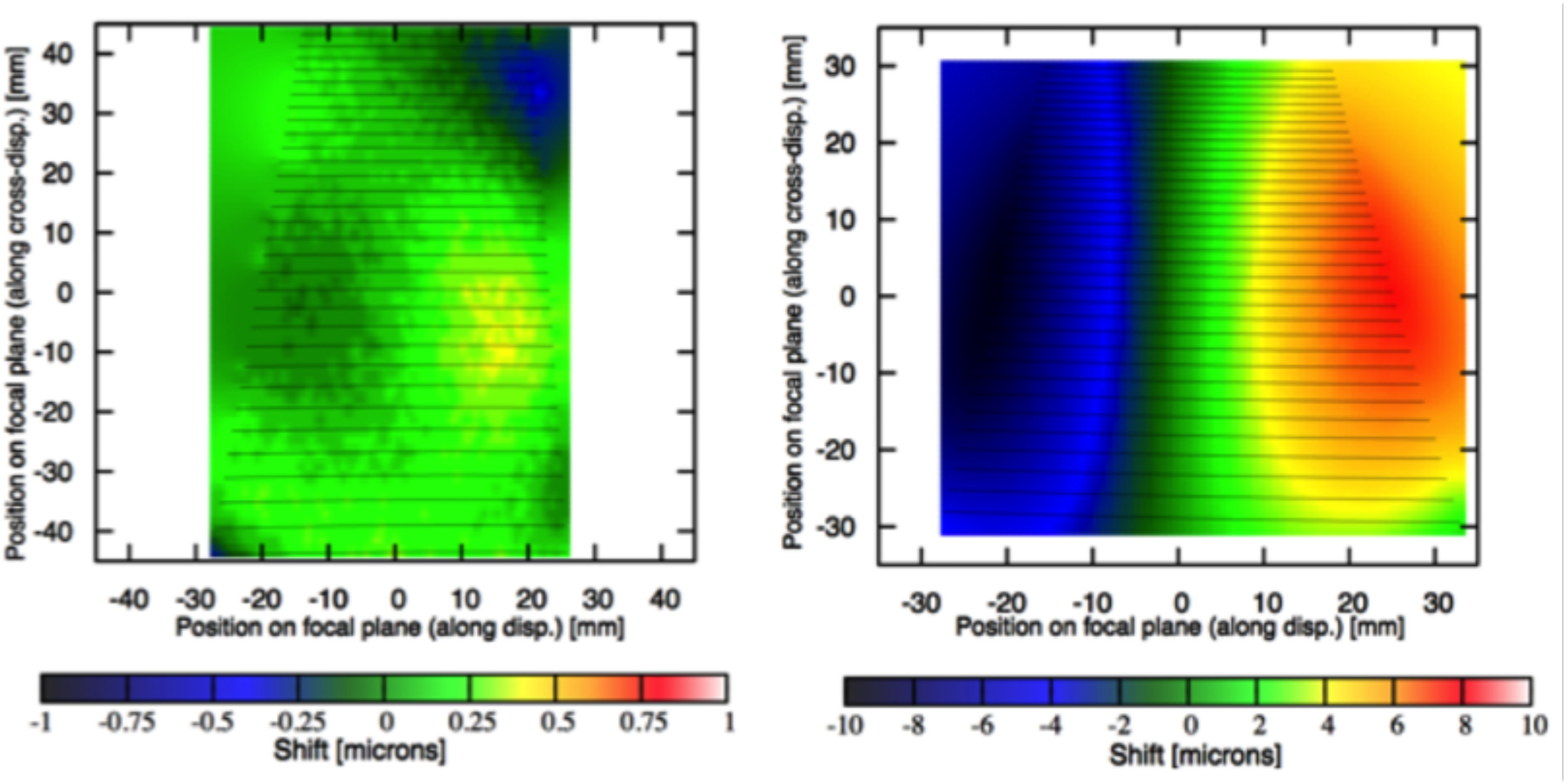

**Figure 50.** *Color coded shifts of the LSF centroid, in microns, for an extreme illumination change of the spectrograph optics (50% inner area of the pupil vs. 50% of the outer area of the pupil). Real changes in the far field illumination are not as extreme, of course, so above results proportionally scale down accordingly, resulting sub-micron shifts of the spectral lines (which are still significant, hundreds of m/s in RV space). Still, it is clear that a thoughtful spectrograph design (G-CLEF blue arm, left) can significantly decrease the false RV signals due to pupil illumination changes. NOTE that the similar figure for HARPS has a color scale that is 10x larger. Also note that HARPS was not designed with pupil illumination insensitivity in mind since back then its importance was not yet recognized.*

While non-circular fiber geometries and optical double-scramblers stabilize and even out the near and far field illuminations, factors ***D*** and ***E*** are still varying in time due to physical changes in the fiber as temperature changes and/or the telescope moves. Since the seeing limited image of a star is at the order of several tens to several hundred microns (reaching mm scale with the ELTs), the optical fibers are inherently multi-mode. Even at a single wavelength light can propagate through such large core fiber in thousands of ways (modes). Looking at the intensity distribution at any point within or after the fiber (near field or far field) what we see is essentially the superposition (or interference) of these multiple modes. Such speckle pattern is clearly apparent when illuminating a fiber with a coherent source, like a laser. Such pattern changes as the fiber moves and the overall barycenter can also be affected by this, which translates to a small but disturbing instrumental RV signal. Agitating (shaking) the fiber can decrease this modal noise (see Figure 51), although constant mechanical stress/disturbance of the fiber decreases the lifetime of the fiber link and results in some quality loss (increased focal ratio degradation, meaning that over time some fraction, up to 10%, of the light can be lost due to internal scattering as the fiber, that is constantly agitated, ages). Note, that at high resolutions the wavelength across the size of the slit (or in other words a spectral line) is so small that the light is almost monochromatic within the LSF.

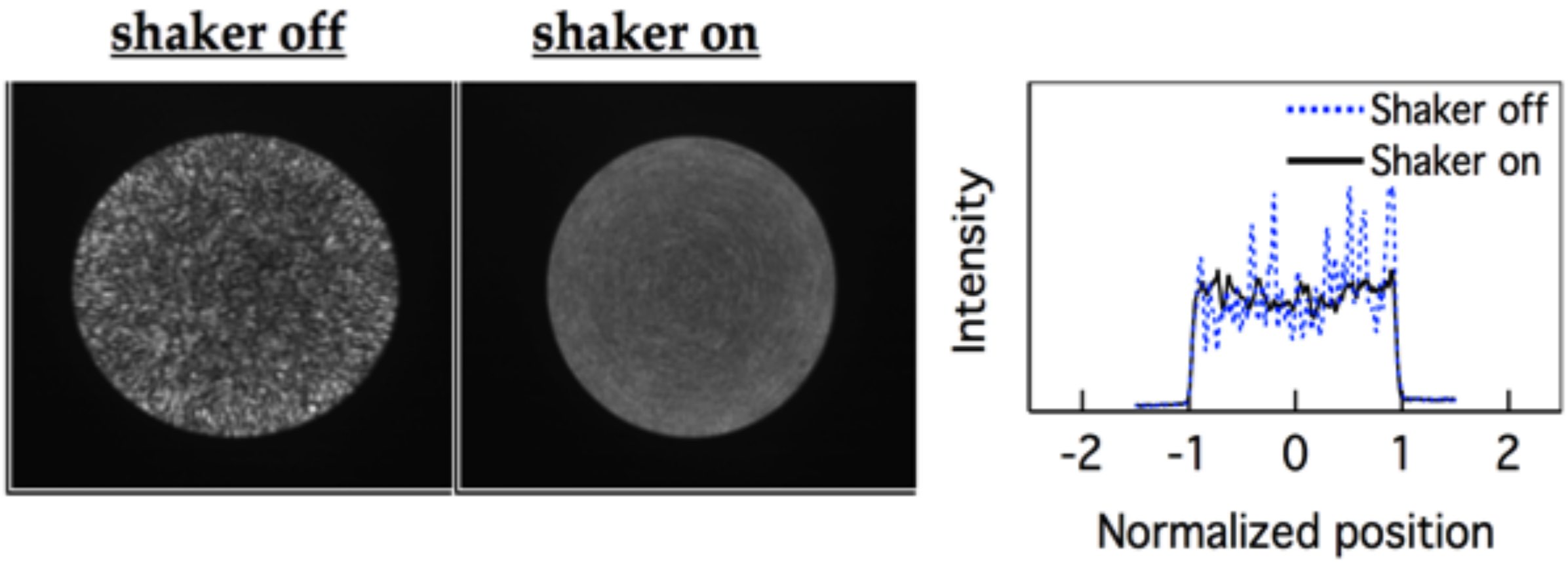

**Figure 51.** *The speckle pattern, the superposition of modes (left) changes during an exposure and can alter the LSF barycenter. To prevent this shaking the fiber (middle) is a solution that can smooth the slit and pupil illuminations, as clearly shown by the cross-sectional intensity plot on the right.*

Another way of solving the modal noise problem is to utilize photonic devices. Recent developments of photonic lanterns and multicore fibers hold the promise of transferring light from a large core, multi-mode fiber into a set of single mode fibers (photonic lantern) or into a fiber that has a large diameter but light is essentially guided in many small fibers embedded within (multi-core fibers). Similarly to an optical double scrambler these devices can perform modal scrambling, further stabilizing the slit and pupil illumination. The total throughput of these devices is not sufficient yet, and the modal scrambling properties also need to be improved, but the technology is promising.

Last but not at least the optical transfer function (***F***) of the spectrograph is relevant for PRV stability since decoupling instrumental effects from astrophysical phenomena is only possible if we understand how and why the LSF is changing over time. Instrumental and environmental effects can alter the OTF (and thus the LSF), and unless calibration exposures can clearly show that these changes are internal to the instrument there is no way of knowing if the change was in the instrument or in the stellar source.

The problem is that in a realistic optical system the OTF is not known, since manufacturing and alignment errors can significantly alter it compared to the target optical design/model. Therefore modeling of the OTF is only possible by fitting real calibration data, e.g. emission line profiles of a ThAr lamp. But the illumination of a calibration source does not fully mimic the illumination of a stellar source - E.g. the number of modes excited in a fiber are very different for a steady and Gaussian or top-hat like calibration illumination, compared to the dancing speckle pattern of a star produced by atmospheric seeing. Thus changes seen in a calibration exposure LSF might not fully resemble the changes of the LSF of a stellar source.

The iodine cell (or other gas cell) technique is not sensitive to such de-coupling, since the calibration source is illuminated by starlight, essentially the science exposure carrying the calibration information. However, due to the relevant absorption, the limited wavelength range, and the complex modeling of iodine lines this technique has its own limitations.

Since every calibration and correction leaves residuals behind, it is a good strategy to minimize the need for the corrections or number of parameters that describe them. If the OTF of an optical system is highly varying along the spectrum (with wavelength and spatial position on the detector) then a large number of parameters are required to describe it, and also a large number of calibration features are required to provide sufficient data for fitting these parameters. It is possible, however, to design optical systems in which the OTF is a very smooth function of wavelength or spatial position on the detector (see Figure 52). By creating a PSF that is highly uniform across the entire echellogram the number of parameters describing the LSF in a calibration or stellar spectrum becomes minimal, and thus fitting is more robust, the residuals are smaller, resulting better PRV performance. Note, that it is only true, if such behavior (uniform PSF) is a property of the as-built optical system, thus not just the optical design process but also the tolerance analysis has to incorporate guidelines of PSF/OTF uniformity.

In any optical system that needs to cover a wide wavelength range there will be a dependence of the OTF with wavelength. But if the FWHM of the PSF of the spectrograph optics can be small compared to the geometrical image of the slit, then the shape of the LSF will be more dependent on the geometrical shape/size of the slit, ***C***, which is constant in time and not dependent on wavelength. Thus in such case the recorded spectral lines will be very uniform across any given spectral order, or the entire recorded spectrum (see Figure 53). Of course this is much easier to achieve when the image of the slit is several tens of microns, since typical imaging cameras of spectrographs render images of zero extent sources to a size of 2-20 microns (the FWHM of a camera optic PSF is at the order of a detector pixel). So for the 8-10m class telescopes or ELTs, where the spectrograph entrance slit is naturally large due to the large plate scale (1” per mm for the GMT), it is not hard to achieve a small PSF compared to the slit image. Still, for smaller telescopes and spectrographs it is advantageous to not just minimize the FWHM of the PSF of the spectrograph, but as pointed out above, to make it uniform even at the expense of not forcing the PSF to be as small as possible.

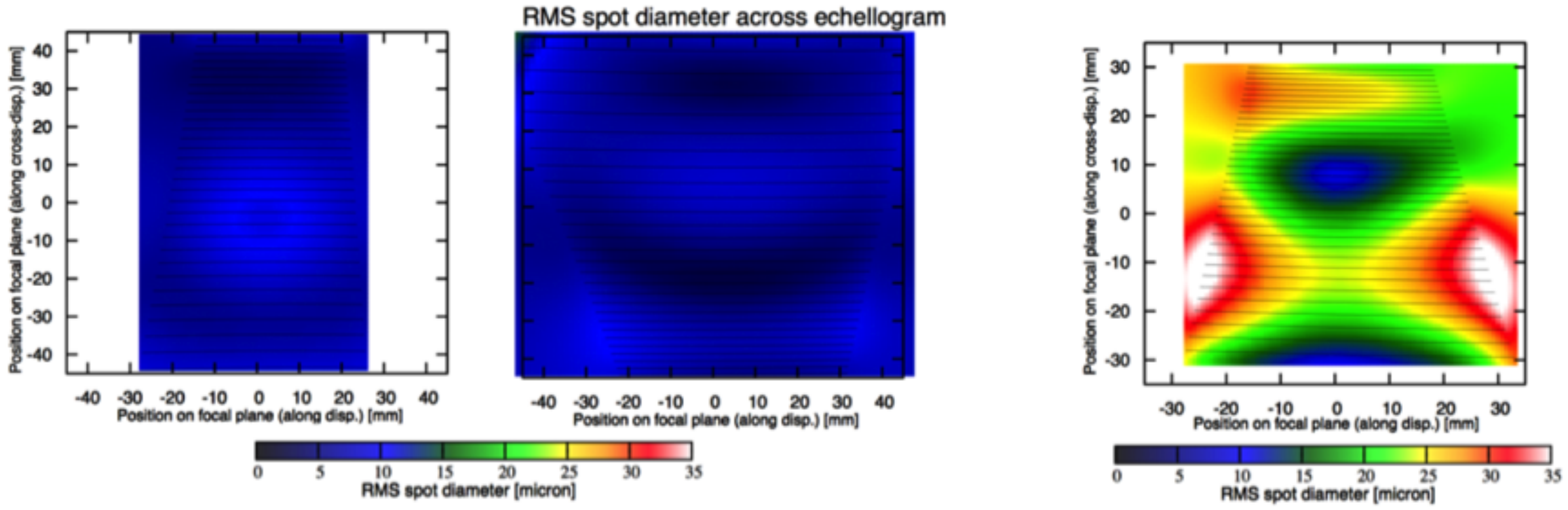

**Figure 52.** *Color-coded map showing the FWHM of the PSF, in microns, over the entire blue and red echellograms of the G-CLEF spectrograph design (left), and comparing it to the same quantity for HARPS (right, single echellogram since HARPS has just one camera). Note how small the size of the PSF and its variation is, thanks to the optical design, for G-CLEF. NOTE that HARPS was not designed to such requirement, although they consciously enforced left and right symmetry, which luckily also brought the same symmetry for sensitivity in changes of pupil illumination (see Figure 50).*

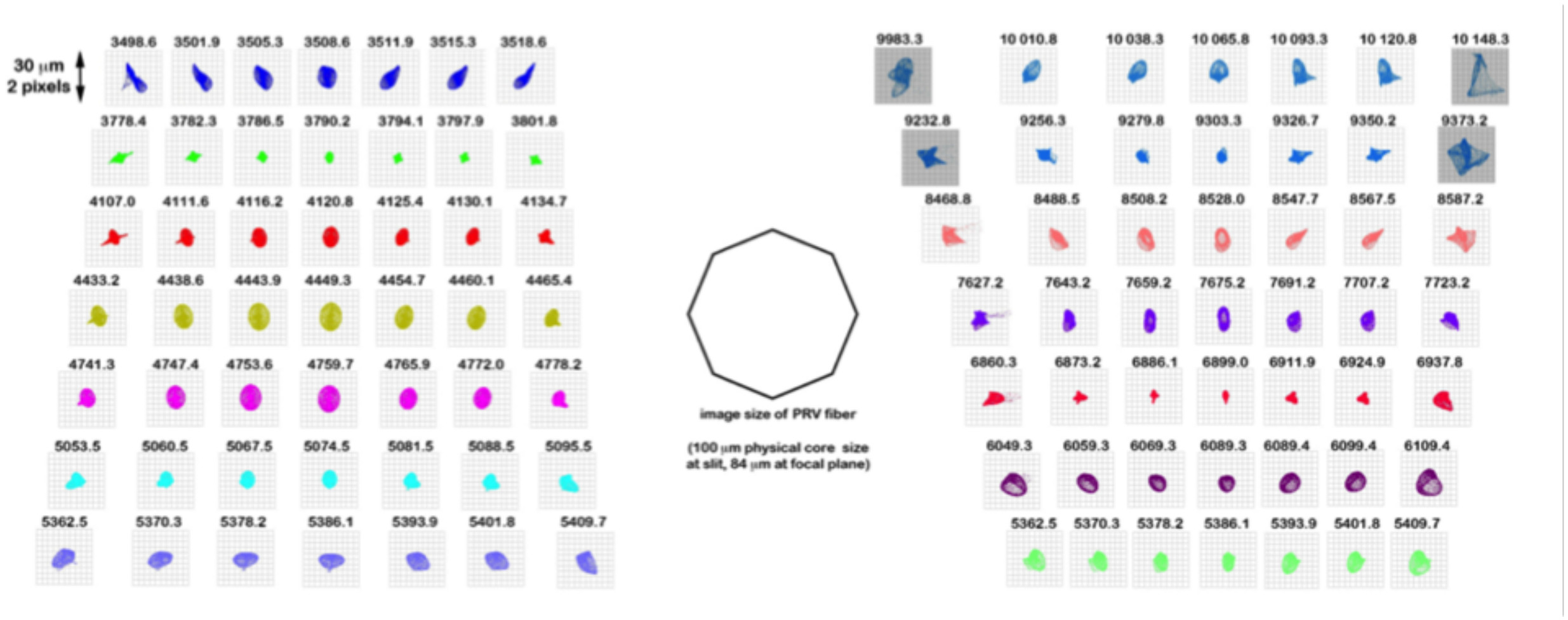

**Figure 53.** *Spot diagrams for the G-CLEF camera designs, following the structure of the echellograms: each spot is placed according to its location on the focal plane. The boxes are 2x2 pixels (30x30 microns), and wavelengths are noted for each box, in Ångstroms. The spots are not just sharp, concentrated, mostly round and smooth, but very uniform across a spectral order (horizontally) or at least display a left-to-right symmetry within an order. In the middle, for comparison, the geometrical image of the slit (a fiber) is shown, to scale. It is clearly apparent that the pure geometrical image of the fiber end face is much larger than any of the spots. The shape of the recorded spectral lines, the LSF, is a convolution of this optical transfer function (the PSF) of the spectrograph and the slit input function (the fiber end face). Since latter is dominant, the resulting LSF is uniform across the echellogram, the intrinsic line width and shape is not a function of wavelength.*

*[1] In order to clearly distinguish between the point spread function (PSF) of the standalone spectrograph optics, (which is only dependent on the optical system itself) and the line profiles of the recorded spectrum (which depends not just on the PSF but also on the slit shape, slit illumination and pupil illumination) we refer to the PSF of the optics as the optical transfer function or OTF. This is valid since the OTF the Fourier transform of the PSF. We refer to the shape of the extracted one-dimensonal spectral line as the line spread function or LSF, which, again, is the result of the convolution and product of multiple instrumental properties/effects.*

## 6.4 Diffraction Limited PRV Spectrographs

The motivation for developing new PRV spectrometers at NIR wavelengths is many-fold (Section 4.3). However, there exists an additional relevant threshold that is crossed when designing PRV instruments for wavelengths beyond $\lambda$=1.0 micron: it becomes possible to correct for Earth's turbulent atmosphere using adaptive optics to reach the diffraction limit of large aperture telescopes. Designing a spectrograph at the diffraction-limit behind an AO system works in favor of improved RV precision. Upon creating a spatially compact input image, gains in spatial resolution translate directly into higher spectral resolution and also higher SNR observations (see below). While seeing-limited instruments can achieve adequate resolution to sample stellar absorption lines, finer sampling yields higher Doppler precision and enables the study of more subtle effects involving line asymmetries, which are important for false-positive analysis involving starspots (Figueira et al. 2013). Furthermore, seeing-limited spectrometers require large and expensive gratings to measure precise Doppler shifts. In addition to a large beam diameter, the focal length of individual optics must be commensurately long to help mitigate aberrations. In comparison, the physical dimensions of a diffraction-limited spectrometer are an order of magnitude smaller in all directions, allowing for a small observatory footprint (mechanical, electrical), high spectral resolution using off-the-shelf components (R>100,000), higher optical quality components, faster design and development timeline, improved thermal stability, and much lower overall cost.

Diffraction-limited designs can also completely eliminate spurious RV variations caused by modal noise. The pure Gaussian beam of a single-mode fiber provides a stable illumination of the spectrograph optics irrespective of changes to wave-guide boundary conditions. Single-mode design options (whether a single small fiber or multiplex) are not available to seeing-limited instruments given the large number of electromagnetic modes (thousands) excited by much larger optical fibers (Chazelas et al. 2010). Recent developments in the astro-photonics show promise for employing so-called "photonic-lanterns" to collect large amounts of starlight using multi-mode fibers, which then output an array of single-mode fibers, thus maximizing throughput while maintaining single-mode operation at the expense of detector real estate.

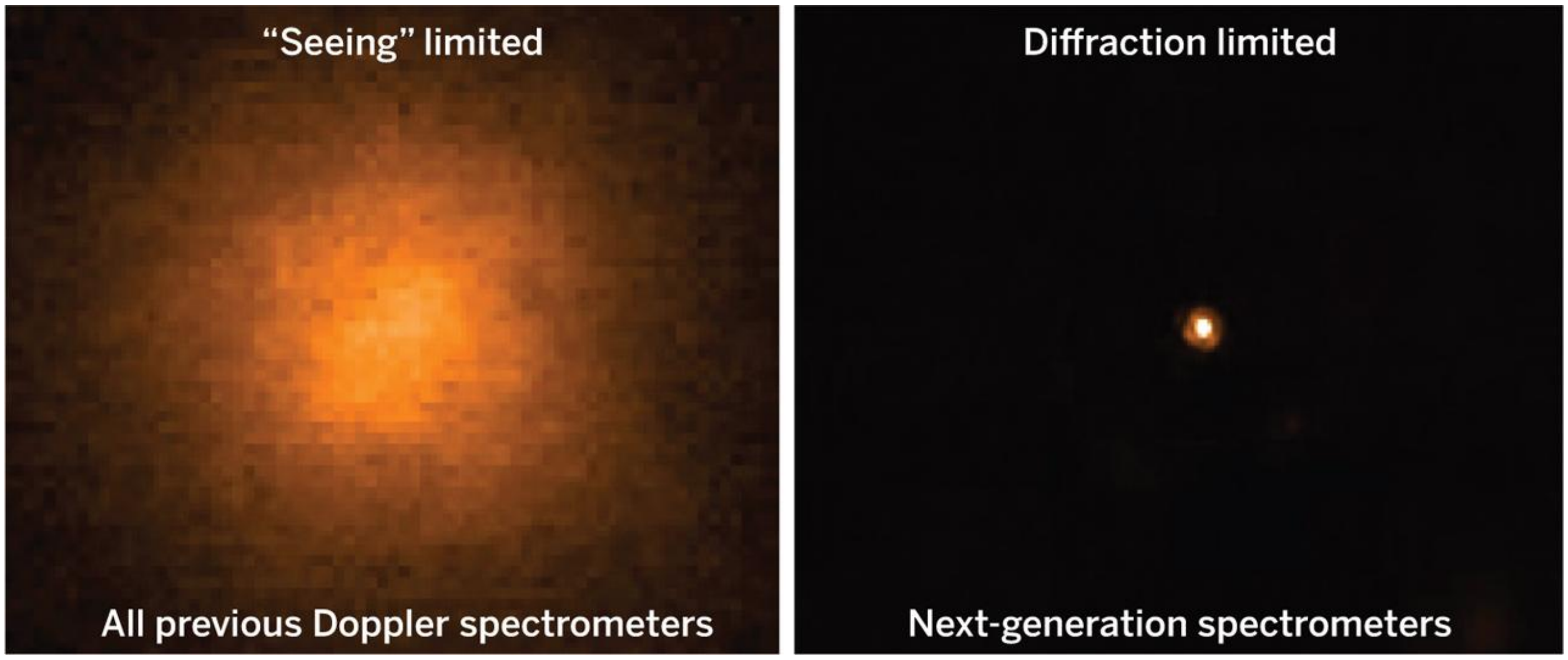

**Figure 54.** *Figure reproduced from Crepp (2014). Comparison input image size for telescopes limited by atmospheric turbulence and those using adaptive optics. As a result of this order of magnitude effect, a diffraction-limited spectrometer will achieve high spectral resolution, high spatial resolution, high throughput, and a compact optical design, while naturally operating in a wavelength range ideal for detecting extrasolar planets.*

Improved spatial resolution not only benefits spectral resolution but also permits the identification of nearby stellar companions, either physically associated or background contaminants, which can easily confuse RV signals. In fact, the fiber acts to effectively "block out" starlight as viewed by the acquisition camera thus acting as a coronagraph. For example, the fore-optics in a diffraction-limited spectrometer could achieve high contrast to directly image nearby companion stars that dilute transit observations or create false positive signals. Such capabilities would offer a powerful tool for reconnaissance Kepler and TESS observations and save telescope time by instead maintaining focus on the highest priority candidate planets.

A diffraction-limited input image also reduces the intensity of bright OH-emission lines by 2-3 orders of magnitude compared to seeing-limited designs. While initially envisioned as a possible wavelength reference, the intrinsic stability of OH-lines limits RV precision to several tens of meters per second, even when NIR observations are coupled with detailed atmospheric modeling (Blake et al. 2007; see also Section 6.2.1.1). By reducing the signal of OH-lines, AO-fed Doppler spectrometers thus achieve a higher SNR ratio. Further, internal Bragg gratings used to suppress the brightest OH-lines in multi-mode fibers may also be used in tandem with much smaller fibers, including single mode.

In addition to the numerous practical benefits of developing diffraction-limited PRV instruments, AO-fed spectrographs provide access to science cases completely inaccessible to seeing-limited designs, such as the ability to study planets orbiting close-separation binary stars (Figures 55). Binary stars constitute 50% of all stellar systems, yet our understanding of planet formation and evolution is currently based on single stars alone. Due to a selection bias, seeing-limited spectrometers cannot spatially separate the light from two nearby sources. A diffraction-limited spectrometer can circumvent this issue by using AO to prevent spectral contamination and enable the first systematic study of planets in close-separation binaries.

Studies of the binary planet population also have important theoretical implications for our understanding of how planets form and evolve:

- How does the occurrence rate of planets change as a function of binary star separation?
- How does the occurrence rate of planets change as a function of binary star mass ratio?
- How does the eccentricity distribution of planets in binaries differ from that of single stars?

The presence of a second star changes the properties of a protoplanetary disk (Artymowicz & Lubow 1994) and, depending on its separation and relative inclination, may either promote or inhibit the growth of dust grains, planetesimals, and embryos (Payne et al. 2009). Models must be able to account for this extra source of radiation and gravity and would benefit significantly from observational constraints in addition to those imposed by single stars alone.

Closely spaced binaries (< 100 AU) can offer unique testing grounds to help discriminate between various planet formation models. A diffraction-limited spectrograph would enable the first RV survey that systematically targets systems with progressively smaller separations (Figure 56). Benefiting from an order of magnitude improvement in spatial resolution, it becomes possible to explore an expansive, important, and as yet unstudied parameter space. In other words, PRV measurements using an AO-fed design will pave the road towards a much deeper understanding of planet formation and evolution.

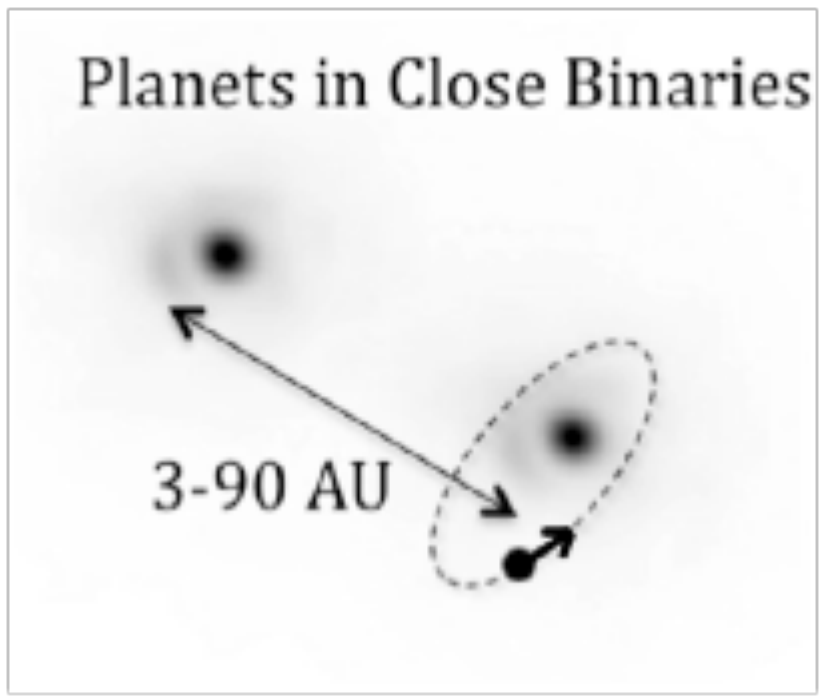

**Figure 55**. *Diffraction-limited imaging using adaptive optics provides access to close-separation binaries by isolating the light from each star, thus minimizing contamination and allowing for the first RV studies in dynamically interesting regimes. AO-fed spectrographs can quantify the frequency of Jovian planets as a function of binary star properties.*

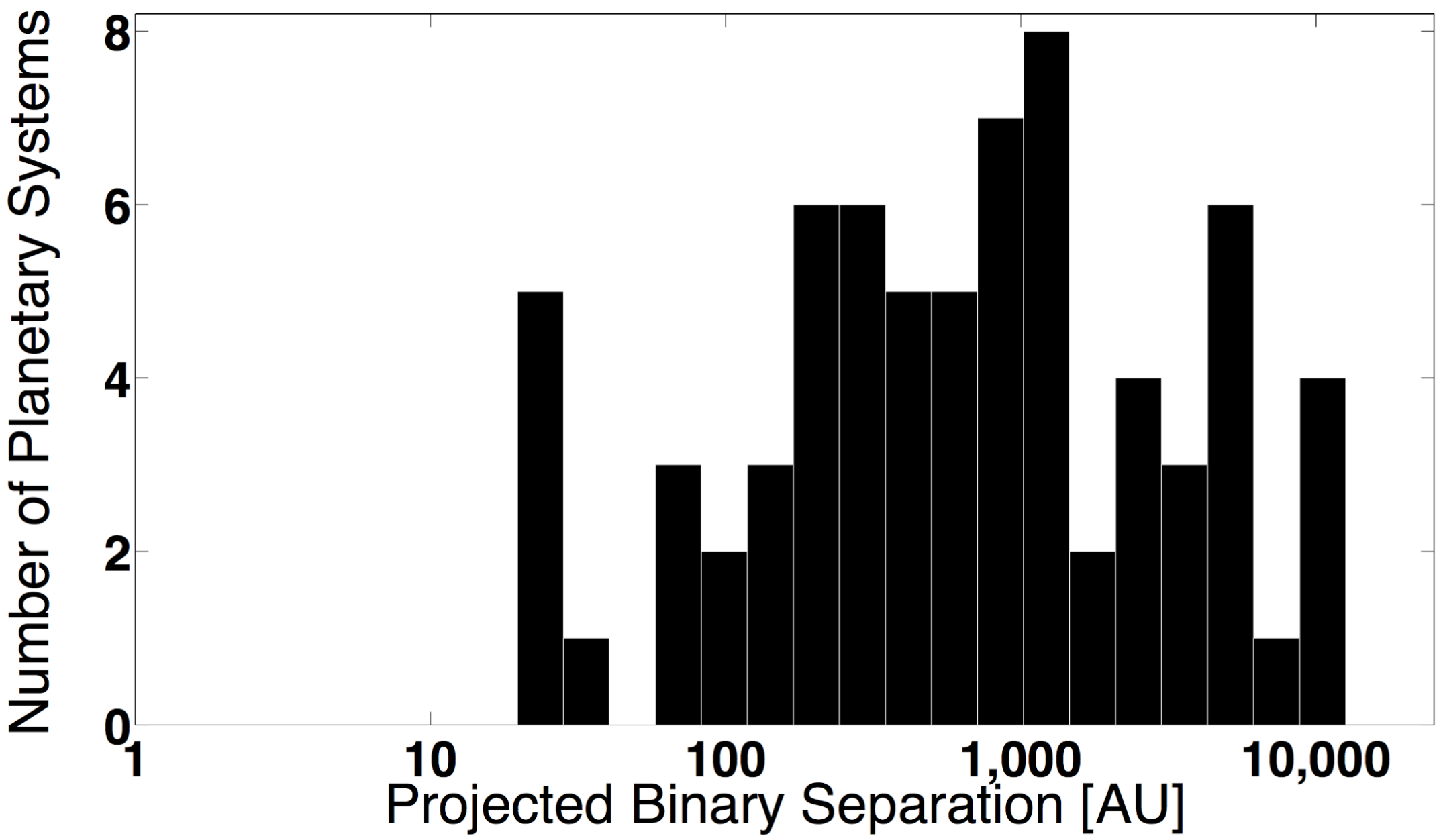

**Figure 56**. *Number of planetary systems discovered orbiting binary stars (plot made using exoplanets.org). Only systems with large projected separation have been searched to date because all Doppler spectrometers are seeing-limited.*

| Noise Source | Seeing-Limited Spectrometer | Diffraction-limited |
|---|---|---|
| Background contamination | Radiation from sky, telescope, and moon | Contamination reduced by factor of 1,000 |
| Binary star interlopers | Contamination from nearby star pollutes spectrum | Immediately identified by acquisition camera |
| Star spots and granulation | Effects are maximized at visible wavelengths (Section 5.2) | Effects reduced by factor of several at near-infrared wavelengths |
| Pressure-mode oscillations | Significant amplitude in Sun-like stars (Section 5.2.6) | Effect reduced dramatically for M-stars |
| Spectral resolution | Moderate/insufficient sampling of stellar spectrum; expensive grating | High spectral resolution; affordable gratings |
| Pressure fluctuations | Vacuum difficult to maintain; expensive; out-gassing | High-quality vacuum seal; inexpensive; low out-gassing |
| Temperature fluctuations | Large instruments are susceptible to temperature fluctuations | Small instruments may be controlled with exquisite precision (1 mK) |
| Detector imperfections | CCD's are inexpensive; achieve requisite noise levels | NIR arrays are expensive; achieve requisite noise levels |
| Wavelength calibration | ThAr lamps proven highly effective | ThAr lamps even brighter at near-infrared wavelengths |
| Fiber modal noise | Must use large (multi-mode) fiber | Modal noise eliminated entirely |

**Table 5**. *List of RV noise sources identified by Lovis & Fischer (2011) in their review chapter (Editor S. Seager) of the Doppler method. Red indicates performance degraded by seeing-limited spectrographs; yellow is acceptable; and green represents superb performance. Diffraction-limited designs address each issue.*

# 7. U.S. Community Input

In this Appendix, we summarize PRV specific text and recommendations from major U.S. community reports over the past decade, much of which is still relevant today.

## 7.1 NASA Science Objectives

The U.S. National Space Policy of June 28, 2010 includes among NASA's responsibilities to "search for planetary bodies and Earth-like planets in orbit around other stars." This flows down to Strategic Outcome (Goal) 2.4 in NASA's 2011 Strategic Plan, which states: "Discover how the Universe works, explore how it began and evolved, and search for Earth-like planets." Per the 2010 Science Mission Directorate (SMD) Strategic Plan one of the key questions is: "What are the characteristics of planetary systems orbiting other stars, and do they harbor life?" and one of the key objectives is "Generate a census of extra-solar planets and measure their properties."

## 7.2 ExEP Program Goals and Objectives

The ExEP Charter states, as it pertains to PRVs:

*"The ExEP will include space and ground-based projects and supporting research and technology, as needed, to accomplish program objectives. In particular, the program will support precursor science and technology activities that contribute to future exoplanet space mission objectives."*

*"The ExEP consists of a series of complementary projects that build upon the technological and scientific legacy of their predecessors to support NASA in its search for and characterization of habitable planets outside the solar system. Each mission measures unique properties of exoplanets that, together, build a comprehensive picture of exoplanets and planetary systems, which no single project can do."*

## 7.3 2010 Decadal Survey

PRVs are highlighted as one of the top three scientific objectives for the next decade in the NRC Decadal Survey "New Worlds: Seeking Nearby, Habitable Planets". The charge from the decadal survey is to "discover planets within a few times the mass of Earth as potential targets for future space□based direct□detection missions." This is listed as a "Mid-Scale Innovations Project" top three priority for ground-based work in the $12 million to $40 million range.

Page 20:
*"One of the fastest-growing and most exciting fields in astrophysics is the study of planets beyond our solar system. The ultimate goal is to image rocky planets that lie in the habitable*

*zone—at a distance from their central star where water can exist in liquid form—and to characterize their atmospheres. To prepare for this endeavor, the committee recommends a program to lay the technical and scientific foundations for a future space imaging and spectroscopy mission. NASA and NSF should support an aggressive program of ground-based precise radial velocity surveys of nearby stars to identify potential candidates."*

Page 7-7 (191):
*"This survey is recommending a program to explore the diversity and properties of planetary systems around other stars, and to prepare for the long-term goal of discovering and investigating nearby, habitable planets."*

Page 7-8 (193):

*"The first task on the ground is to improve the precision radial velocity method by which the majority of the close to 500 known exoplanets have been discovered. … Using existing large ground-based or new dedicated mid-size ground-based telescopes equipped with a new generation of high-resolution spectrometers in the optical and near-infrared, a velocity goal of 10 to 20 centimeters per second is realistic."*

Page 1-8:

*"To prepare for direct imaging, "NASA and NSF should support an aggressive program of ground-based high-precision radial velocity surveys of nearby stars to identify potential candidates"*

Page 263:
*"It will be important to make strategic investments in new ground-based capabilities during the coming decade. One important component will be the aggressive development of ground-based high-precision radial velocity surveys of nearby stars at optical and near-infrared wavelengths (including efforts to determine the effects of stellar activity on these measurements). These surveys will need new spectrometers and significant time allocation on 8- to 10-meter-class telescopes."*

## 7.4 NSF Portfolio Review

Subsequent to the 2010 Decadal Survey, NSF undertook a portfolio review in which PRVs were the highest ranked recommendation in Planetary Science and Star Formation:

*Below we list the rank-ordered technical capabilities in AST-supported areas that are needed to address the highest-ranked PSSF (Planetary Systems & Star Formation) scientific priorities from NWNH…*

*1. Extreme-precision OIR Doppler spectroscopy (TC--T): The detection of Earth-mass planets in*

*habitable zone orbits requires radial-velocity (Doppler shift) precision of 0.1-0.2 m/s at optical wavelengths and somewhat lesser precision for NIR studies of cool, low-mass stars. While shot-noise statistics provide a fundamental limit, coupling of light to the instrument, opto-mechanical stability and optimal wavelength calibration are all areas that still merit work. Thus, substantial instrument and analysis development will likely be needed, over the course of several years. Once an understanding of extreme-precision Doppler techniques is in hand, substantial observational resources would be required to carry out the requisite surveys.*

*To make further progress, NWNH recommends the development of new spectrometers capable of achieving 0.1-0.2 m/s precision, and adequate allocation of observing time on 4-m to 10-m telescopes. This is a challenging goal, and one that will likely require investment in technology development. Near-infrared spectrometers may be advantageous for stars cooler than spectral type M4V, which emit their peak flux in the near-infrared.*

*Critically needed technical capabilities to address this research question are high precision (≤1 m/s) radial velocity programs for Doppler planet detection and Kepler follow-up (of larger, higher mass planets), high resolution spectroscopy to characterize the properties of host stars, small-to-moderate telescopes for photometric follow-up of microlensing events/ground-based transit surveys, and instrument development for extreme-precision optical spectrometers (to reach 10 cm/s) and high-resolution near-IR spectrographs to detect planets orbiting cool stars (later than M4V).*

## 7.5 ExoPAG 6 and 7 meetings

At the ExoPAG 6 and 7 meetings, held October 13-14, 2012, and January 5-6, 2013 in Reno, NV and Long Beach CA, respectively, two mini-workshops were held to discuss PRVs and the relevance to NASA. Presentations by a number of US-based astronomers were given covering the topics discussed herein this report. The presentation materials are available at the NASA Exoplanet Exploration Website (http://exep.jpl.nasa.gov/exopag/). In particular, we heard from Andrew Szentgyorgyi on the design of the G-CLEF PRV spectrometer for the GMT, Phil Muirhead on general PRV instrumentation and survey design requirements, Suvrath Mahadevan on the design of the NIR PRV Habitable Planet Finder Spectrometer, Andreas Quirrenbach on the design of CARMENES, an Optical+NIR PRV spectrometer, and from Valeri Makarov on the stellar activity limits to PRV detection sensitivity. Much of their presentations and effort have been incorporated and reflected in this white paper.

## 7.6 2011 Penn State RV Workshop

In 2011, over 200 international PRV scientists gathered at Penn State for a workshop. Taking place shortly after the 2010 National Research Council Astro Decadal Survey, 50 US PRV scientists endorsed a series of recommendations written by a self-appointed committee of the workshop attendees -- William Cochran, Dawn Gelino, Sara Heap, John Johnson and David Latham. This “PRV response” to the 2010 decadal survey noted:

*“We note that the European Southern Observatory (ESO) is developing ESPRESSO for the VLT*

*and CODEX for the ELT, which will provide important new capabilities to European astronomers. Without appropriate investments in instruments and telescope time, the US will lose competitiveness in this critical field. To achieve the most challenging goals, e.g. to detect planets in the habitable zones of solar type stars, will require some technology development to achieve the highest levels of precision. … Kepler will continue to be a pioneer in the field of super-Earth and Earth-sized exoplanets making PRV data on faint targets an absolute necessity to validate the Kepler results and to provide primary physical characterization of the transiting systems. JWST will also need supporting PRV data to validate targets and characterize planets for spectroscopic follow-up observations. Supporting PRV data will also be essential to validate and characterize planets discovered in future, space-based, whole-sky transiting planet surveys. NASA and NSF currently support a large number of observatories that could benefit current and future PRV work (Keck, IRTF, NOAO, Gemini). Both a new generation of optical and/or near-infrared spectrometers, and significant amounts of telescope time are essential to meet the objectives of the decadal survey."*

And then continued to make the following recommendations:

1. *Initiate an open, competitive program to develop advanced precise radial velocity (PRV) spectrographs for existing U.S.-affiliated observatories both in the optical to achieve the 10 cm/s goal set by the Decadal report, and in the near-infrared to provide initial PRV capabilities of 1-10 m/s1;*
2. *Augment the existing program of PRV observations of Kepler candidates by obtaining observing time at existing facilities having the requisite instrumentation;*
3. *Support upgrades of existing spectrographs and the building of new PRV spectrographs (~1 m/s) on NASA-, NSF-, or privately-funded telescopes. These instruments would provide the cadence, precision, and wide sky coverage needed for surveys to find rocky planets orbiting nearby stars. Some of these upgrades or new instruments using existing technology could proceed on a rapid enough timescale to be valuable for Kepler follow-up;*
4. *Accelerate the development of PRV spectrographs by explicitly encouraging instrumentation and/or technology fellowships in this area;*
5. *NSF, NASA, and other government agencies should collaborate to develop the requisite wavelength calibration capabilities to reach the ASTRO2010 goals in both the optical and near-infrared;*
6. *Develop a comprehensive calibration and archiving plan to ensure data obtained via these initiatives are available to the astronomical community in a timely fashion.*

## 7.7 2008 Exoplanet Community Report RV Chapter

In 2008, the exoplanet community gathered to assess the field in preparation for the 2010 Decadal Survey. The Exoplanet Community report contains chapters on different exoplanet detection methods, including a seventh chapter on PRVs led by Guillermo Torres and Dawn Gelino. The chapter details many excellent science objectives and recommendations that are still relevant today. In this sense, this white paper should be considered as an updated addendum to that text. In particular, the authors highlight the relevance of PRVs to NASA mission objectives, required telescope resources, the M dwarf opportunity, and present current and future plans for instrumentation and techniques. Among the concluding recommendations:

*"Support efforts to design Doppler measurement instruments for 20–30-m class telescopes, intended for high precision."*

*"Support efforts to obtain the observations needed to understand stellar jitter. This will require a focused and dedicated effort run by teams that have demonstrated sub-m/s precision capability."*

*"Support research to significantly improve the stability of the wavelength reference used for Doppler measurements in the optical, and to extend it to the IR. One avenue would be the implementation of femtosecond laser frequency combs, both in the optical and the IR. Another would be to extend the applicability of gas□absorption cells already used in the optical to longer wavelengths."*

*"Support the development of multi-object high-precision optical and IR Doppler instruments for wide field telescopes (such as the SDSS 2.5-m telescope with a 7 square degree field of view, or China's LAMOST 4-m wide-field telescope with a 20 square degree field of view). These instruments would have an enormous multiplexing advantage, and would be a much more efficient way of increasing the number of stars observed compared to simply adding more telescopes with single-object RV instruments."*

## 7.8 2008 Exoplanet Task Force Report

In 2008, an additional exoplanet community gathering was held in anticipation of the 2010 Decadal Survey, the result of which is the 2008 Exoplanet Task Force report chaired by Jonathan Lunine and included input from PRV community members such as Debra Fischer, Greg Laughlin, Josh Winn, Andreas Quirrenbach, David Latham and Didier Queloz. The report and recommendations synthesized the input of over 85 individual white papers submitted to the 2010 Decadal Survey.

The Task Force report recommended bolstering support for both radial velocity and transit discovery and characterization of super Earths orbiting in the habitable zones of M-dwarfs with NIR PRVs (Section 6.1):

*"Access to sufficient time with premier radial velocity instruments is currently the key bottleneck for the follow-up of transiting planets discovered by ground-based surveys and COROT, and the situation will be worse when Kepler begins producing candidates."*

*"A significant investment in ground-based telescopes, equivalent to a dedicated 4-meter facility*

*as well as additional NASA time on the Keck telescopes, could enable radial-velocity surveys to find statistically-significant numbers of exo- planets with M sin i down to 10 M□ around solar-type stars in the habitable zone. A program focussed on a selected set of late-type stars, down to early M-dwarfs, could find one or two dozen planets with M sin i approaching the mass of the Earth.”*

***“****The potential of the near-IR radial velocity technique for late M-dwarfs seems quite high but as yet unproven. ... Additional resources will need to be expended in order to make this technique operational and understand its ultimate capability.”*

***“****The value of the radial velocity technique will remain high in the latter part of the next 15 years, as improvements in accuracy allow for smaller short-period planets to be detected, and as a support mode for photometric and astrometric space-based surveys.”*

Finally, in 2008 the list of PRV spectrometers was relatively small compared to today, demonstrating vast progress and investment in instrumentation development. The report mentioned that PRV spectrometers were already over-subscribed and correctly predicted the lack of PRV resources available for Kepler Follow-up:

*“There are a handful of high resolution optical spectrometers that have demonstrated routine velocity precision approaching or better than 1 m/s (HARPS, Keck-HIRES, AAT-UCLES, VLT-UVES) and additional telescopes armed with high resolution spectrometers have demonstrated RV precision in the 2-7 m/s range (e.g., HET, Lick-Hamilton, Magellan-MIKE, Subaru-HDS, Thuringia State Observatory). The Doppler planet search surveys at all of these telescopes are currently oversubscribed and demand for high resolution spectroscopy is expected to increase, both because the detection of low mass planets requires a substantially larger number of observations and because of an increase in demand for verification from other techniques. For example, it is not clear how high precision Doppler programs will be able to follow up the flood of transit candidates expected from the Kepler mission.”*

# 8. Acronyms and Acknowledgments

| | |
|---|---|
| AAT | Anglo-Australian Telescope |
| AFTA | Astrophysically Focused Telescope Assets |
| AO | Adaptive Optics |
| AST | NSF Astrophysics Division |
| ASTRO2010 | NRC 2010 Decadal Survey in Astronomy and Astrophysics |
| CARMENES | Calar Alto high-Resolution search for M dwarfs with Exo-earths with Near-infrared and optical Échelle |
| CODEX | COsmic Dynamics and EXo-earth experiment |
| EDI | Externally Dispersed Interferometer or Interferometry |
| ELT | Extremely Large Telescope |
| ESO | European Southern Observatory |
| ESPRESSO | Echelle SPectrograph for Rocky Exoplanet and Stable Spectroscopic Observations |
| ExEP | Exoplanet Exploration Program |
| Exo-C | Exoplanet Coronagraph Probe Class mission |
| Exo-S | Exoplanet Starshade Probe Class mission |
| HARPS-N | High Accuracy Radial velocity Planet Searcher - North |
| HARPS-S | High Accuracy Radial velocity Planet Searcher - South |
| HDS | High Dispersion Spectrograph at Subaru |
| HET | Hobby Eberly Telescope |
| HIRES | High Resolution Echelle Spectrograph on Keck |
| HPF | Habitable Planet Finder |
| IR | Infrared |
| IRTF | NASA Infrared Telescope Facility |
| LAMOST | Large Sky Area Multi-Object Fibre Spectroscopic Telescope |
| LFC | Laser Frequency Comb |
| LSF | Line Spread Function |
| K2 | NASA Kepler Two-Wheel Extended Mission |
| MIKE | Magellan Inamori Kyocera Echelle Spectrograph |
| MOS | Multi-Object Spectrograph |
| NASA | National Aeronautics and Space Administration |
| NIR | Near-Infrared |
| NOAO | National Optical Astronomy Observatory |
| NWNH | New Worlds, New Horizons; See ASTRO2010 |
| NSF | National Science Foundation |
| OIR | Optical-Infrared |
| PRV | Precise Radial Velocity |
| PSF | Point Spread Function |
| PSSF | Planetary Systems & Star Formation |
| RMS | Root Mean Square (often used in quantify Doppler precision) |
| RV | Radial Velocity |
| SDSS | Sloan Digital Sky Survey |
| SED | Spectral Energy Distribution |
| SMD | Science Mission Directorate at NASA |
| SNR | Signal-to-Noise Ratio |
| TESS | Transiting Exoplanet Survey Satellite |
| TRENDS | TaRgetting bENchmark-objects with Doppler Spectroscopy |
| UCLES | University College London Echelle Spectrograph |
| UVES | Ultraviolet and Visual Echelle Spectrograph at the VLT |
| VLT | Very Large Telescope |
| WFIRST | Wide-Field InfraRed Survey Telescope |

This research has made use of the NASA Exoplanet Archive, which is operated by the California Institute of Technology, under contract with the National Aeronautics and Space Administration under the Exoplanet Exploration Program. This research has made use of the Exoplanet Orbit Database and the Exoplanet Data Explorer at exoplanets.org. Peter Plavchan and Dave Latham would like to thank the participation of the PRV community in the preparation of this report.

# 9. References

Aigrain, S., et al., 2012. "A simple method to estimate radial velocity variations due to stellar activity using photometry," MNRAS, 419, 3147

Akeson, R. L., et al., 2013. "The NASA Exoplanet Archive: Data and Tools for Exoplanet Research," PASP, 125, 989

Allard, F., Homeier, D., Freytag, B., & Sharp, C. M. 2012. "Atmospheres From Very Low-Mass Stars to Extrasolar Planets," EAS Publications Series, 57, 3

Anglada-Escude, G., & Butler, R. P., 2012a. "The HARPS-TERRA Project. I. Description of the Algorithms, Performance, and New Measurements on a Few Remarkable Stars Observed by HARPS," ApJS, 200, 15

Anglada-Escude, G., et al., 2012b. "Design and Construction of Absorption Cells for Precision Radial Velocities in the K Band Using Methane Isotopologues," PASP, 124, 586

Arentoft, T., et al., 2008. "A Multisite Campaign to Measure Solar-like Oscillations in Procyon. I. Observations, Data Reduction, and Slow Variations," ApJ, 687, 1180

Artigau, É., et al., 2014. "SPIRou: the near-infrared spectropolarimeter/high-precision velocimeter for the Canada-France-Hawaii telescope," SPIE, 9147, 15A

Artymowicz, P., & Lubow, S., 1994. "Dynamics of binary-disk interaction. 1: Resonances and disk gap sizes," ApJ, 421, 651

Asplund, M., et al., 2000. "The effects of numerical resolution on hydrodynamical surface convection simulations and spectral line formation," A&A, 359, 669

Audard, M., et al., 2000. "Extreme-Ultraviolet Flare Activity in Late-Type Stars," ApJ, 541, 396

Bahng, J., & Schwarzschild, M., 1961. "Lifetime of Solar Granules," ApJ, 134, 312

Bailey, J. I., et al., 2012. "Precise Infrared Radial Velocities from Keck/NIRSPEC and the Search for Young Planets," ApJ, 749, 16

Balthasar, H., et al.,1982. "Terrestrial O2 lines used as wavelength references - Comparison of measurements and model computations," A&A, 114, 357

Barnes, J. R., et al., 2012. "Red Optical Planet Survey: a new search for habitable earths in the southern sky," MNRAS, 424, 591

Barnes, S. A., et al., 2010. “A Simple Nonlinear Model for the Rotation of Main-sequence Cool Stars. I. Introduction, Implications for Gyrochronology, and Color-Period Diagrams,” ApJ, 722, 222

Basri, G., et al., 2010. “Photometric Variability in Kepler Target Stars: The Sun Among Stars---a First Look,” ApJL, 713, 155

Basri, G., et al., 2011. “Photometric Variability in Kepler Target Stars. II. An Overview of Amplitude, Periodicity, and Rotation in First Quarter Data,” AJ, 141, 20

Basri, G., et al., 2013. “Comparison of Kepler Photometric Variability with the Sun on Different Timescales,” ApJ, 769, 37

Bastien, F., et al., 2013. “An observational correlation between stellar brightness variations and surface gravity,” Natur, 500, 427

Bastien, F., et al., 2014. “Radial Velocity Variations of Photometrically Quiet, Chromospherically Inactive Kepler Stars: A Link between RV Jitter and Photometric Flicker.” AJ, 147, 29

Bean, J. L., et al, 2010. “The CRIRES Search for Planets Around the Lowest-mass Stars. I. High-precision Near-infrared Radial Velocities with an Ammonia Gas Cell,” ApJ, 713, 410

Bean, J. L., et al., 2011. “The Optical and Near-infrared Transmission Spectrum of the Super-Earth GJ 1214b: Further Evidence for a Metal-rich Atmosphere,” ApJ, 743, 92

Bedding, T. R., & Kjeldsen, H., 2003. “Solar-like Oscillations,” PASA, 20, 203

Bedding, T. R., & Kjeldsen, H., 2007. “Observations of solar-like oscillations,” CoAst, 150, 106

Bender, C. F., et al., 2012. “The SDSS-HET Survey of Kepler Eclipsing Binaries: Spectroscopic Dynamical Masses of the Kepler-16 Circumbinary Planet Hosts,” ApJ, 751, 31

Berta, Z. K., et al., 2012. “Transit Detection in the MEarth Survey of Nearby M Dwarfs: Bridging the Clean-first, Search-later Divide,” AJ, 144, 145

Bertaux, J. L., et al., 2014. “TAPAS, a web-based service of atmospheric transmission computation for astronomy,” A&A, 564, 46

Blake, C. H., et al., 2007. “Multiepoch Radial Velocity Observations of L Dwarfs,” ApJ, 666, 1198

Blake, C. H., et al., 2010. “The NIRSPEC Ultracool Dwarf Radial Velocity Survey,” ApJ, 723, 684

Blake, C. H. & Shaw, M. M., 2011. “Measuring NIR Atmospheric Extinction Using a Global Positioning System Receiver,” PASP, 123, 1302

Boisse, I., et al., 2011. “Disentangling between stellar activity and planetary signals,” A&A, 528, 4

Bonfils, X., et al., 2007. “The HARPS search for southern extra-solar planets. X. An *m* sin *i* = 11 $M_{Earth}$ planet around the nearby spotted M dwarf GJ 674,” A&A, 474, 293

Bonfils, X., et al., 2013. “The HARPS search for southern extra-solar planets. XXXI. The M-dwarf sample,” A&A, 549, 109

Bottom, Michael., et al., 2013. “Optimizing Doppler Surveys for Planet Yield,” PASP, 125, 240

Borgniet, S., et al., 2014. “Extrasolar planets and brown dwarfs around A-F type stars. VIII. A giant planet orbiting the young star HD 113337,” A&A, 561, 65

Bouchy, F., & Carrier, F., 2003. “Present Observational Status of Solar-type stars,” Ap&SS, 284, 21

Brown, R. A., 2009. “On the Completeness of Reflex Astrometry on Extrasolar Planets Near the Sensitivity Limit,” ApJ, 699, 711

Broomhall, A.-M., et al., 2011. “Solar-cycle variations of large frequency separations of acoustic modes: implications for asteroseismology,” MNRAS, 413 2978

Butler, R. P., et al., 1996. “Attaining Doppler Precision of 3 m/s,” PASP, 108, 500

Butler, R. P., et al., 1999. “Evidence for Multiple Companions to upsilon Andromedae,” ApJ, 631, 1215

Caccin, B., et al., 1985. “Terrestrial O2 lines used as wavelength references Experimental profiles and asymmetries vs. model computations,” A&A, 149, 357

Cegla, H. M., et al., 2012. “Stellar jitter from variable gravitational redshift: implications for RV confirmation of habitable exoplanets,” MNRAS, 421, 54

Chaplin, W. J., et al., 2011. “Ensemble Asteroseismology of Solar-Type Stars with the NASA Kepler Mission,” Sci, 332, 213

Chapman, G. A., et al., 2001. "An Improved Determination of the Area Ratio of Faculae to Sunspots," ApJ, 555, 462

Charbonneau, D. et al., 2002. "Detection of an Extrasolar Planet Atmosphere," ApJ, 568, 377

Charbonneau, D., et al., 2007. "When Extrasolar Planets Transit Their Parent Stars," prpl.conf, 701-716

Chazelas, B., 2010, "New scramblers for precision radial velocity: square and octagonal fibers," SPIE, 7739, 47

Christensen-Dalsgaard, J., 2004. "Physics of solar-like oscillations," SoPh, 220, 137

Cochran, W. D., 1998. "Confirmation of radial velocity variability in Arcturus," ApJ, 334, 339

Cosentino, R., et al., 2012. "Harps-N: the new planet hunter at TNG," SPIE, 8446, 1V

Cosentino, R., et al., 2014. "HARPS-N @ TNG, two year harvesting data: performances and results," SPIE, 9147, 8C

Cranmer, S. R., et al., 2014. "Stellar Granulation as the Source of High-Frequency Flicker in Kepler Light Curves," ApJ, 781, 124

Crepp, J. R., et al., 2012. "The TRENDS High-contrast Imaging Survey. I. Three Benchmark M Dwarfs Orbiting Solar-type Stars," ApJ, 761, 39

Crepp, J. R., et al., 2013a. "The TRENDS High-contrast Imaging Survey. II. Direct Detection of the HD 8375 Tertiary," ApJ, 771, 46

Crepp, J. R., et al., 2013b. "The TRENDS High-contrast Imaging Survey. III. A Faint White Dwarf Companion Orbiting HD 114174," ApJ, 774, 1

Crepp, J., 2014, "Improving Planet-Finding Spectrometers," Science 346, 6211

Crepp, J., et al., 2014, "The TRENDS High-contrast Imaging Survey. V. Discovery of an Old and Cold Benchmark T-dwarf Orbiting the Nearby G-star HD 19467," ApJ, 781, 29

Crockett, C. J., et al., 2012. "A Search for Giant Planet Companions to T Tauri Stars," ApJ, 761, 164

Crossfield, I., et al., 2015. "A nearby M star with three transiting super-Earths discovered by K2," ApJ, submitted

Cunha, D., et al., 2014. “Impact of micro-telluric lines on precise radial velocities and its correction,” A&A, 568, 35

Delabre, B., & Manescau, A., 2010. “CODEX optics,” SPIE, 7735, 5J

Deming, D., et al., 2005. “Infrared radiation from an extrasolar planet,” Natur, 434,740

Deshpande, R., et al., 2013. “The SDSS-III APOGEE Radial Velocity Survey of M Dwarfs. I. Description of the Survey and Science Goals,” AJ, 146, 156

Desort, M., et al., 2007. “Search for exoplanets with the radial-velocity technique: quantitative diagnostics of stellar activity,” A&A, 473, 983

Donati, J.-F., et al., 2014. “Modelling the magnetic activity and filtering radial velocity curves of young Suns : the weak-line T Tauri star LkCa 4,” MNRAS, 444, 3220

Dravins, D., et al., 1981. “Solar granulation - Influence of convection on spectral line asymmetries and wavelength shifts,” A&A, 96, 345

Dressing, C. D., et al., 2015. “The Mass of Kepler-93b and the Composition of Terrestrial Planets”, ApJ, accepted.

Dumusque, X., et al., 2011a. “Planetary detection limits taking into account stellar noise. I. Observational strategies to reduce stellar oscillation and granulation effects,” A&A, 525, 140

Dumusque, X., et al., 2011b. “Planetary detection limits taking into account stellar noise. II. Effect of stellar spot groups on radial-velocities,” A&A, 527, 82

Dumusque, X., et al., 2011c. “The HARPS search for southern extra-solar planets. XXX. Planetary systems around stars with solar-like magnetic cycles and short-term activity variation,” A&A, 535, 55

Dumusque, X., et al., 2012. “An Earth-mass planet orbiting alpha Centauri B,” Natur, 491, 207

Dumusque, X., et al., 2014. “SOAP 2.0: A Tool to Estimate the Photometric and Radial Velocity Variations Induced by Stellar Spots and Plages,” ApJ, 796, 132

Eastman, J. D., et al., 2014. "NRES: the network of robotic Echelle spectrographs,” SPIE, 9147, 16

Figueira, P., et al., 2010a. “Evidence against the young hot-Jupiter around BD +20 1790,” A&A, 513, 8

Figueira, P., et al., 2010b. “Radial velocities with CRIRES. Pushing precision down to 5-10 m/s,” A&A, 511, 55

Figueira, P., et al., 2012. “Comparing radial velocities of atmospheric lines with radiosonde measurements,” MNRAS, 420, 287

Figuera, P., et al., 2013. “Line-profile variations in radial-velocity measurements. Two alternative indicators for planetary searches,” A&A, 557, 93

Fortney, J. J., et al., 2008. “Synthetic Spectra and Colors of Young Giant Planet Atmospheres: Effects of Initial Conditions and Atmospheric Metallicity,” ApJ, 683, 1104

Frazier, E. N., 1971. “Multi-Channel Magnetograph Observations. III: Faculae,” SoPh, 21 42

Fressin, F., et al., 2013. “The False Positive Rate of Kepler and the Occurrence of Planets,” ApJ, 766, 81

Frohlich, C., & Lean, J., 2004. “Solar radiative output and its variability: evidence and mechanisms,” A&ARv, 12, 273

Galvagni, M., et al., 2012. “The collapse of protoplanetary clumps formed through disc instability: 3D simulations of the pre-dissociation phase,” MNRAS, 427, 1725

Gaudi., B. S., & Winn, J., N., 2007. “Prospects for the Characterization and Confirmation of Transiting Exoplanets via the Rossiter-McLaughlin Effect” ApJ, 655, 550

Gaudi., B. S., et al., 2008. “Discovery of a Jupiter/Saturn Analog with Gravitational Microlensing,” Sci, 319, 927

Gilliland, R. L., et al., 2011. “Kepler Mission Stellar and Instrument Noise Properties,” ApJS, 197, 6

Gomes da Silva, J., et al., 2011. “Long-term magnetic activity of a sample of M-dwarf stars from the HARPS program. I. Comparison of Activity Indices,” A&A, 534, A30

Gomes da Silva, J., et al., 2012. “Long-term magnetic activity of a sample of M-dwarf stars from the HARPS program. II. Activity and radial velocity,” A&A, 541, 9

Gould, A., et al., 2010. “Frequency of Solar-like Systems and of Ice and Gas Giants Beyond the Snow Line from High-magnification Microlensing Events in 2005-2008,” ApJ, 720, 1073

Grasset, O., et al., 2009. “A Study of the Accuracy of Mass-Radius Relationships for Silicate-Rich and Ice-Rich Planets up to 100 Earth Masses,” ApJ, 693, 722

Griffin, R., & Griffin, R., 1973. “Accurate wavelengths of stellar and telluric absorption lines near lambda 7000 Angstroms,” MNRAS, 162, 255

Gullikson, K., et al., 2014. "Correcting for Telluric Absorption: Methods, Case Studies, and Release of the TelFit Code," AJ, 148, 53

Halverson, S., et al., 2014. "Development of Fiber Fabry-Perot Interferometers as Stable Near-infrared Calibration Sources for High Resolution Spectrographs," PASP, 126, 445

Haywood, R. D., et al., 2014. "Planets and Stellar Activity: Hide and Seek in the CoRoT-7 system," MNRAS, 443, 2517

Helled, R., et al., 2013. "Giant Planet Formation, Evolution, and Internal Structure," arXiv1311.1142, published in Protostars and Planets VI

Hernán-Obispo, M., et al., 2010. "Evidence of a massive planet candidate orbiting the young active K5V star BD+20 1790," A&A, 512, 45

Hillenbrand, L., et al., 2014. "Empirical Limits on Radial Velocity Planet Detection for Young Stars," the 18th Cambridge Workshop on Cool Stars, Stellar Systems, and the Sun proceedings edited by G. van Belle & H. Harris, arXiv:1408.3475

Howard, A., et al., 2010. "The California Planet Survey. I. Four New Giant Exoplanets," ApJ, 721, 1467

Howard, A. W., et al., 2012. "Planet Occurrence within 0.25 AU of Solar-type Stars from Kepler," ApJS, 201, 15

Howard, A. W., et al., 2013. "A rocky composition for an Earth-sized exoplanet," Natur, 503, 381

Howard, A. W. & Fulton, B. J., 2014. "Limits on Planetary Companions from Doppler Surveys of Nearby Stars," reported submitted to the NASA Exoplanet Exploration Program Office on October 24th, 2014, https://www.dropbox.com/sh/d1x58leszejt5f3/AACQXgwqLEzT7jMsRWA8_Nxha?dl=0.

Huélamo, N., et al., 2008. "TW Hydrae: evidence of stellar spots instead of a Hot Jupiter," A&A, 489, 9

Huerta, M., et al., 2008. "Starspot-Induced Radial Velocity Variability in LkCa 19," ApJ, 678, 472

Johnson, J. A., et al., 2011. "Retired A Stars and Their Companions. VII. 18 New Jovian Planets," ApJS, 197, 26

Kausch, W., et al., 2014. "Molecfit: A Package for Telluric Absorption Correction," ASPC, 485, 403

Kjeldsen, H., & Bedding, T. R.,1995. “Amplitudes of stellar oscillations: the implications for asteroseismology,” A&A, 293, 87

Kjeldsen, H., & Bedding, T. R., 2011. “Amplitudes of solar-like oscillations: a new scaling relation,” A&A, 529L, 8

Knutson, H., et al., 2007. “A map of the day-night contrast of the extrasolar planet HD 189733b,” Natur, 447, 183

Lagrange, A.-M., et al., 2010. “Using the Sun to estimate Earth-like planets detection capabilities . I. Impact of cold spots,” A&A, 512, 38

Lagrange, A.-M., et al., 2013. “Planets around stars in young nearby associations. Radial velocity searches: a feasibility study and first results,” A&A, 559, 83

Langellier, N., et al., 2014. “Green astro-comb for HARPS-N,” SPIE, 9147, 8N

Latham, D. W., et al.,1989. “The unseen companion of HD114762 - A probable brown dwarf,” Natur, 339, 38

Latham, D. W., et al. 2011. “A First Comparison of Kepler Planet Candidates in Single and Multiple Systems,” ApJL, 732, 24

Laughlin, G. et al., 2005. “The GJ 876 Planetary System: A Progress Report,” APJL, 622, 1182
Léger, A., et al., 2009. “Transiting exoplanets from the CoRoT space mission. VIII. CoRoT-7b: the first super-Earth with measured radius,” A&A, 506, 287

Li, C.-H., et al., 2008. “A laser frequency comb that enables radial velocity measuremnts with a precision of 1 cm $s^{-1}$,” Nature, 452, 610

Lin, D. N. C., et al., 1996. “Orbital migration of the planetary companion of 51 Pegasi to its present location,” Nature, 380, 606

Lin, D. N. C., & Ida, S., 2007. “On the Origin of Massive Eccentric Planets,” ApJ, 477, 781

Livingston, W., & Wallace, L., 1991. “An Atlas of the Solar Spectrum in the Infrared from 1850 to 9000 $cm^{-1}$,” NSO Technical Report (Tucson, AZ: National Solar Observatory, National Optical Astronomy Observatory)

Lockwood, G. W., et al., 1992. “Long-term solar brightness changes estimated from a survey of Sun-like stars,” Natur, 360, 653

Lockwood, G. W., et al., 2007. “Patterns of Photometric and Chromospheric Variation among Sun-like Stars: A 20 Year Perspective,” ApJS, 171, 260

Lockwood, A. C., et al., 2014. “Near-IR Direct Detection of Water Vapor in Tau Boötis b,” ApJ, 738, 29

Lovis, C., et al., 2006. “The exoplanet hunter HARPS: unequalled accuracy and perspectives toward 1 cm s $^{-1}$ precision,” SPIE, 6269, 0P

Lovis, C., et al., 2011. “The HARPS search for southern extra-solar planets. XXXI. Magnetic activity cycles in solar-type stars: statistics and impact on precise radial velocities,” arXiv1107.5325

Lovis, C., & Fischer, D., 2011. “Radial Velocity Techniques for Exoplanets,” Exoplanets, edited by S. Seager. Tucson, AZ: University of Arizona Press, 526 pp. ISBN 978-0-8165-2945-2., p.27-53

McCracken, et al., 2014. “Single-lock: A Stable Fabry-Perot based wavelength calibrator,” SPIE, 9147, 3L

McQuillen, A., et al., 2014. “Rotation Periods of 34,030 Kepler Main-Sequence Stars: The Full Autocorrelation Sample,” ApJS, 211, 24

Mahadevan, S., et al., 2010. “The habitable zone planet finder: a proposed high-resolution NIR spectrograph for the Hobby Eberly Telescope to discover low-mass exoplanets around M dwarfs,” SPIE, 7735, 6X

Mahadevan, S., & Ge, J., 2009. “The Use of Absorption Cells as a Wavelength Reference for Precision Radial Velocity Measurements in the Near-Infrared,” ApJ, 692 1590

Mahadevan, S., et al., 2012. “The habitable-zone planet finder: a stabilized fiber-fed NIR spectrograph for the Hobby-Eberly Telescope,” SPIE, 8446, 1S

Makarov, V. V., et al., 2009. “Starspot Jitter in Photometry, Astrometry, and Radial Velocity Measurements,” ApJL, 707, 73

Makarov, V.V., et al., 2010. “Astrometric Jitter of the Sun as a Star,” ApJ, 717, 1202

Mamajek, E. E., & Hillenbrand, L. A., “Improved Age Estimation for Solar-Type Dwarfs Using Activity-Rotation Diagnostics,” 2008, 687, 1264

Mandell, A. M., et al., 2011. “Non-detection of L-band Line Emission from the Exoplanet HD189733b,” ApJ, 728, 18

Marcus, R. A., 2010. “Minimum Radii of Super-Earths: Constraints from Giant Impacts,” ApJ, 712, 73

Marcy, G. W., et al., 2014. “Masses, Radii, and Orbits of Small Kepler Planets: The Transition from Gaseous to Rocky Planets,” ApJS, 210, 20

Marcy, G. W., & Butler, R. P., 1996. “A Planetary Companion to 70 Virginis,” ApJL, 464, 147

Marcy, G. W., et al. 2002. “A Planet at 5 AU around 55 Cancri,” ApJ, 581, 1375

Marley, M. S., et al., 2007. “On the Luminosity of Young Jupiters,” ApJ, 655,541

Marois, C., et al., 2008. “Direct Imaging of Multiple Planets Orbiting the Star HR 8799,” Science, 322, 1348

Marois, C., et al., 2010. “Images of a fourth planet orbiting HR 8799,” Natur, 468, 1080

Mayor, M., et al., 2011. “The HARPS search for southern extra-solar planets XXXIV. Occurrence, mass distribution and orbital properties of super-Earths and Neptune-mass planet,” arXiv eprint: 1109.2497

Mayor, M., et al., 2003. “Setting new standards with HARPS,” Msngr, 114, 20

Mayor, M., & Queloz, D., 1995. “A Jupiter-mass companion to a solar-type star,” Natur, 378, 355

McQuillan, A., et al., 2012. “Statistics of stellar variability from Kepler. I. Revisiting Quarter 1 with an astrophysically robust systematics correction,” A&A, 539, 137

Meunier, N., et al., 2010a. “Using the Sun to estimate Earth-like planets detection capabilities. II. Impact of plages,” A&A, 512 , 39

Meunier, N., et al., 2010b. “Reconstructing the solar integrated radial velocity using MDI/SOHO,” A&A, 519, 66

Meunier, N., & Lagrange, A.M. 2013. “Using the Sun to study the impact of stellar activity on exoplanet detectability,” AN, 334, 141

Nelder, J., & Mead, R., 1965, Computer Journal 7, 308

Osterman, S., et al., 2011. “Laser Frequency Comb Supported Stellar Radial Velocity Determination in the NIR: Initial Results,” ESS, 20205

Pasquini, L., et al., 2008. “CODEX: the high-resolution visual spectrograph for the E-ELT,” SPIE, 7014, 1I

Pasquini, L., et al., 2010a, “CODEX: An Ultra-stable High Resolution Spectrograph for the E-ELT,” Msngr, 140, 20

Pasquini, L., et al., 2010b. “Codex,” SPIE, 7735, 2F

Paulson, D. B., et al., 2004. “Searching for Planets in the Hyades. V. Limits on Planet Detection in the Presence of Stellar Activity,” AJ, 127, 3579

Paulson, D. B., & Yelda, S., 2006. “Differential Radial Velocities and Stellar Parameters of Nearby Young Stars,” PASP, 118, 706

Payne, M. J., et al., 2009. “Outward migration of terrestrial embryos in binary systems,” MNRAS, 400, 1936

Pepe, F., et al., 2000. “HARPS: a new high-resolution spectrograph for the search of extrasolar planets,” SPIE, 4008, 582

Pepe, F., et al., 2002. “The CORALIE survey for southern extra-solar planets VII. Two short-period Saturnian companions to HD 108147 and HD 168746,” A&A, 388, 632

Pepe, F., et al., 2003. “Performance verification of HARPS: first laboratory results,” SPIE, 4841, 1045

Pepe, F., et al., 2011. “The HARPS search for Earth-like planets in the habitable zone. I. Very low-mass planets around HD 20794, HD 85512, and HD 192310,” A&A, 534, 58

Pepe, F., et al., 2010. “ESPRESSO: the Echelle spectrograph for rocky exoplanets and stable spectroscopic observations,” SPIE, 7735, 0F

Pepe, F., 2013. “An Earth-sized planet with an Earth-like density,” Nature, 503, 377

Pepe, F. A., & Lovis, C., 2008. “From HARPS to CODEX: exploring the limits of Doppler measurements,” PhST, 130, 014007

Perruchot, S., et al., 2008. “The SOPHIE spectrograph: design and technical key-points for high throughput and high stability,” SPIE, 7014, 0J

Plavchan, P., Chen, Xi., & Pohl, G., 2015. “What is the Mass of alpha Centauri B b?” ApJ, submitted

Plavchan, P., et al., 2013a. “Precision near-infrared radial velocity instrumentation I: absorption gas cells,” SPIE, 8864, 1J

Plavchan, P., et al., 2013b. “Precision near-infrared radial velocity instrumentation II: noncircular core fiber scrambler,” SPIE, 8864, 0G

Prato, L., et al., 2008. “A Young-Planet Search in Visible and Infrared Light: DN Tauri, V836 Tauri, and V827 Tauri,” ApJ, 687,103

Queloz, D., et al., 1999. “The CORALIE survey for southern extra-solar planets. I. A planet orbiting the star Gliese 86,” A&A, 354, 99

Queloz, D., et al., 2000. “Detection of a spectroscopic transit by the planet orbiting the star HD209458,” A&A, 359, 13

Queloz, D., et al., 2001. “No planet for HD 166435,” A&A, 379, 279

Queloz, D., et al., 2009. “The CoRoT-7 planetary system: two orbiting super-Earths,” A&A, 506, 303

Quinn, S. N., et al., 2012. “Two "b"s in the Beehive: The Discovery of the First Hot Jupiters in an Open Cluster,” ApJL, 756, 33

Quinn, S. N., et al., 2014. “HD 285507b: An Eccentric Hot Jupiter in the Hyades Open Cluster,” ApJ, 787, 27

Quirrenbach, A., et al., 2012. “CARMENES. I: instrument and survey overview,” SPIE, 8446, 0R

Radick, R. R., et al., 1998. “Patterns of Variation among Sun-like Stars,” ApJS, 118, 239

Radick, R. R., 2003. “Variability of sunlike stars,” JASTP, 65,105

Radovan, M. V., et al., 2010. “A radial velocity spectrometer for the Automated Planet Finder Telescope at Lick Observatory,” SPIE, 7735, 4K

Rayner, J., et al., 2012. “iSHELL: a 1-5 micron cross-dispersed R=70,000 immersion grating spectrograph for IRTF,” SPIE, 8446, 2C

Reiners, A., et al., 2010. “Detecting Planets Around Very Low Mass Stars with the Radial Velocity Method,” ApJ, 710, 432

Ricker, G., et al., 2014. “Transiting Exoplanet Survey Satellite (TESS),” SPIE, 9143, 20

Robertson, P., & Mahadevan, S., 2014. “Disentangling Planets and Stellar Activity for Gliese 667C,” ApJL, 793, 24

Robertson, P., et al., 2014. “Stellar activity masquerading as planets in the habitable zone of the M dwarf Gliese 581,” Science, 345, 440

Rodler, F., et al., 2012a. “Weighing the Non-transiting Hot Jupiter tau Boo b,” ApJL 753, 25

Rodler, F., et al., 2012b. “Search for radial velocity variations in eight M-dwarfs with NIRSPEC/Keck II,” A&A, 538, 141

Rupprecht, G., et al., 2004. “The exoplanet hunter HARPS: performance and first results,” SPIE, 5492, 148

Saar, S. H., et al.,1998. “Magnetic Activity-related Radial Velocity Variations in Cool Stars: First Results from the Lick Extrasolar Planet Survey,” ApJL, 498, 153

Saar, S. H., 2009. “The Radial Velocity Effects of Stellar Surface Phenomena,” AIPC, 1094, 152

Sanchis-Ojeda, R., et al., 2012. “Alignment of the stellar spin with the orbits of a three-planet system,” Natur, 487, 449

Schrijver, C. J., & Zwann, C., 2000. “Solar and stellar magnetic activity,” Cambridge Univ. Press

Seager, S., et al., 2007. “Mass-Radius Relationships for Solid Exoplanets,” ApJ, 669, 1279

Seifahrt, A., et al., 2010. “Synthesising, using, and correcting for telluric features in high-resolution astronomical spectra . A near-infrared case study using CRIRES,” A&A, 524, 11

Seifahrt, A., & Kaufl, H. U. 2008. “High precision radial velocity measurements in the infrared. A first assessment of the RV stability of CRIRES,” A&A, 491, 929

Setiawan, J., et al., 2007. “Evidence for a Planetary Companion around a Nearby Young Star,” ApJ, 660, 145

Skumanich, A., 1972. “Time Scales for CA II Emission Decay, Rotational Braking, and Lithium Depletion,” ApJ, 171, 565

Smith, M. A.,1982. “Precise radial velocities. I - A preliminary search for oscillations in Arcturus,” ApJ, 253, 727

Snellen, I. A. G., 2004. “A new method for probing the atmospheres of transiting exoplanets,” MNRAS, 353, 1

Snellen, I. A. G., et al., 2010. “The orbital motion, absolute mass and high-altitude winds of exoplanet HD209458b,” Natur, 465, 1049

Snellen, I. A. G., et al. 2013. “Finding Extraterrestrial Life Using Ground-based High-dispersion Spectroscopy,” ApJ, 764, 182

Spanò, P., et al., 2008. “New design approaches for a very high resolution spectrograph for the combined focus of the VLT,” SPIE, 7014, 0M

Spanò, P., et al., 2012. “Very high-resolution spectroscopy: the ESPRESSO optical design,” SPIE, 8446, 7V

Swift, J., et al., 2015. “Miniature Exoplanet Radial Velocity Array (MINERVA) I. Design, Commissioning, and First Science Results,” SPIE/JATIS, submitted

Szentgyorgyi, A. H., & Furesz, G. 2007. “Precision Radial Velocities for the Kepler Era,” in The 3rd Mexico-Korea Conference on Astrophysics: Telescopes of the Future and San Pedro Martir, edited by S. Kurtz, vol. 28 of Rev. Mexicana Astron. Astrofis. Conf. Ser., 129

Szentgyorgyi, A., et al., 2012. “The GMT-CfA, Carnegie, Catolica, Chicago Large Earth Finder (G-CLEF): a general purpose optical echelle spectrograph for the GMT with precision radial velocity capability,” SPIE, 8446, 1H

Tanner, A., et al., 2012. “Keck NIRSPEC Radial Velocity Observations of Late-M Dwarfs,” ApJS, 203, 10

Thibault S., et al., 2012. “SPIRou @ CFHT: spectrograph optical design,” SPIE, 8446, 30T

Title, A. M., et al., 1989. “Statistical properties of solar granulation derived from the SOUP instrument on Spacelab,” ApJ, 336, 475

Tokovinin, A., et al., 2013. “CHIRON - A Fiber Fed Spectrometer for Precise Radial Velocities,” PASP, 125, 13336

Triaud, A.H., et al., 2009. “The Rossiter-McLaughlin effect of CoRoT-3b and HD 189733b,” A&A, 506, 377

Tull, R. G., et al., 1998. “High Resolution Spectrograph for the Hobby-Eberly Telescope,” SPIE, 193, 1008

Unruh, Y. C., et al., 1999. “The spectral dependence of facular contrast and solar irradiance variations,” A&A, 345, 635

van Eyken, J. C., et al., 2012. “The PTF Orion Project: A Possible Planet Transiting a T-Tauri Star,” ApJ, 755, 42

Vanderburg, A., et al., 2015. In preparation

Vogt, S. S., et al., 1986. “The Lick Observatory Hamilton Echelle Spectrometer,” PASP, 99, 1214

Vogt, S. S., et al., 1994. “HIRES: the high-resolution echelle spectrometer on the Keck 10-m Telescope,” SPIE, 2198, 362

Vogt, S. S., et al., 2014. “APF-The Lick Observatory Automated Planet Finder,” PASP, 126, 359

Winn, J. N., et al., 2005. “Measurement of Spin-Orbit Alignment in an Extrasolar Planetary System,” ApJ, 631, 1215

Winn, J. N., 2014. “Requirements on telescope time for measuring planet masses,” TESS Science Memo No. 12, Version 3, April 17th, 2014

Wright, J. T., 2005. “Radial Velocity Jitter in Stars from the California and Carnegie Planet Search at Keck Observatory,” PASP, 117, 657

Wright, J. T., et al., 2011. “The Exoplanet Orbit Database,” PASP, 123, 412

Yi, X., et al., 2015. “Demonstration of a Near-IR Laser Comb for Precision Radial Velocity Measurements in Astronomy,” in prep.

Yuk, I.-S., et al., 2010. “Preliminary design of IGRINS (Immersion GRating INfrared Spectrograph),” SPIE, 7735, 1M